\documentclass[showpacs,twocolumn,superscriptaddress,pra,aps,%
longbibliography]{revtex4-1}

\usepackage{amsmath}
\usepackage{amssymb}
\usepackage{maybemath}
\usepackage{xcolor}
\usepackage{graphicx}
\usepackage{bm}
\usepackage{comment}
\usepackage{maybemath}

\usepackage{dcolumn}
\newcolumntype{.}{D{.}{.}{-1}}

\usepackage[%
colorlinks=true,
urlcolor=blue,
linkcolor=blue,
citecolor=blue
]{hyperref}

\definecolor{kellygreen}{rgb}{0.3, 0.73, 0.09}

\definecolor{garrosgreen}{rgb}{0.1, 0.4, 0.1}
\definecolor{dartmouthgreen}{rgb}{0.05, 0.5, 0.06}

\definecolor{duelferred}{rgb}{0.7, 0.2, 0.1}
\definecolor{cambridgeblue}{rgb}{0.1, 0.3, 1.0}
\definecolor{oxfordblue}{rgb}{0.05, 0.2, 0.7}

\definecolor{gold}{rgb}{0.85,.66,0}

\def\dd{{\mathrm{d}}}
\def\ii{{\mathrm{i}}}
\def\ee{{\mathrm{e}}}

\def\vdw{van der Waals}
\def\cp{Casimir--Polder}

\def\calL{{\mathcal L}}
\def\calF{{\mathcal F}}

\begin{document}

\newcommand{\addrROLLA}{Department of Physics,
Missouri University of Science and Technology,
Rolla, Missouri 65409-0640, USA}

\newcommand{\addrHDphiltheo}{Institut f\"ur Theoretische Physik,
Universit\"{a}t Heidelberg,
Philosophenweg 16, 69120 Heidelberg, Germany}

\newcommand{\addrLEBEDEV}{P. N. Lebedev Physics
Institute, Leninsky prosp.~53, Moscow, 119991 Russia}

\newcommand{\addrMUC}{Max--Planck--Institut f\"ur 
Quantenoptik, Hans--Kopfermann-Stra\ss{}e~1, 
85748 Garching, Germany}

\newcommand{\addrBUDKER}{Budker Institute of Nuclear Physics, 
630090 Novosibirsk, Russia}

\newcommand{\addrRQC}{Russian Quantum Center, 
Business-center ``Ural'', 100A Novaya street,
Skolkovo, Moscow, 143025 Russia}

\title{Long-range interactions of hydrogen atoms in excited states.~I.\\
\texorpdfstring{$\bm{2S}$--$\bm{1S}$}{2S--1S} interactions and 
Dirac--\texorpdfstring{$\bm{\delta}$}{delta} perturbations}

\author{C. M. Adhikari}
\affiliation{\addrROLLA}

\author{V. Debierre}
\affiliation{\addrROLLA}

\author{A. Matveev}
\affiliation{\addrLEBEDEV}
\affiliation{\addrMUC}

\author{N. Kolachevsky}
\affiliation{\addrLEBEDEV}
\affiliation{\addrMUC}
\affiliation{\addrRQC}

\author{U. D. Jentschura}
\affiliation{\addrROLLA}

\begin{abstract}
The theory of the long-range interaction of metastable excited atomic states with 
ground-state atoms is analyzed. 
We show that the long-range interaction is essentially
modified when quasi-degenerate states are available for virtual
transitions.
A discrepancy in the literature regarding the \vdw{} coefficient 
$C_6(2S;1S)$ describing the interaction of metastable atomic hydrogen 
($2S$ state) with a ground-state hydrogen atom is resolved.
In the the \vdw{} range $a_0 \ll R \ll a_0/\alpha$,
where $a_0= \hbar/(\alpha m c)$ is the Bohr radius and
$\alpha$ is the fine structure constant,
one finds the symmetry-dependent result
$E_{2S;1S}(R) \approx (-176.75 \pm 27.98) \, E_h \, 
(a_0/R)^6$ ($E_h$ denotes the Hartree energy).
In the \cp{} range $a_0/\alpha \ll R \ll \hbar c/\calL$,
%DEFINING LAMB SHIFT HERE
where $\calL\equiv E\left(2S_{1/2}\right)-E\left(2P_{1/2}\right)$ is the Lamb shift
energy, one finds 
$E_{2S;1S}(R) \approx (-121.50 \pm 46.61) \, E_h \, (a_0/R)^6$.
In the the Lamb shift range $R \gg \hbar c/{\cal L}$,
we find an oscillatory tail with a negligible interaction energy 
below $10^{-36} \, {\rm Hz}$.
Dirac--$\delta$ perturbations to the 
interaction are also evaluated and results are 
given for all asymptotic distance ranges;
these effects describe the 
hyperfine modification of the interaction,
or, expressed differently, the shift of the hydrogen
$2S$ hyperfine frequency due to interactions with 
neighboring $1S$ atoms. The $2S$ hyperfine frequency has recently 
been measured very accurately in atomic beam experiments.
\end{abstract}


\maketitle

% \tableofcontents

\newpage

%
% Introduction
%
\section{Introduction}
\label{sec1}

The purpose of this paper is twofold.  First, we aim to revisit the calculation
of the long-range (\vdw{} and \cp{} interaction)
for ground-state hydrogen interacting with an excited-state 
atom in an $2S$ state. Second,
we aim to study the perturbation of the \vdw{} interactions by a Dirac-$\delta$
potential perturbing the metastable excited state which participates in the
interaction. Such a Dirac-$\delta$ potential
can be due to the electron-nucleus (hyperfine) interaction 
in one of the atoms~\cite{JeYe2006} or due to a
self-energy radiative correction~\cite{Be1947}.  Special emphasis is laid on
the role of quasi-degenerate levels and on the exchange term, which is due to
the possibility of $1S$--$2S$ atoms becoming a $2S$--$1S$ pair after the
exchange of two virtual photons~\cite{Ch1972}. 

It is interesting to notice that two results given in the literature for the
so-called van der Waals coefficient of the $1/R^6$ nonretarded interaction
between $1S$ and $2S$ states, are in significant mutual disagreement
(numerically, the authors of Refs.~\cite{DeYo1973} obtain 
a value of roughly $177$ in atomic units, 
while a result of about $57$ has been derived
in Refs.~\cite{Ch1972,TaCh1986}). We attempt
a thorough analysis of the discrepancy.  Two different methods of calculation
were employed in Ref.~\cite{DeYo1973} (direct sum over virtual atomic states,
including the continuum) and Refs.~\cite{Ch1972,TaCh1986} (integration over analytic
expressions representing the polarizability). 

The role of the virtual, quasi-degenerate $2P$ states deserves special
attention.  For $2S$ reference states, the $2P_{1/2}$ and $2P_{3/2}$
levels are displaced only by the Lamb shift and fine structure,
respectively.
Significant modifications of the long-range interactions result from the
presence of the quasi-degenerate states. 

Recently, precision measurements of the $2S$ hyperfine splitting have been
carried out using an atomic beam consisting of a mixture of ground-state $1S$
hydrogen atoms, and metastable $2S$ atoms \cite{KoFiKaHa2004,KoEtAl2009}.  To
leading order, the van der Waals interaction shifts all hyperfine structure
components equally.  However, there is a correction to the \vdw{} interaction
due to the the hyperfine structure (HFS), which depends on the total
(electron$+$nucleus) angular momentum quantum number $F$.  This correction
shifts HFS components closer to each other.  This latter effect is analyzed in
the current paper; it is of phenomenological significance because of \vdw{}
interactions inside the atomic beam.  Again, special attention is required in
the treatment of the quasi-degenerate atomic levels.

Let us recall here that the general subject of long-range interactions of
simple atoms is very well known to the physics community, and a 
few investigations on simple atomic systems can be found in
Refs.~\cite{DaDa1966,Da1967,RaLyDa1968a,RaLyDa1968b,RaDa1969,RaIkDa1970,%
DaDrVi1968,DuDuDa1970,%
DeYo1973,TaCh1986,YaBaDaDr1996,YaDaBa1997}.  Various aspects of the problem
have been studied in depth: e.g., the importance of multipole mixing effects,
and of perturbations by hyperfine effects, has been stressed in
Ref.~\cite{RaLyDa1968a,RaLyDa1968b,RaDa1969,RaIkDa1970}. 
Higher-order effects such as dipole-octupole mixing
terms were discussed in detail for hydrogen in Ref.~\cite{TaCh1986} and for
helium in Ref.~\cite{YaBaDaDr1996}.  The dipole-dipole interaction potential of
helium, including retardation, has been discussed in great detail in
Refs.~\cite{JaDrDa1995,ChCh1996}, including a number of numerical examples.
More complex alkali-metal dimers have been considered in
Refs.~\cite{MaSaDa1994,MaDa1995}.

Throughout this article, 
we work in SI mksA units and keep all factors of 
$\hbar$ and $c$ in the formulas. 
With this choice, 
we attempt to enhance the accessibility of the 
presentation to two different communities, 
namely, the quantum electrodynamics (QED) community which in general uses the natural 
unit system, and the atomic physics community where the
atomic unit system is canonically employed.
In the former, one sets $\hbar = c = \epsilon_0 = 1$, 
and the electron mass is denoted as $m$.
The relation $e^2 = 4\pi\alpha$ then allows to 
identify the expansion in the number of quantum 
electrodynamic corrections with powers of the 
fine structure constant $\alpha \approx 1/137.036$.
This unit system is used, 
e.g., in the investigation reported in Ref.~\cite{Pa2005longrange}
on relativistic corrections to the Casimir--Polder 
interaction (with a strong overlap with QED).
In the atomic unit system, we have 
$\left|e\right| = \hbar = m = 1$, and $4 \pi \epsilon_0 = 1$.
The speed of light, in the atomic unit system, is 
$c = 1/\alpha \approx 137.036$. This system of units 
is especially useful for the analysis of purely 
atomic properties without radiative effects.
As the subject of the current study lies in between 
the two mentioned fields of interest, we choose the 
SI unit system as the most appropriate reference frame
for our calculations. The formulas do not become unnecessarily 
complex, and can be evaluated with ease for any 
experimental application.

We organize this paper as follows. The problem is somewhat involved, as such,
we attempt to orient ourselves in Sec.~\ref{orient}. The direct term in the
$2S$--$1S$ interaction is analyzed in Sec.~\ref{long-range}.  In
Sec.~\ref{subsec:CloseD}, we study that interaction in the \vdw{} range. The
very-large-distance limit is discussed in Sec.~\ref{subsec:AwayD} (atomic
distance larger than the wavelength of the Lamb shift transition), and the
intermediate Casimir--Polder range in Sec.~\ref{subsec:MidD}.  The mixing term
in the $2S$--$1S$ interaction is analyzed in Sec.~\ref{exchange}.  We then
analyze a Dirac-$\delta$ (HFS induced) induced modification both for the
$2S$--$1S$ interaction as well as for the $1S$--$1S$ interaction in
Sec.~\ref{delta}. In Sec.~\ref{sec:Numerics}, we numerically evaluate the shift
of the $2S$ hyperfine frequency due to the long-range interaction with a ground
state hydrogen atom.  Conclusions are drawn in Sec.~\ref{conclu}.

%
% Orientation
%
\section{Orientation}
\label{orient}

In order to evaluate the \vdw{} correction to the $2S$--$1S$ 
hyperfine frequency, one needs to diagonalize the total Hamiltonian 
\begin{equation} \label{eq:TotH}
H_{\rm total} = H_S + H_{\rm FS} + H_{\rm LS} + H_{\rm HFS} + H_{\rm vdW} \,.
\end{equation}
Here, $H_S$ is the Schr\"{o}dinger Hamiltonian,
$H_{\rm FS}$ is the fine structure Hamiltonian, which can be approximated as
(see Chap.~34 of Ref.~\cite{BeLiPi1982vol4})
\begin{align}
\label{eq:FineSH}
H_{\rm FS}= & \; 
\sum_{i=A,B} \left[ -\frac{\vec{p}_i^{\,4}}{8m^3c^2}
+ \frac{1}{2} \alpha\left(\frac{\hbar^2 g_s}{2m^{2} \, c}\right)
\frac{\vec{L}_i\cdot\vec{S}_i}{| \vec r_i |^3}\right.
\nonumber\\
& \; \left.
+ \frac{\hbar^{3}}{8m^{2}c}\,4\pi\alpha \, 
\delta^{(3)}(\vec{r}_i) \right],
\end{align}
where $m$ is the electron mass, the $\vec{p}_i$ denote the 
momenta of the two atomic electrons
relative to their nuclei
($i$ runs over the atoms $A$ and $B$), and the 
$\vec{r}_i = \vec x_i - \vec R_i$ denote the 
coordinates relative to the nuclei
(the electron and nucleus coordinates 
are $\vec x_i$ and $\vec R_i$, respectively).
We restrict the discussion to neutral 
hydrogen atoms and thus assume a nuclear charge number of $Z = 1$. 
We shall use the following approximation to the
``Lamb shift Hamiltonian'', which constitutes an 
effective Hamiltonian useful in the 
evaluation of the leading radiative 
correction to dynamic processes~\cite{KaIv1997optspec,Je2004rad},
\begin{equation} 
\label{HLS}
H_{\mathrm{LS}}=\sum_{i=A,B}\frac{4}{3}\,\alpha^2mc^2
\left(\frac{\hbar}{mc}\right)^3\ln\left(\alpha^{-2}\right) \,
\delta^{\left(3\right)}\left(\vec{r}_i\right) \,.
\end{equation}
We shall use this Hamiltonian later in the 
analysis of the radiative correction to the 
long-range interatomic interaction.
The Hamiltonian for the hyperfine interaction~\cite{ItZu1980,JeYe2006}
reads as
\begin{align} 
\label{eq:HyperFSH}
H_{\rm HFS}= & \; 
\frac{\mu_0}{4\pi} \;
\mu_B \, \mu_N \, g_s \, g_p
\sum_{i=A,B}
\left[ \frac{8\pi}{3} \,
\vec{S}_i\cdot\vec{I}_i\,
\delta^{(3)}(\vec{r}_i) \right.
\nonumber\\
& \; \left. + 
\frac{3 \, ( \vec{S}_i \cdot \hat{r}_i )
( \vec{I}_i \cdot \hat{r}_i ) - 
( \vec{S}_i \cdot \vec{I}_i )}{| \vec r_i|^3}
+ \frac{\vec{L}_i\cdot\vec{\mu}_i}{\hbar|\vec{r}_i|^3}
\right] \,.
\end{align}
Here, the unit vectors are $\hat r_i = \vec r_i / | \vec r_i |$.
The spin operator for the electron $i$ is $\vec{S}_i = \vec\sigma_i/2$,
while $\vec{I}_i$ is the spin operator for
proton $i$ (both spin operators are dimensionless).
The electronic and protonic $g$ factors
are $g_s\simeq2.002\,319$ and $g_p\simeq5.585\,695$, 
while $\mu_B\simeq9.274\,010\,\times10^{-24}\,\mathrm{A m}^2$ is the Bohr magneton 
and $\mu_N\simeq5.050\,784\,\times10^{-27}\,\mathrm{A m}^2$ is the nuclear 
magneton~\cite{MoTaNe2012}. It is well known that, for $S$ states, the second term in the
fine structure Hamiltonian~(\ref{eq:FineSH}), and the second and third 
terms in the hyperfine structure Hamiltonian~(\ref{eq:HyperFSH}) have vanishing
contributions.  For $S$ states, the relevant term in the hyperfine 
Hamiltonian therefore is of the Dirac-$\delta$ type. Hence, we put special 
emphasis on the modifications occasioned by such Dirac-$\delta$ potentials.

The \vdw{} energy is normally derived as follows.
One first writes the attractive and repulsive
terms that describe the electron-electron,
electron-proton, and proton-proton interactions in the 
two atoms (excluding the intra-atomic terms).
This leads to the total Coulomb interaction
\begin{equation} \label{eq:CoulombInt}
\begin{aligned} [b]
V_{\mathrm{C}} &= 
\frac{e^2}{4\pi\epsilon_0}
\left( \frac{1}{|\vec R_A - \vec R_B|}
+\frac{1}{\left|\vec x_A-\vec x_B\right|}
\right.
\\
& \; \left.
-\frac{1}{|\vec x_A-\vec R_B|} 
-\frac{1}{|\vec x_B - \vec R_A |} \right) \,.
\end{aligned}
\end{equation}
One then uses the fact that the separation
$|\vec{R}_A-\vec{R}_B|$ between the two nuclei (protons) is
much larger than that between a given proton and its respective electron, that
is, much larger than both 
$|\vec r_A| = |\vec x_A - \vec{R}_A|$ and
$|\vec r_B| = |\vec x_B - \vec{R}_B|$.
One then writes $\vec x_A-\vec R_B = \vec r_A + (\vec R_A - \vec R_B)$
and $\vec x_B-\vec R_A = \vec r_B + (\vec R_B - \vec R_A)$.
Expanding in $\vec r_A$ and $\vec r_B$, one obtains
\begin{align}
\label{HVDW}
H_{\rm vdW} =& \; \frac{e^2}{4\pi\epsilon_0} \,
\frac{  \vec r_A \cdot \vec r_B - 3 \,
(\vec r_A \cdot \hat R) \,
(\vec r_B \cdot \hat R)}{R^3} 
\nonumber\\
=& \; \frac{e^2}{4\pi\epsilon_0 \, R^3} \,
\left( \delta_{k\ell} - 3 \, \hat R_k \hat R_\ell \right)
r_{Ak} \, r_{B\ell} \,,
\end{align}
where $\vec R = \vec R_A - \vec R_B$, $R = | \vec R|$ and $\hat{R}=\vec{R}/R$.
The indices $k$ and $\ell$ corresponding to the Cartesian coordinates
are summed over (Einstein summation convention).
The \vdw{} interaction term, for a $2S$--$1S$ 
system, has vanishing elements in first-order 
perturbation theory.
Both atoms $A$ and $B$ have to undergo a virtual 
dipole transition to a $P$ state for a nonvanishing 
effect, and the leading-order \vdw{} interaction is 
obtained in second-order perturbation theory,
leading to a $1/R^6$ interaction energy.
The propagator denominator in the standard Rayleigh--Schr\"{o}dinger
expression for the second-order energy shift due to 
$H_{\rm vdW}$ is 
equal to the sum of the virtual
excitation energies of both atoms~\cite{DeYo1973}.
The close-range asymptotics
of the interatomic interaction energy 
thus goes as $1/R^6$~\cite{DeYo1973}.
For an interatomic distance of $R \sim 30 a_0 \ldots 100 \, a_0$
(one hundred atomic units), the energy shift 
is on the order of $10^{-8} \ldots 10^{-12}$ atomic units (Hartrees).
The hierarchy
\begin{equation}
\left< H_{\rm vdW} \right> \ll
\left< H_{\rm HFS} \right> \ll
\left< H_{\rm LS} \right> \ll
\left< H_{\rm FS} \right>
\end{equation}
is thus fulfilled for $R \gtrsim 30 \, a_0$.
For sufficiently large interatomic distance, the Dirac $\delta$
potential of the HFS acts as a perturbation
and can be treated as such, and we shall 
focus on this regime in the current manuscript.

For clarity, we should point out that the 
Hamiltonian~\eqref{HVDW} remains valid in the nonretardation
approximation. One can understand retardation as follows:
When the phase of the atomic oscillation during a virtual 
transition changes appreciably on the time scale 
it takes light to travel the interatomic separation 
distance $R$, then the retarded form of the 
\vdw{} interaction has to be used.
The criterion for the validity of the nonretardation
approximation thus is
\begin{equation}
\frac{R}{c} \ll \frac{\hbar}{E_h} = \frac{a_0}{\alpha \, c} \,,
\end{equation}
or, more precisely,
\begin{equation}
\label{vdWrange}
a_0 = \frac{\hbar}{\alpha m c} \ll
R \ll
\frac{\hbar}{\alpha^2 m c} = \frac{a_0}{\alpha} \,,
\end{equation}
if we take into account that substantial overlap of the 
electronic wavefunctions is to be avoided.

The retarded interatomic interaction cannot be 
obtained on the basis of Eq.~\eqref{HVDW} alone;
one has to use the atom-field interaction term
[see Eq.~(85.4) of Ref.~\cite{BeLiPi1982vol4}],
\begin{equation} 
\label{eq:IntPot}
V(t) =  -\vec E(\vec R_A, t) \cdot \vec d_A(t)
-\vec E(\vec R_B, t) \cdot \vec d_B(t) \,,
\end{equation} 
where $\vec d_i = e \, \vec r_i$ is the
dipole operator for atom $i$ (for atoms with more
than one electron, one has to sum over all the electrons
in the atoms $i=A,B$). The $\vec R_A$ and $\vec R_B$
are the positions of the atomic nuclei, and $\vec E$ denotes the 
operator of the quantized electric field.
An elegant way of deriving the retarded Casimir--Polder 
interaction, described in Eq.~(85.4) of Ref.~\cite{BeLiPi1982vol4},
then consists in the matching of the scattering 
amplitude obtained from quantum electrodynamics,
against the effective interatomic interaction Hamiltonian. 
Alternative derivations use time-ordered perturbation
theory~\cite{PoTh1995}.

The functional form of the interaction depends on the distance range
In the \vdw{} range~\eqref{vdWrange} of interatomic distances,
the interaction of 
ground-state atoms is of the usual $R^{-6}$ functional form. 
This remains valid if one atom is in a metastable excited state.
In the so-called Casimir--Polder range,
\begin{equation}
\label{CPrange1}
R \gg \frac{\hbar}{\alpha^2 m c}\,,%MODIFIED HERE = \frac{a_0}{\alpha} 
\end{equation}
the interatomic distance is much larger than the 
wavelength of an optical transition, and the interaction of 
\emph{ground-state} atoms has an $R^{-7}$ function form.
For the long-range interaction involving excited metastable atoms, however,
we have to distinguish a third range of very large 
interatomic distances, 
\begin{equation}
\label{LSrange}
\mbox{Casimir--Polder~II (or Lamb shift):} \quad 
R \gg \frac{\hbar c}{\cal L} \,,
\end{equation}
which we would like to refer to as the Lamb shift range.
Here, $\cal L$ is the Lamb shift energy. 
For metastable atoms,
the Casimir--Polder range is bounded from  above by the 
Lamb shift range, and the condition~\eqref{CPrange1} 
should be modified to read
\begin{equation}
\label{CPrange}
\mbox{Casimir--Polder~I:} \quad 
\frac{\hbar c}{\cal L} \gg
R \gg \frac{\hbar}{\alpha^2 m c}\,,%MODIFIED HERE = \frac{a_0}{\alpha} 
\end{equation}
For the $1S$--$1S$ interaction, 
the interaction energy reaches the 
Casimir--Polder asymptotic form, 
proportional to $1/R^7$, 
in both regimes described by 
Eqs.~\eqref{LSrange} and~\eqref{CPrange}.
For the $2S$--$1S$ interaction,
it is only in the very long-range regime~(\ref{LSrange}) 
that we have a $R^{-7}$ interaction, with competing 
oscillatory terms~\cite{SaKa2015,DoGuLa2015,JeDe2016vdW} proportional to 
$(\calL^4/R^2) \cos[\calL R/(\hbar c)]$.

A further complication arises.
The state with atom $A$ in an excited state and atom $B$ in 
the ground state, $| 2S \rangle_A \, | 1S \rangle_B$,
is degenerate with the state $| 1S \rangle_A \, | 2S \rangle_B$
with the quantum numbers reversed among the atoms.
There is no direct first-order coupling between 
$| 2S \rangle_A \, | 1S \rangle_B$ and 
$| 1S \rangle_A \, | 2S \rangle_B$ due 
to the \vdw{} interaction~\eqref{HVDW},
but in second order, an off-diagonal term
is obtained which is of the same order-of-magnitude as the 
diagonal term, i.e., the term with the same 
in and out states. The Hamiltonian matrix in the 
basis of the degenerate states $| 2S \rangle_A \, | 1S \rangle_B$ and
$| 1S \rangle_A \, | 2S \rangle_B$ has off-diagonal 
(exchange) terms
of second order in the \vdw{} interaction~\cite{Ch1972}.
The energy eigenvalues and eigenstates are easily 
found in the degenerate basis and are studied here 
in Sec.~\ref{exchange}.

%
% $\bf{1S}$--$\bf{2S}$ Long--Range Interaction
%
\section{\texorpdfstring{$\maybebm{2S}$}{2S}--\texorpdfstring{$\maybebm{1S}$}{1S} 
Direct Interaction}
\label{long-range}

%
% Formalism
%
\subsection{Formalism}
\label{formlism}

According to Eq.~(85.17) in Chap.~85 of Ref.~\cite{BeLiPi1982vol4},
the interaction energy between two atoms $A$ and $B$ in states 
$\left| A \right>$ and $\left| B \right>$ is given by
\begin{align}
\label{osc}
& E_{A;B}^{\left(\mathrm{dir}\right)}(R) = 
{\rm Re} \frac{\ii \hbar}{\pi  c^4  (4 \pi\epsilon_0)^2} 
\int\limits_0^\infty {\rm d}\omega 
\alpha_A(\omega) 
\alpha_B(\omega) 
{\rm e}^{2 \ii \omega R/c} \, \frac{\omega^4}{R^2}\,
\nonumber\\[0.133ex]
& \times \left[ 1 
+ 2 \ii \frac{c}{\omega R}
- 5 \left( \frac{c}{\omega R} \right)^2 
- 6 \ii \left( \frac{c}{\omega R} \right)^3 
+ 3 \left( \frac{c}{\omega R} \right)^4 
\right] \,.
\end{align}
Here the superscript $\left(\mathrm{dir}\right)$ stands for ``direct'', as we
anticipate that this interaction energy is to be supplemented by the so-called
exchange interaction, to be discussed in Sec.~\ref{exchange}.  
The integral~\eqref{osc} constitutes the generalization 
of the second-order \vdw{} shift given by the 
application of Eq.~\eqref{HVDW}, to the long-range limit,
where retardation sets in.
% EXPLAINING TO REFEREE 2 THAT "THE ATOM-FIELD INTERACTION" IS INDEED CONTAINED HERE
Eq.~(\ref{osc}) contains the atom-field interaction at the lowest
relevant order in the elastic scattering case, where the initial and final
states are identical (\emph{e.g.} all photons emitted are reabsorbed
and \emph{vice versa}). We here restrict
the discussion to the leading effect in the multipole expansion, given by the
dipole polarizability $\alpha_i$ ($i = A,B$).  The designation of the real part
of the energy shift is necessary because the integrand constitutes a complex
rather than real quantity, and the poles of the integrand are displaced from
the real axis according to the Feynman prescription. For the
dipole polarizability $\alpha_A(\omega)$ (of atom $A$), we have
\begin{align}
\label{defalpha}
\alpha_A(\omega) =& \; P_A(\omega) + P_A(-\omega) \,,
\nonumber\\[0.133ex]
P_A(\omega) =& \; \frac{e^2}{3} \, 
 \left< \psi_A \left| \vec r \, 
\frac{1}{H - E_A + \hbar \omega - \ii \, \epsilon} \, \vec r
\right| \psi_A \right>
\nonumber\\[0.133ex]
=& \; \frac{e^2}{3} \, \sum_n \sum_{i = 1}^3
\frac{\left| \left< \psi_A \left| \vec r \right| \psi_n \right> \right|^2}
{E_n - E_A + \hbar \omega - \ii \, \epsilon} \,,
\end{align}
where $H$ in the propagator denominator denotes the 
Schr\"{o}dinger Hamiltonian of the relevant atom.
The $\epsilon$ parameter in Eq.~(\ref{defalpha}),
ensures that the integration (\ref{osc}) is carried along the Feynman contour;
the limit $\epsilon \rightarrow 0^+$ is taken 
after the integration is carried out.
Under appropriate conditions, which are discussed in detail below, we may
perform a Wick rotation $\dd \omega \to \ii \, \dd \omega$ in the 
integral~(\ref{osc}). The resulting Wick (W) rotated expression is the familiar one
which is usually taken as the starting point of the investigations (see,
e.g.,~Ref.~\cite{Pa2005longrange}),
\begin{align}
\label{exp}
E^{\left(\mathrm{dir}\right)\mathrm{W}}_{A;B}(R) =& \;
-\frac{\hbar}{\pi \, c^4 \, (4 \pi \epsilon_0)^2} \, 
\int\limits_0^\infty {\rm d}\omega\, 
\alpha_A(\ii \omega) 
\alpha_B(\ii \omega) 
\nonumber\\[0.133ex]
& \; \times
{\rm e}^{-2\omega R/c}\,
\frac{\omega^4 \, }{R^2}
\left[ 1 
+ 2\left(\frac{c}{\omega R}\right)
+ 5\left(\frac{c}{\omega R}\right)^2
\right.
\nonumber\\[0.133ex]
& \; \left. + 6\left(\frac{c}{\omega R}\right)^3
+ 3\left(\frac{c}{\omega R}\right)^4
\right].
\end{align}
We do not explicitly indicate the ``real part'' on the right-hand side of the
above equation, because the polarizability $\alpha_A(\ii \omega)$ is manifestly
real if we set $\epsilon = 0$ in Eq.~\eqref{defalpha}, and there are no poles
near the integration contour in Eq.~\eqref{exp} to be considered.  If both
atoms are in their $\left|1S\right>$ ground state, then the
expressions~\eqref{osc} and~\eqref{exp} are equal
[$E_{1S;1S}^{\left(\mathrm{dir}\right)}(R) =
E^{\left(\mathrm{dir}\right)\mathrm{W}}_{1S;1S}(R)$], and the Wick rotation
is permissible.

Let us now study the case $\left|A\right> = \left|2S\right>$ and
$\left|B\right> = \left|1S \right>$ for atomic hydrogen as a paradigmatic
example of a long-range interaction involving a metastable excited state. In
this case, the Wick rotated integral~(\ref{exp}) is not equal to~(\ref{osc}),
and extra care is needed [see also App.~\ref{plots}].  The dipole
polarizability $\alpha_{2S}$ can naturally be split into two contributions, the
first of which is due to the quasi-degenerate 
$\left|2P_{1/2}\right>$ and
$\left|2P_{3/2}\right>$ states which are displaced from $\left|2S\right>$ only
by the Lamb shift and by the fine structure, respectively. 
The second contribution  
is due to $nP$ states with principal quantum number $n \geq 3$.  After
doing the angular algebra for the $\left|2P_{1/2}\right>$ and
$\left|2P_{3/2}\right>$ states whose oscillator strengths~\cite{BeSa1957} with
respect to $2S$ are distributed in a ratio $\tfrac13 : \tfrac23$, we obtain
\begin{subequations}
\label{pol}
\begin{align}
\label{alpha2S}
\alpha_{2S}(\omega) =& \; 
\overline\alpha_{2S}(\omega)+\widetilde \alpha_{2S}(\omega)\,,
\\[0.133ex]
\label{alpha2Sbar}
\overline\alpha_{2S}(\omega) =& \;
\overline P_{2S}(\omega) + \overline P_{2S}(-\omega) \,,
\\[0.133ex]
\label{alpha2Stilde}
\widetilde \alpha_{2S}(\omega) =& \;
\widetilde  P_{2S}(\omega) + \widetilde  P_{2S}(-\omega) \,,
\\[0.133ex]
\label{P2Sbar}
\overline P_{2S}(\omega) =& \; \frac{e^2}{9} \, 
 \sum_{\mu}
\frac{\left| \left< 2S \left| \vec{r} \right| 2P (m\!=\!\mu)\right> \right|^2}
{-{\cal L} + \hbar \omega - \ii \, \epsilon} 
\\[0.133ex]
& \; + \frac{2 e^2}{9} \,
 \sum_{\mu}
\frac{\left| \left< 2S \left| \vec{r} \right| 2P (m\!=\!\mu)\right> \right|^2}
{{\cal F} + \hbar \omega - \ii \, \epsilon } 
\nonumber\\[0.133ex]
=& \; 3 e^2 \, a_0^2 \, 
\left( \frac{1}{-{\cal L} + \hbar\omega -\ii \epsilon} +
\frac{2}{{\cal F} + \hbar\omega -\ii \epsilon } \right)\,,
\nonumber\\[0.133ex]
\label{P2Stilde}
\widetilde  P_{2S}(\omega) =& \; \frac{e^2}{3} \, 
\sum_{n \geq 3}  \sum_{\mu}
\frac{\left| \left< 2S \left| \vec{r} \right| nP (m\!=\!\mu) \right> \right|^2}
{E_n - E_{2S} + \hbar \omega - \ii \, \epsilon} \,.
\end{align}
\end{subequations}
The nondegenerate contribution to the $2S$ 
polarizability is denoted as $\widetilde \alpha_{2S}$
(the quasi-degenerate $2P$ levels are excluded).
The quasi-degenerate $2P$ levels are contained in 
$\overline \alpha_{2S}$.
All sums are over the nonrelativistic $P$ states with magnetic 
projection quantum numbers $\mu = -1,0,1$.
The Lamb shift energy ${\cal L}$ and the 
fine structure energy ${\cal F}$ are
defined as
\begin{align} \label{eq:HyperLamb}
& E(2S_{1/2}) - E(2P_{1/2}) \equiv {\cal L} \,,
\nonumber\\[0.133ex]
& E(2P_{3/2}) - E(2S_{1/2}) \equiv {\cal F} \,.
\end{align}
The leading-order expressions for ${\cal L}$ and 
${\cal F}$ read as ${\cal L} = \frac{\alpha}{6 \pi} 
\alpha^4 m c^2 \, \ln[\alpha^{-2}]$ and 
${\cal F} = \alpha^4 m c^2/32$, respectively~\cite{ItZu1980}
[see also Eq.~\eqref{HLS}].

%
% van der Waals range of interatomic distance}
%
\subsection{van der Waals range \texorpdfstring{$\bm{a_0\ll R \ll a_0/\alpha}$}{}} 
\label{subsec:CloseD}

We investigate the $2S$--$1S$ interaction in the 
\vdw{} regime~\eqref{vdWrange}. 
There is no exponential or oscillatory suppression of any 
atomic transition in this regime, but we can approximate
\begin{align}
& E^{(\mathrm{dir})}_{A;B}(R) = 
{\rm Re} \frac{\ii \hbar}{\pi \, c^4 \, (4 \pi \epsilon_0)^2}
\int\limits_0^\infty {\rm d}\omega \, 
\alpha_A(\omega) \, 
\alpha_B(\omega) \,
{\rm e}^{2 \ii \omega R/c} \, 
\nonumber\\[0.133ex]
& \times 
\frac{\omega^4}{R^2}\, \left[ 1 
+ 2 \ii \frac{c}{\omega R}
- 5 \left( \frac{c}{\omega R} \right)^2 
- 6 \ii \left( \frac{c}{\omega R} \right)^3 
+ 3 \left( \frac{c}{\omega R} \right)^4 
\right] 
\nonumber\\[0.133ex]
& \approx 
\frac{3}{\pi} \frac{\hbar}{(4 \pi \epsilon_0)^2\,R^6} \, 
{\rm Re}\,\ii \int\limits_0^\infty {\rm d}\omega \, 
\alpha_A(\omega) \, \alpha_B(\omega) \,.
\end{align}
The functional form therefore is of the van der Waals type
\begin{equation}
\label{C62S1Sshort}
E^{(\mathrm{dir})}_{A;B}\left( R \right)
\approx - \frac{D_6(A;B)}{R^6} \,,
\end{equation}
with the van der Waals coefficient
\begin{equation} \label{eq:PlainD6}
D_6(A;B) =
\frac{3}{\pi} \frac{\hbar}{(4 \pi \epsilon_0)^2}
{\rm Re} \left( -\ii \int\limits_0^\infty {\rm d}\omega \, 
\alpha_A(\omega) \, \alpha_B(\omega) \right) \,.
\end{equation}
For the $2S$--$1S$ interaction, this implies that
\begin{subequations}
\begin{align}
\label{defC6}
& D_6(2S;1S) = 
\frac{3}{\pi} \frac{\hbar}{(4 \pi \epsilon_0)^2}
\nonumber\\[0.133ex]
& \; \times {\rm Re} \left( -\ii \int\limits_0^\infty {\rm d}\omega \, 
\left[ \overline\alpha_{2S}(\omega) +
\widetilde \alpha_{2S}(\omega) \right] \, 
\alpha_{1S}(\omega) \right) \\
&={\overline D}_6(2S;1S)+\widetilde{D}_6(2S;1S)\,.
\end{align}
\end{subequations}
For $\left|A\right> = \left|2S\right>$ and 
$\left|B\right> = \left|1S\right>$,
$D_6$ therefore is the sum of two contributions
${\overline D}_6$ and $\widetilde{ D}_6$,
which correspond to the degenerate
$\overline\alpha_{2S}$ and nondegenerate
$\widetilde \alpha_{2S}$  contributions to the $2S$ polarizability,
respectively. 
The degenerate contribution to $D_6$ can be handled analytically. 
We use the integral identity
\begin{equation} 
\label{eq:CUInt}
-\frac{\mathrm{i}}{\pi}
\int\limits_{-\infty}^{\infty}
\dd x \,
\frac{a \, b}{\left[\left(a-\mathrm{i}\epsilon\right)^2-x^2\right] \,
\left[\left(b-\mathrm{i}\epsilon\right)^2-x^2\right]}
\underset{\epsilon\to0^+}{\rightarrow}\frac{1}{a+b},
\end{equation}
which is valid for $a$ and $b$ real (regardless of their sign).
A change in integration limits to the 
interval $(0, \infty)$ can be absorbed in a prefactor $2$.
The result for ${\overline D}_6(2S;1S)$ reads
\begin{align} 
\label{overlineC6}
{\overline D}_6(2S;1S) =& \;
\frac{3}{\pi} \frac{\hbar}{( 4 \pi \epsilon_0 )^2 } \;
\frac{2 e^2}{3}
\sum_{k}\left|\left<1S\right|\vec{r}\left|k\right>\right|^2\nonumber\\
&\times\frac{2 e^2}{9}\sum_\mu 
\left| \left< 2S \left| \vec{r} \right| 2P (m\!=\!\mu) \right> \right|^2
\nonumber\\
&\times\frac{\pi}{2\hbar}
\left(\frac{1}{E_k-E_{1S}-\mathcal{L}}+\frac{2}{E_k-E_{1S}+\mathcal{F}}\right)
\nonumber\\
\approx & \;
\frac{3}{\pi} \frac{\hbar}{( 4 \pi \epsilon_0 )^2 } \;
\left\{ \frac{2 e^2}{3}
\sum_{k} \frac{ \left|\left<1S\right|\vec{r}\left|k\right>\right|^2 }{ E_k-E_{1S} } 
\right\} 
\nonumber\\
&\times\left(\frac{2 e^2}{9} \left( 27 \, a_0^2 \right)\right)
\times 3 \frac{\pi}{2\hbar} 
\nonumber\\
\approx & \;
\frac{3}{\pi} \frac{\hbar \, e^2 \, a_0^2}{( 4 \pi \epsilon_0 )^2 } \;
\left\{ \alpha_{1S}\left(0\right) \right\} \,
\times\left( 6\right) \times 3\frac{\pi}{2\hbar}
\nonumber\\
=& \; \frac{243}{2} E_h\, a_0^6 \,,
\end{align}
where we took the limit $\mathcal{L}\rightarrow0$, $\mathcal{F}\rightarrow0$ at
the end of the calculation.
We have used the known result 
\begin{equation}
\label{static1S}
\alpha_{1S}\left(0\right) = \frac92 \frac{e^2 a_0^2}{E_h} \,,
\end{equation}
where $E_h = \alpha^2 m c^2$ is the Hartree energy.
We can now give a more thorough analysis of the 
discrepancy of the results for the 
$(2S; 1S)$ \vdw{} coefficient reported in Refs.~\cite{Ch1972,DeYo1973,TaCh1986}.
Namely, the denominator $a + b$ in Eq.~\eqref{eq:CUInt} 
just corresponds to the sum of the excitation energies 
of the two atoms in the calculation of the \vdw{} coefficient;
the contribution of a virtual $P$ state in one of the atoms is seen
to be nonvanishing even if it is displaced from the 
reference state only by an infinitesimal shift 
$a = \calL, \calF \to 0$.
By contrast, if one takes the limit $\calL, \calF \to 0$ 
too early, i.e., before evaluating the integral~\eqref{eq:CUInt},
then in Eq.~\eqref{alpha2Sbar}, one obtains 
$\overline \alpha_{2S}(\omega) = 0$,
because the two terms $\overline P_{2S}(\pm \omega)$
just cancel each other.
Or, expressed more concisely, because of the 
exact energetic degeneracy of the $2S$ and $2P$ states
in the nonrelativistic theory,
the virtual $2P$ states are excluded from the 
sum over virtual states in the nonrelativistic 
expression of the polarizability,
which leads to the erroneous result reported in Refs.~\cite{Ch1972,TaCh1986}. 
Only if the formulation of the nonrelativistic
expression of the polarizability is enhanced
by the fine structure and Lamb shift denominators, 
as in Eq.~\eqref{pol}, can we obtain the missing contribution
${\overline D}_6(2S;1S) $ given in Eq.~\eqref{overlineC6}.
The contribution of the quasi-degenerate levels 
is more obvious in the sum-over-states approach 
chosen in Ref.~\cite{DeYo1973}, where according to 
Eq.~\eqref{eq:CUInt}, the sum of the excitation 
energies of both atoms enters the propagator 
denominator [see also Eqs.~(12a) and~(12b) of Ref.~\cite{DeYo1973}].

For the nondegenerate contribution, %MODIFICATION HERE
we can perform the Wick rotation and obtain
the following integral representation
\begin{equation}
\label{WickforD6Tilde}
\begin{aligned} [b]
\widetilde {D}_6(2S; 1S) &=
\frac{3}{\pi} \frac{\hbar}{(4 \pi \epsilon_0)^2}
{\rm Re} \left( -\ii \int\limits_0^\infty {\rm d}\omega \, 
 \widetilde \alpha_{2S}(\omega) \, \alpha_{1S}(\omega) \right) \\
 &=
\frac{3}{\pi} \frac{\hbar}{(4 \pi \epsilon_0)^2}
\int_0^\infty {\rm d}\omega\, 
\widetilde \alpha_{2S}(\ii \omega) \, 
\alpha_{1S}(\ii \omega) \,,
\end{aligned}
\end{equation}
which is convenient for a numerical evaluation.
Namely, according to Eq.~\eqref{defalpha}, one can write the 
corresponding polarizabilities as the sum over two 
matrix elements $P(\omega)$ and $P(-\omega)$ of a resolvent operator,
where the $P$ matrix elements can be written in terms
of hypergeometric functions. 
The calculation of a 
convenient representation of the 
polarizability of low-lying $S$ 
states~\cite{GaCo1970,Pa1993,JePa1996} becomes 
easier if one uses a 
coordinate-space integration based on the Sturmian 
decomposition of the radial hydrogen Green function in terms
of Laguerre polynomials~\cite{SwDr1991b}. 
After the radial integrals,
one evaluates the sum over the Sturmian integrals in terms of hypergeometric
functions with the help of formulas contained in Ref.~\cite{Ba1953vol1}.
The result of this calculation for the ground state is 
{\allowdisplaybreaks
\begin{subequations}
\label{analytic}
\begin{align}
\label{P1S}
& P_{1S}(\omega) = -
\frac{e^2 \, a_0^2}{E_h} \,
\left[ \frac{2t^2}{3 (1-t)^5 (1 + t)^4}
\left(38 t^7
+ 26 t^6 \right.\right.\nonumber\\[0.133ex]
&\left.\left. + 19 t^5 - 19 t^4 - 12 t^3 + 12 t^2 + 3 t - 3\right)
\right.
\nonumber\\[0.133ex]
& \left. +
\frac{256 \, t^9}{3 \, (t - 1)^5 \, (t + 1)^5} \,
{}_2 F_1\left(1, -t, 1 - t, \left( \frac{1-t}{1+t} \right)^2 \right) \right],
\nonumber\\[0.133ex]
& t = \left( 1 + \frac{2 \hbar \omega}{\alpha^2 m c^2} \right)^{-1/2} \,,
\end{align}
where
\begin{equation}
P_{1S}(\omega) = \frac{e^2}{3} \,
\sum_{n \geq 3} \sum_{\mu}
\frac{\left| \left< 1S \left| \vec r \right| nP (m\!=\!\mu) \right> \right|^2}
{E_n - E_{1S} + \hbar \omega - \ii \, \epsilon} \,,
\end{equation}
and the sum includes the continuum.
We here take the opportunity to correct a typographical error
in Eq. (3a) of Ref.~\cite{AdKaJe2016} which led to an inconsistent sign
of the term involving the hypergeometric function.
For the $2S$ state, one obtains the nondegenerate matrix element
\begin{align}
\label{tildeP2S}
&  \widetilde P_{2S}(\omega) =
\frac{e^2 \, a_0^2}{E_h} \,
\bigg[ \frac{16\tau^2}{3 (\tau-1)^6 (1 + \tau)^4}
(1181 \tau^8 - 314 \tau^7 
\nonumber\\[0.133ex]
& 
- 16 \tau^6 - 166 \tau^5 
+ 14 \tau^4 + 138 \tau^3 - 48 \tau^2 - 42 \tau + 21)
\nonumber\\[0.133ex]
& - 
\frac{16384 \, \tau^9 \, (4 \tau^2 - 1)}{3 \, (\tau - 1)^6 \, (\tau + 1)^6} \,
{}_2 F_1\left(1, -2 \tau, 1 - 2 \tau, \left( \frac{1- \tau}{1+\tau} \right)^2 
\right) 
\nonumber\\[0.133ex]
& - \frac{72 \tau^2}{1 - \tau^2} \bigg], 
\qquad \qquad
\tau = \left( 1 + \frac{8 \hbar \omega}{\alpha^2 m c^2} \right)^{-1/2} \,.
\end{align}
\end{subequations}
}
Indeed, the $2P$ state is excluded from the sum over states in 
Eq.~\eqref{tildeP2S} by the subtraction of the term 
$72 \tau^2/(\tau^2 - 1)$: one can verify that the 
expression~\eqref{tildeP2S}
is finite in the limit $\tau \to 1$, which is equivalent to 
vanishing photon energy $\omega \to 0$.

A numerical integration of Eq.~\eqref{WickforD6Tilde} then yields
the following value for $ \widetilde{D}_6(2S;1S)$,
\begin{equation}
\label{widetildeC6}
 \widetilde{D}_6(2S; 1S) = 55.252\,266\,285 \,E_h\, a_0^6\,.
\end{equation}
We have verified this result using discrete numerical 
methods~\cite{SaOe1989}, 
where the radial Schr\"{o}dinger equation is evaluated on a lattice,
and a discrete pseudospectrum (due to the finite size of the
lattice) represents the continuum spectrum.
The result for $ D_6(2S; 1S)$
according to Table~VI of Ref.~\cite{TaCh1986}
reads $56.7999\, E_h\, a_0^6$, 
while according to Table~2 of Ref.~\cite{Ch1972},
it is $(56.5 \pm 0.5) \, E_h\, a_0^6$.
Both results are not in 
perfect agreement with our 
result, though numerically close.
This observation is consistent with 
the derivations outlined in Refs.~\cite{Ch1972,TaCh1986},
which suggest that the results reported in the 
cited investigation 
may correspond to the nondegenerate contribution. 
The total \vdw{} coefficients $D_6$ is 
obtained as the sum of the contributions given in 
Eqs.~\eqref{overlineC6} and~\eqref{widetildeC6},
\begin{align}
\label{C62S1Sshortnum}
D_6(2S; 1S) =& \;  \widetilde{D}_6(2S; 1S) +\overline D_6(2S; 1S)
\nonumber\\[0.133ex]
=& \; 176.752\,266\,285 \, E_h\, a_0^6 \,,
\end{align}
where we confirm all significant digits of the 
previously reported result~\cite{DeYo1973} of 
$176.752$. For the $1S$--$1S$ interaction, we confirm the known 
result~\cite{Ko1967,DeYo1973} of
$D_6(1S;1S) = 6.499\,026\,705\, E_h\, a_0^6$, 
and add a few digits of numerical significance.
In particular, this result shows that the result for
$D_6(1S;1S)$  is numerically close to $\tfrac{13}{2}$,
but not exactly equal to a rational number.
We should add that the numerical accuracy of the 
strictly nonrelativistic results given in 
Eqs.~\eqref{widetildeC6} and~\eqref{C62S1Sshortnum}
extends to all digits indicated.
However, reduced-mass, relativistic and radiative corrections 
contribute on the level of $10^{-4} \ldots 10^{-3}$. 
For definiteness, we should also clarify that the 
electron mass $m$ is used as the mass of the 
hydrogen atom, not the reduced mass of the 
electron-proton system (see also the discussion in 
Sec.~\ref{sec:Numerics}).

An alternative treatment is possible in the present \vdw{} range. There exists
an integral identity similar to (\ref{eq:CUInt}), namely
\begin{equation} 
\label{eq:CDInt}
\frac{1}{\pi}a'b'
\int_{-\infty}^{\infty}\frac{\mathrm{d}x}
{\left(a'^2+x^2\right)\left(b'^2+x^2\right)}=
\frac{\mathrm{sgn}\left(a'\right) \,
\mathrm{sgn}\left(b'\right)}{\left|a'\right|+\left|b'\right|} \,.
\end{equation}
The two integrals~\eqref{eq:CUInt} and~\eqref{eq:CDInt}
are thus equal for $a+b=a'+b'$; \emph{if and only if} $a'$
and $b'$ are both positive.

Notice from~(\ref{defalpha}) and~(\ref{defC6}) that 
$D_6\left(2S;1S\right)$ is given by an integral of the type (\ref{eq:CUInt}), 
namely, by
\begin{align} \label{eq:MakeItClear}
D_6(2S;1S) \equiv & \; \mathrm{Re}\,-\mathrm{i}\frac{4\hbar}{3\pi}
\frac{e^4}{(4 \pi \epsilon_0)^2}
\sum_{mn}
\int\limits_0^{\infty}\mathrm{d}\omega\,
\nonumber\\
& \; \times
\frac{ (E_m-E_{1S}) \, \left<1S\right| \vec r \left|m\right> \cdot
\left< m\right| \vec r \left|1S\right>}% 
{\left[\left(E_m-E_{1S}-\mathrm{i}\epsilon\right)^2-
\left(\hbar\omega\right)^2\right]}
\nonumber\\
& \; \times
\frac{(E_n-E_{2S}) \, \left<2S\right| \vec r \left|n\right> \cdot 
\left< n\right| \vec r \left|2S\right>}%
{\left[\left(E_n-E_{2S}-\mathrm{i}\epsilon\right)^2-\left(\hbar\omega\right)^2\right]}.
\end{align}
At this point we may not perform the Wick rotation that 
takes us from an integral of the type (\ref{eq:CUInt}) 
to an integral of the type (\ref{eq:CDInt}). 
Indeed, for $n=2P_{1/2}$, we have $b=E_n-E_{2S}=-\mathcal{L}<0$ 
and the conditions for the equality of (\ref{eq:CUInt}) and 
(\ref{eq:CDInt}) is not fulfilled.
However, as was noticed by Deal and Young in \cite{DeYo1973}, 
any integral of the type (\ref{eq:CDInt}) is equivalent to an 
integral of the type (\ref{eq:CUInt}) provided 
we are able to replace the (possibly negative) 
quantities $a$ and $b$ by two \emph{positive} quantities $a'$ and $b'$ 
so that $a + b = a' + b'$.
Hence, we can rewrite (\ref{eq:MakeItClear}) as
\begin{align} \label{eq:PreTrick}
& D_6\left(2S;1S\right)= 
\mathrm{Re}-\mathrm{i}\frac{4\hbar}{3\pi}
\frac{e^4}{(4 \pi \epsilon_0)^2}
\sum_{mn} \int\limits_0^{\infty}\mathrm{d}\omega\,
\nonumber\\
& \qquad \times
\frac{\left(E_m-\frac{1}{2}\left(E_{1S}+E_{2S}\right)\right) \,
\left<1S\right|\vec r \left|m\right> \cdot 
\left< m\right| \vec r \left|1S\right>}%
{\left[\left(E_m-\frac{1}{2}\left(E_{1S}+E_{2S}\right)-\mathrm{i}\epsilon\right)^2-
\left(\hbar\omega\right)^2\right]}
\nonumber\\
& \qquad \times\frac{\left(E_n-\frac{1}{2}\left(E_{1S}+E_{2S}\right)\right) \, 
\left<2S\right|\vec r \left|n\right> \cdot
\left< n\right|\vec r \left|2S\right>}%
{\left[\left(E_n-\frac{1}{2}\left(E_{1S}+E_{2S}\right)-\mathrm{i}\epsilon\right)^2-
\left(\hbar\omega\right)^2\right]} \,.
\end{align}
In what follows we will make use of the space-saving notation
\begin{equation} \label{eq:AvgEn}
E_{\overline{1S2S}}\equiv\frac{1}{2}\left(E_{1S}+E_{2S}\right).
\end{equation}
Notice that for all single-atom hydrogen eigenstates [except for $1S$, which
never enters as a virtual state in the expression of $2S$ polarizabilities],
we have $E_m, E_n>E_{\overline{1S2S}}$. In other words, identifying
(\ref{eq:PreTrick}) with the model integral (\ref{eq:CUInt}), we have $a$ and
$b$ positive. Hence the condition for the equality of (\ref{eq:CUInt}) and
(\ref{eq:CDInt}) is fulfilled.
We then perform the Wick rotation and rewrite (\ref{eq:PreTrick}) as
\begin{align} \label{eq:PostTrick}
& D_6\left(2S;1S\right)=
\frac{4\hbar}{3\pi}
\frac{e^4}{(4 \pi \epsilon_0)^2}
\sum_{mn} \int\limits_0^{\infty}\mathrm{d}\omega\,
\nonumber\\
& \qquad \times
\frac{\left(E_m-\frac{1}{2}\left(E_{1S}+E_{2S}\right)\right) \,
\left<1S\right|\vec r \left|m\right> \cdot
\left< m\right| \vec r \left|1S\right>}%
{\left[\left(E_m-\frac{1}{2}\left(E_{1S}+E_{2S}\right)\right)^2 +
\left(\hbar\omega\right)^2\right]}
\nonumber\\
& \qquad \times\frac{\left(E_n-\frac{1}{2}\left(E_{1S}+E_{2S}\right)\right) \,
\left<2S\right|\vec r \left|n\right> \cdot
\left< n\right|\vec r \left|2S\right>}%
{\left[\left(E_n-\frac{1}{2}\left(E_{1S}+E_{2S}\right)\right)^2 +
\left(\hbar\omega\right)^2\right]} \,.
\end{align}

We introduce the following polarizabilities,
which have the mean energy $E_{\overline{1S2S}}$ in the
propagator denominators,
\begin{subequations} \label{eq:PolGalore}
\begin{align}
\alpha_{1S\left(2S\right)}\left(\omega\right)
=& \; \frac{e^2}{3}\sum_{\pm}
\left< 1S\right| \vec r  \, 
\frac{1}{H-E_{\overline{1S2S}}\pm\hbar\omega} \,
\vec r \left|1S\right>
\nonumber\\
=& \;
P_{1S\left(2S\right)}\left(\omega\right)+P_{1S\left(2S\right)}\left(-\omega\right)\,,
\\
\alpha_{2S\left(1S\right)}\left(\omega\right) =& \;
\frac{e^2}{3}\sum_{\pm}\left< 2S\right| \vec r  \,
\frac{1}{H-E_{\overline{1S2S}}\pm\hbar\omega} \, 
\vec r \left|2S\right>
\nonumber\\
=& \; P_{2S\left(1S\right)}\left(\omega\right) +
P_{2S\left(1S\right)}\left(-\omega\right)\,.
\end{align}
\end{subequations} 
We finally obtain
\begin{align} \label{eq:WickInt}
D_6\left(2S;1S\right) =& \;
\frac{3}{\pi} \frac{\hbar}{(4 \pi \epsilon_0)^2}
\int\limits_0^{\infty} 
\dd \omega\,
\alpha_{1S\left(2S\right)}\left(\mathrm{i}\omega\right)
\alpha_{2S\left(1S\right)}\left(\mathrm{i}\omega\right)
\nonumber\\
=& \; 176.752\,266\,285 \, E_h\, a_0^6 \,.
\end{align}
This matches the value (\ref{C62S1Sshortnum}) found by the previously followed
method. With such a choice of the reference energies in the denominators, we
have shown that the Wick rotation is made automatically valid by the inequality
$E_m>E_{\overline{1S2S}}$ for the virtual $P$ states with energies $E_m$.
This procedure also results in the automatic inclusion
of the quasi-degenerate states.

%
% Limit of Large Distance
%
\subsection{Very large interatomic distance 
\texorpdfstring{$\bm{R\gg \hbar c/\mathcal{L}}$}{}}
\label{subsec:AwayD}

For very large interatomic separations, the classic result is that of Casimir
and Polder \cite{CaPo1948}, and it is given, when both atoms are in the ground
state, by
\begin{equation} 
\label{eq:LongCPGround}
E^{(\mathrm{dir})}_{1S;1S}(R) \approx \; -\frac{23}{4 \pi}
\frac{\hbar c}{(4\pi\epsilon_0)^2}\frac{1}{R^7} \; 
\alpha_{1S}(0) \, \alpha_{1S}(0) \,.
\end{equation}
which can be obtained by the Wick-rotated version (\ref{exp}) of the integral.
When one of the atoms sits in an excited state, however (here, the $2S$ state),
there is an extra term coming from the contribution of the pole that is picked
up when carrying out the Wick rotation. The pole corresponds to the $2P_{1/2}$
level. We thus have two competing contributions in the 
very-long-range limit, the first being the 
generalization of Eq.~\eqref{eq:LongCPGround} to the 
$2S$--$1S$ interaction,
\begin{equation} 
\label{eq:LongCPExcI}
E^{(\mathrm{dir}) \, \rm I}_{2S;1S}\left(R\right) \approx \;
-\frac{23}{4 \pi}\frac{\hbar c}{(4\pi\epsilon_0)^2}\frac{1}{R^7} \;
\alpha_{1S}(0) \, \alpha_{2S}(0) \,,
%OFF \; R \to \infty \,
\end{equation}
the other being an oscillatory term~\cite{SaKa2015,DoGuLa2015,JeDe2016vdW}
of the functional form
\begin{multline} 
\label{eq:LongCPExcII}
E^{(\mathrm{dir}) \, \rm II}_{2S;1S}\left(R\right) \sim 
\frac{e^2}{(4 \pi \epsilon_0)^2 \, R^2} \,
\left( \frac{\calL}{\hbar c} \right)^4 \,
\cos\left( \frac{2 \calL R}{\hbar c} \right) \,
\\[0.1133ex]
\times \sum_\mu | \langle 2S | \vec r | 2P(m \! = \! \mu) \rangle |^2 \,
\alpha_{1S}(0) \,.
\end{multline} 
The term $E^{(\mathrm{dir}) \, \rm I}_{2S;1S}$ is the Wick-rotated term
(\ref{exp}) in the long-range limit. The term $E^{(\mathrm{dir}) \, \rm II}_{2S;1S}$
is the pole contribution from the $2P_{1/2}$ level, which lies lower
than the $2S$ level. In the van der Waals
range (\ref{vdWrange}), both the Wick-rotated and pole contribution
decay as $1/R^6$. However, in the present large separation regime (\ref{LSrange}),
we see that the pole term exhibits a long-range tail proportional to $R^{-2}$.
For the $2S$--$1S$ interaction, it is the ratio
$\left(\mathcal{L}R\right)/\left(\hbar c\right)$ that determines which
one of these powers yields the dominant contribution. Hence, 
we have a the regime
change around $R=\hbar c/\mathcal{L}$, with long-range tails
extending beyond such separations.
Parametrically, using $\calL \sim \alpha^5 \, m \, c^2$,
and $\hbar c/\calL \sim a_0/\alpha^4$,
one obtains the following estimates,
\begin{subequations}
\label{estimates}
\begin{align}
\label{estimatesI}
E^{(\mathrm{dir}) \, \rm I}_{2S;1S}\left(R\right) \sim & \;
\frac{E_h}{\alpha^4 \, (R/a_0)^7} \,,
\\[0.1133ex]
\label{estimatesII}
E^{(\mathrm{dir}) \, \rm II}_{2S;1S}\left(R\right) \sim & \;
\frac{\alpha^{16} \, \cos(2 \alpha^4 R/a_0) \, E_h}{(R/a_0)^2} \,.
\end{align}
\end{subequations}
Both of these estimates are relevant for $R \gtrsim \hbar c/\calL$.
The transition region where $E^{\rm I}_{2S;1S}\left(R\right)$ 
becomes commensurate with $E^{\rm II}_{2S;1S}\left(R\right)$
is thus reached for 
\begin{equation}
\label{IeqII}
R \sim \frac{\hbar c}{\cal L} \sim \frac{a_0}{\alpha^4} \,,
\qquad
E^{\rm I}_{2S;1S}\left(R\right) \sim 
E^{\rm II}_{2S;1S}\left(R\right) \sim 
\alpha^{24} E_h \,.
\end{equation}
The frequency shift in this region is 
of the order of $10^{-36} \, {\rm Hz}$,
and thus far too small to be of any relevance
for experiments.
In view of the prefactor $\calL^4$ in 
Eq.~\eqref{eq:LongCPExcII}, the same conclusion is reached
as recently found in Ref.~\cite{Je2015rapid}
for atom-surface interactions:
Namely, for long-range interactions
involving the metastable $2S$ state, 
a potentially interesting  oscillating long-range is found, 
but its numerical coefficient is too small 
to be of significance.

Our very-long-range regime %MODIFICATION
is given by (\ref{LSrange}).
Expressed in units of the Hartree energy $E_h$,
the physical values of the Lamb shift and fine structure 
energies are
\begin{subequations} \label{eq:LFValues}
\begin{align}
{\cal L} =& \; 1.61 \times 10^{-7} \, E_h \,, 
\\[0.133ex]
{\cal F} =& \; 1.67 \times 10^{-6} \, E_h \approx 10 \, {\cal L} \,.
\end{align}
\end{subequations}
The long-range approximation is thus valid in the 
region
\begin{equation}
R \gg \frac{\hbar c}{\cal L} =
\frac{a_0}{\alpha} \frac{E_h}{\cal L} =
8.206 \times 10^8 a_0 =
0.0434 \, {\rm m}  \,.
\end{equation}
According to Eq.~\eqref{IeqII}, the oscillatory tail and the $1/R^7$ \cp{} term
have comparable magnitude as we enter the very-long-range
regime~\eqref{LSrange}, %MODIFIED EQ. REF. TO GO ALONG WITH CHANGES
but the oscillatory tail given in
Eq.~\eqref{estimatesII} could be assumed to dominate for distances exceeding
the Lamb shift transition wavelength.  This consideration, though, should be
taken with a grain of salt.  Namely, in the long-range limit, one has to take
into consideration the fact that the width of the $2P_{1/2}$ state is of the
same order-of-magnitude ($\alpha^5 m c^2$) as the Lamb shift itself
\cite{BeSa1957}. For $R \gg \hbar c/\calL$,
the oscillatory tails are thus exponentially suppressed
according to the factor $\exp[2 \ii (\calL + \ii \Gamma/2) R)/\left(\hbar
c\right)] \sim \exp(-2 \Gamma R/\left(\hbar c\right))$, where $\Gamma$ is the
natural energy width of the $2P_{1/2}$ state. 
Still, it is of Academic interest to note that 
the oscillatory long-range tail exists.

%
% Intermediate Distances
%
\subsection{Intermediate distance 
\texorpdfstring{$\bm{a_0/\alpha \ll R\ll \hbar c/\mathcal{L}}$}{}} 
\label{subsec:MidD} 

It is very interesting indeed to also investigate the
intermediate range of interatomic distances, given by (\ref{CPrange}).
The treatment becomes a little sophisticated.
Namely, as far as virtual transitions with a change in the 
principal quantum number are concerned, we are 
in the Casimir--Polder regime where the result 
is given by an $R^{-7}$ interaction [only the 
virtual $2P_{1/2}$ state gives rise to an 
oscillatory tail, and this occurs--without any 
change in the principal quantum number---only for the 
$2S$--$1S$ interaction].
The $2S$--$1S$ interaction would therefore be proportional to 
$R^{-7}$ if the $2S$ polarizability were restricted to the 
term $\widetilde \alpha_{2S}$.
However, the frequency range corresponding to the 
intermediate distance range~\eqref{CPrange} %MODIFIED EQ. REF. TO GO ALONG WITH CHANGES
is so low that the frequency-dependent
quasi-degenerate polarizability $\overline \alpha_{2S}$
in the integral~\eqref{osc} is not
exponentially suppressed. We thus have
\begin{align} \label{eq:HereOnly}
E_{2S;1S}^{\left(\mathrm{dir}\right)}\left(R\right)
\approx & \; {\overline E}_{2S;1S}^{\left(\mathrm{dir}\right)}\left(R\right)
\nonumber\\
= & \;  3 \frac{ \hbar }{(4 \pi \epsilon_0)^2}  
{\rm Re} \left( \ii \int\limits_0^\infty {\rm d}\omega \, 
\alpha_{1S}(0) \, {\overline \alpha}_{2S}(\omega) \right) \,.
\end{align}
The static ground-state 
polarizability $\alpha_{1S}(0)$ is given in Eq.~\eqref{static1S}.
Furthermore, on the scale of distances in the 
intermediate range, we may approximate the Lamb shift and 
the fine structure energy by zero after doing the integrals.
This yields
%TURNING THIS INTO SUBEQ.
\begin{subequations} \label{eq:PiOverHbar}
\begin{align}
\label{MASTER1}
& \lim_{{\cal L} \to 0} 
{\rm Re} \left( \ii \int\limits_0^\infty {\rm d}\omega \,
\frac{2 {\cal L}}%
{(-{\cal L} - \ii\epsilon)^2 - (\hbar \omega)^2} \right)
= \frac{\pi}{\hbar} \,.
\end{align}
Due to the different pole structure under the 
sign change from the Lamb shift as compared to the 
fine structure transition ($-{\cal L} < 0$, but ${\cal F} > 0$),
it is nontrivial to check that 
\begin{align}
\label{MASTER2}
& \lim_{{\cal F} \to 0} 
{\rm Re} \left( -\ii \int\limits_0^\infty {\rm d}\omega \, 
\frac{2 {\cal F}}%
{({\cal F} - \ii\epsilon)^2 - (\hbar \omega)^2} \right)
= \frac{\pi}{\hbar} \,,
\end{align}
\end{subequations}
The result for the asymptotics in the intermediate 
range thus reads as 
\begin{align}
\label{C62S1Smedium}
\overline E_{2S; 1S}^{\left(\mathrm{dir}\right)}\left(R\right) 
=& \; - \frac{\overline D_6(2S; 1S)}{R^6} 
= - \frac{243}{2} E_h \left( \frac{a_0}{R} \right)^6 \,.
\end{align}
The interaction is thus still of the $R^{-6}$ form, as it is in the \vdw\,range, but 
the coefficient is reduced in magnitude as compared to 
Eq.~\eqref{C62S1Sshortnum}.

A few words on the precise formulation of the 
intermediate distance range are perhaps in order.
Namely, in principle, one might argue that the intermediate range should be bounded 
from above by $\hbar c/\mathcal{F}$, 
instead of $\hbar c/\mathcal{L}$, as the former quantity is
smaller than the latter. 
In the rather narrow window where 
$\hbar c/\mathcal{F} < R < \hbar c/\mathcal{L}$, 
transitions between $2S$ and $2P_{3/2}$
states are suppressed while those between $2S$ and $2P_{1/2}$ states are not.
We do not dwell further on the details of this regime, 
because an order-of-magnitude estimate of the 
frequency shifts, analogous to the one carried out 
in Sec.~\ref{subsec:AwayD}, reveals that they do not exceed $10^{-21} \, {\rm Hz}$
in the discussed distance range.
Mathematically speaking, the inequality $R \ll \hbar c/\mathcal{L}$ implies 
$R \ll \hbar c/\mathcal{F}$ because $\calF$ and $\calL$ are
apart by only a single order-of-magnitude [see Eq.~\eqref{eq:LFValues}]. 
The regime $\hbar c/\mathcal{F} < R < \hbar c/\mathcal{L}$
can only be accessed reliably by a numerical
calculation (see Sec.~\ref{sec:Numerics}).

%
% $\bf{1S}$--$\bf{2S}$ Exchange Interaction
%
\section{\texorpdfstring{$\bm{2S}$--$\bm{1S}$}{(2S--1S)} 
Exchange Interaction}
\label{exchange}

%
% Formalism
%
\subsection{Formalism}
\label{formlismlso}

We now consider the $2S$--$1S$ exchange interaction. The states
$\left|1S\right>_A\left|2S\right>_B$ and
$\left|2S\right>_A\left|1S\right>_B$ are energetically degenerate,
which induces the need for special care in the treatment of 
the \vdw{} interaction. 
The general eigenvalue problem reads as 
follows,
\begin{equation} \label{eq:EVEq}
\left(H_S+H_{\rm vdW}\right)\left|\Psi\right>=E\left|\Psi\right> \,,
\end{equation}
where $H_S$ is the Schr\"{o}dinger Hamiltonian 
(sum over both atoms). In what follows we shall 
attempt to give a somewhat streamlined 
derivation of the \vdw{} mixing term resulting from the
energetic degeneracy, which confirms the results
obtained in Ref.~\cite{Ch1972}. The basis states are
\begin{subequations}
\begin{align}
|\Psi_1 \rangle =& \; \left|1S\right>_A\left|2S\right>_B  \,,
\\[0.133ex]
|\Psi_2 \rangle =& \; \left|2S\right>_A\left|1S\right>_B  \,.
\end{align}
\end{subequations}
The first-order perturbations to these wave functions are
\begin{equation}
| \delta \Psi_{j=1,2} \rangle = 
\left( \frac{1}{E_0 - H_S} \right)' \, 
H_{\rm vdW} |\Psi_{j=1,2} \rangle \,,
\end{equation}
where $E_0 = E_{1S} + E_{2S}$ is the unperturbed
energy of the metastable, noninteracting two-atom system. 
The prime on the Green function indicates that the degenerate
states have been excluded from the sum over virtual states.
One calculates the Hamiltonian matrix with elements
\begin{equation}
H_{ij} = 
\left( \langle \Psi_i | + \langle \delta \Psi_i | \right) 
\left( H_S + H_{\rm vdW} \right)
\left( | \Psi_j \rangle + | \delta \Psi_j \rangle \right) \,,
\end{equation}
with $i,j = 1,2$. The result has the structure
\begin{equation}
H = \left( \begin{array}{cc} 
E_0 + X & Y \\
Y & E_0 + X 
\end{array} \right) \,,
\end{equation}
where
\begin{subequations} \label{eq:MakeItSimpler}
\begin{align}
X =& \; {\sum_{mn}}'
\frac{\left|\left<1S\,2S\right| H_{\rm vdW}\left| m\,n\right>\right|^2}%
{E_{1S}+E_{2S}-\left(E_m+E_n\right)},
\\
Y =& \; {\sum_{mn}}'
\frac{\left<2S\,1S\right| H_{\rm vdW}\left| m\,n\right> \, 
\left< m\,n\right| H_{\rm vdW}\left|1S\,2S\right>}%
{E_{1S}+E_{2S}-\left(E_m+E_n\right)} \,.
\end{align}
\end{subequations}
Again, the prime on the sum denotes the exclusion
of the reference state. This matrix thus assumes the form 
\begin{equation} \label{eq:MatrixAssume}
H = \left( \begin{array}{cc}
E_0 - \dfrac{D_6(2S;1S)}{R^6} & - \dfrac{M_6(2S;1S)}{R^6} \\[2.133ex]
-\dfrac{M_6(2S;1S)}{R^6}  & E_0 - \dfrac{D_6(2S;1S)}{R^6} 
\end{array} \right) \,,
\end{equation}
where we define the two coefficients 
\begin{subequations}
\begin{align}
\label{eq:C1D6}
& D_6(2S;1S) =
\frac23 \frac{e^4}{\left(4\pi\epsilon_0\right)^2}
{\sum_{mn}}'
\frac{\left|\left<1S\right|\vec r \left|m\right>|^2 \,
|\left<2S\right|\vec r \left|n\right>\right|^2}%
{E_m+E_n - (E_{1S}+E_{2S})} \,,
\\
\label{eq:C2M6}
& M_6(2S;1S) = 
\frac23 
\frac{e^4}{\left(4\pi\epsilon_0\right)^2}
{\sum_{mn}}' \frac{1}{E_m+E_n - (E_{1S}+E_{2S})} 
\nonumber\\
& \qquad  \times
\left<1S\right|\vec r \left|n\right> \cdot
\left<n\right|\vec r \left|2S\right> \,
\left<2S\right|\vec r \left|m\right> \cdot
\left<m\right|\vec r \left|1S\right>. 
\end{align}
\end{subequations}
It can be shown that $D_6(2S;1S)$, as defined by (\ref{eq:C1D6}), 
agrees with the earlier expression (\ref{eq:PlainD6}). 
The eigenenergies and corresponding eigenvectors of matrix (\ref{eq:MatrixAssume}) are 
\begin{subequations}
\label{EMIXED}
\begin{align}
E_\pm =& \; E_0 - \frac{D_6 \pm M_6}{R^6} \,, \label{eq:EigenE}
\\[0.133ex]
\label{eq:EigenV} 
|\psi_\pm \rangle =& \;
\frac{1}{\sqrt{2}} \, 
\left( | \Psi_1 \rangle \pm | \Psi_2 \rangle \right)  \,,
\end{align}
\end{subequations}
so that we obtain a symmetry-dependent van der Waals coefficient
\begin{equation}
C_6 = D_6 \pm M_6
\end{equation}
which is obtained from a direct and a mixing term,
depending on the sign in the coherent superposition (\ref{eq:EigenV}).
Using the integral representation~\eqref{eq:CUInt},
one can bring $M_6$ into the form~\eqref{eq:MakeItClear}
\begin{multline} 
M_6(2S;1S) \equiv \mathrm{Re}\,-\mathrm{i}\frac{4\hbar}{3\pi}
\frac{e^4}{(4 \pi \epsilon_0)^2}
\sum_{mn}
\int\limits_0^{\infty}\mathrm{d}\omega\,
\\
\times
\frac{ (E_m-E_{1S}) \, \left<1S\right| \vec r \left|m\right> \cdot
\left< m\right| \vec r \left|2S\right>}%
{\left[\left(E_m-E_{1S}-\mathrm{i}\epsilon\right)^2-
\left(\hbar\omega\right)^2\right]}
\\
\times
\frac{(E_n-E_{2S}) \, \left<2S\right| \vec r \left|n\right> \cdot
\left< n\right| \vec r \left|1S\right>}%
{\left[\left(E_n-E_{2S}-\mathrm{i}\epsilon\right)^2-\left(\hbar\omega\right)^2\right]}.
\end{multline}
Expressed in terms of polarizabilities, 
one obtains
\begin{subequations} 
\label{eq:NowPolar}
\begin{align}
D_6(2S;1S) =& \; \mathrm{Re}
\frac{-3 \ii \hbar}{\pi (4 \pi \epsilon_0)^2} 
\int\limits_0^{\infty}\dd\omega\,
\alpha_{1S}\left(\omega\right)\alpha_{2S}\left(\omega\right) \,,
\\
M_6(2S;1S) =& \; \mathrm{Re}
\frac{-3 \ii \hbar}{\pi (4 \pi \epsilon_0)^2} 
\int\limits_0^{\infty}\mathrm{d}\omega \,
\alpha_{\underline{1S}2S}\left(\omega\right) 
\alpha_{1S\underline{2S}}^*\left(\omega\right) \,,
\end{align}
\end{subequations}
where we define the mixed polarizabilities via
\begin{subequations}
\label{eq:AlphaMixedPol}
\begin{align}
\alpha_{\underline{A}B}\left(\omega\right) =& \;
\frac{e^2}{3}\sum_{\pm}\left< A\right|\vec r 
\frac{1}{H-E_A-\mathrm{i}\epsilon\pm\hbar\omega}
\vec r \left|B\right>\\
&=P_{\underline{A}B}\left(\omega\right) + P_{\underline{A}B}\left(-\omega\right)\,,\\
\alpha_{A\underline{B}}\left(\omega\right) =& \;
\frac{e^2}{3}\sum_{\pm}\left< A\right|\vec r 
\frac{1}{H-E_B-\mathrm{i}\epsilon\pm\hbar\omega}
\vec r \left|B\right>\\
&=P_{A\underline{B}}\left(\omega\right)+P_{A\underline{B}}\left(-\omega\right)\,.
\end{align}
\end{subequations}
For the $(2S; 1S)$ system, one obtains
\begin{multline} \label{eq:Q1s2s}
P_{1S2S}\left(\omega\right) =
\frac{e^2\,a_0^2}{E_h}\frac{512\sqrt{2}\,\nu^2}%
{729\left(-1+\nu^2\right)^2\left(-4+\nu^2\right)^3}\\
\times\left(\vphantom{\left(1,-\nu;1-\nu;\frac{1-\nu}{1+\nu}\frac{2-\nu}{2+\nu}\right)}128-
272\nu^2+120\nu^4+253\nu^6+972\nu^7+419\nu^8\right.\\
\left.-1\,944\nu^7\,
{}_2F_1\left(1,-\nu;1-\nu;\frac{1-\nu}{1+\nu}\frac{2-\nu}{2+\nu}\right)\right),
\\
\nu = n_{\mathrm{eff}}\left( 1 + 
\frac{2n_{\mathrm{eff}}^2 \hbar \omega}{\alpha^2 m c^2} \right)^{-1/2} \,.
\end{multline}
Here we will typically choose the effective quantum number $n_{\mathrm{eff}}$
to be either $1$ (which yields $P_{\underline{1S}2S}$) or $2$ (which yields
$P_{1S\underline{2S}}$), as required for input into
Eq.~(\ref{eq:NowPolar}). Another possibility,
less physically transparent but quite handy for numerical calculations, is to
choose $n_{\mathrm{eff}}$ such that the reference energy
$E_{n_{\mathrm{eff}}}=-\alpha^2\,m\,c^2/\left(2n_{\mathrm{eff}}^2\right)$ in
the propagator corresponds to the average (\ref{eq:AvgEn}) of the energies of
the $n=1$ and $n=2$ levels (see Secs.~\ref{subsec:CloseD} and~\ref{subsec:CloseM}). 
The latter choice corresponds to $n_{\mathrm{eff}}=2\sqrt{2/5}$.

Taking retardation into account, the generalization of Eq.~\eqref{EMIXED}
(minus the unperturbed energy $E_0$)
to the Casimir-Polder energy is 
\begin{align} \label{eq:EntangledFinal}
&E_\pm =
{\rm Re} \frac{\mathrm{i}}{\pi} \frac{\hbar}{c^4\left(4 \pi \epsilon_0\right)^2} \, 
\int\limits_0^\infty {\rm d}\omega \, 
{\rm e}^{2 \ii \omega R/c} \, \frac{\omega^4}{R^2}\,
\nonumber\\[0.133ex]
&\times \left[\alpha_{1S}\left(\omega\right) \, 
\alpha_{2S}\left(\omega\right)\pm
\alpha_{\underline{1S}2S}\left(\omega\right) \, 
\alpha_{1S\underline{2S}}^*\left(\omega\right)\right]
\nonumber\\[0.133ex]
& \times \left[ 1 
+ 2 \ii \frac{c}{\omega R}
- 5 \left( \frac{c}{\omega R} \right)^2 
- 6 \ii \left( \frac{c}{\omega R} \right)^3 
+ 3 \left( \frac{c}{\omega R} \right)^4 
\right] 
\nonumber\\[0.133ex]
&=E_{2S;1S}^{\left(\mathrm{dir}\right)}\left(R\right)\pm 
E_{2S;1S}^{\left(\mathrm{mxd}\right)}\left(R\right) \,.
\end{align}
This result generalizes Eq.~(\ref{eq:EigenE}) to the 
Casimir--Polder regime.
It involves the mixed polarizabilities defined in Eq.~(\ref{eq:AlphaMixedPol}). 
We refer to the second summand $E_{2S;1S}^{\left(\mathrm{mxd}\right)}$ 
as the exchange term. 
Diagrammatically, it is obtained from a process in which 
an initial 
$\left|1S\right>_A\left|2S\right>_B$ atoms makes a 
transition to a $\left|2S\right>_A\left|1S\right>_B$ 
state via the exchange of two photons.
As was the case for the direct $2S$--$1S$ interaction term, we can single out
three different distance regimes for the exchange term, which we now
investigate.

%
% van der Waals range of interatomic distance}
%
\subsection{van der Waals range 
\texorpdfstring{$\bm{ a_0\ll R\ll a_0/\alpha}$}{}}
\label{subsec:CloseM}

In the \vdw\,range (\ref{vdWrange}), we proceed in a similar way to
Sec.~\ref{subsec:CloseD}, and have 
% TURNING THIS INTO SUBEQ.
\begin{subequations} \label{eq:EMxdvdW}
\begin{align} \label{eq:vdWMix}
E_{2S;1S}^{\left(\mathrm{mxd}\right)}\left(R\right) \approx & \;
{\rm Re}\, \frac{3\mathrm{i}}{\pi} \frac{\hbar}{(4 \pi \epsilon_0)^2}\frac{1}{R^6}  \, 
 \int\limits_0^\infty {\rm d}\omega \, 
\alpha_{\underline{1S}2S}(\omega) \, \alpha_{1S\underline{2S}}(\omega) 
\nonumber\\
=& \; {\rm Re}\, \frac{3\mathrm{i}}{\pi} \frac{\hbar}{(4 \pi \epsilon_0)^2}\frac{1}{R^6} \, 
 \int\limits_0^\infty {\rm d}\omega \, 
\alpha_{\underline{1S}2S}(\omega) \nonumber\\
& \times
\left[ \widetilde\alpha_{1S\underline{2S}}(\omega) + 
\overline\alpha_{1S\underline{2S}}(\omega)\right].
\end{align}
This can be rewritten as
\begin{equation}
\label{M62S1Sshort}
E_{2S;1S}^{\left(\mathrm{mxd}\right)}\left(R\right) = 
-\frac{M_6\left(2S;1S\right)}{R^6}\,.
\end{equation}
\end{subequations}
where $M_6\left(2S;1S\right)=\widetilde{M}_6\left(2S;1S\right)+
\overline{M}_6\left(2S;1S\right)$ is the sum of the nondegenerate
$\widetilde{M}_6\left(2S;1S\right)$ and degenerate
$\overline{M}_6\left(2S;1S\right)$ contributions to
the mixed van der Waals coefficient, with 
notations obvious from (\ref{eq:vdWMix}). As was done before, we
can, for the nondegenerate contribution, 
perform the Wick rotation. For the degenerate
contribution, 
we follow the same procedure as in
Sec.~\ref{subsec:CloseD}, centered on the integral identity (\ref{eq:CUInt}).
This yields
\begin{align} 
\label{eq:WH}
E_{2S;1S}^{\left(\mathrm{mxd}\right)}\left(R\right) &=
-\left(-18.630\,786\,871+\frac{917504}{19683}\right) E_h
\left(\frac{a_0}{R}\right)^6
\nonumber\\
&=-27.983\,245\,543\,E_h\left(\frac{a_0}{R}\right)^6
\end{align}
where we make use of (\ref{eq:Q1s2s}), whence
\begin{equation} \label{eq:AddingM6}
M_6=27.983\,245\,543\,E_h\,a_0^6\,,
\end{equation}
to be compared to $D_6$ as given by Eq.~(\ref{eq:WickInt}).
The two terms in Eq.~\eqref{eq:WH} correspond to 
the nondegenerate ($-18.630\,786\ldots$) and 
degenerate ($\frac{917504}{19683}$) contributions, respectively.
Their sum matches the results found in
Refs.~\cite{Ch1972,DeYo1973}.

As was the case for the direct interaction (see Sec.~\ref{subsec:CloseD}), an
alternative treatment exists whereby we make use of the integral identities
(\ref{eq:CUInt}) and (\ref{eq:CDInt}). This yields the following expression for
the van der Waals coefficient $M_6$:
\begin{equation} \label{eq:WickIntM}
M_6\left(2S;1S\right) =
\frac{3}{\pi} \frac{\hbar}{(4 \pi \epsilon_0)^2} 
\int_0^{\infty}\dd\omega\,
\left| \alpha_{\overline{1S2S}}\left(\omega\right) \right|^2 \,
\end{equation}
where the mixed polarizability $\alpha_{\overline{1S2S}}$ 
with average reference energy (\ref{eq:AvgEn}) is defined by
\begin{align} \label{eq:PolInt}
\alpha_{\overline{1S2S}}\left(\omega\right)=& \;
\frac{e^2}{3}\sum_{\pm}\left< 1S\right|\vec r 
\frac{1}{H-E_{\overline{1S2S}}\pm\hbar\omega}
\vec r \left|2S\right>
\nonumber\\
=& \;
P_{\overline{1S2S}}\left(\omega\right) +
P_{\overline{1S2S}}\left(-\omega\right)\,.
\end{align}
A numerical calculation based on Eq.~\eqref{eq:WickIntM}
confirms the result given in Eq.~\eqref{eq:WH}.

%
% Limit of Large Distance
%
\subsection{Very large interatomic distance 
\texorpdfstring{$\bm{R \gg \hbar c/\mathcal{L}}$}{}}
\label{subsec:AwayM}

For very large interatomic separations, 
the paradigm of Sec.~\ref{subsec:AwayD} applies.
In particular, the order-of-magnitude estimates
given in Eq.~\eqref{estimates} apply to the 
mixing term as well. 
We do not consider the tiny frequency shifts 
of order $10^{-36} \, {\rm Hz}$ or less
in this range any further, here.

%
% Intermediate Distances
%
\subsection{Intermediate distance 
\texorpdfstring{$\bm{a_0/\alpha \ll R\ll \hbar c/\mathcal{L}}$}{}}
\label{subsec:MidM}

In the intermediate range of interatomic distances,
the treatment follows that of Sec.~\ref{subsec:MidD}. 
Namely, only quasi-degenerate intermediate states contribute 
non-negligibly to the interaction, and we find
\begin{align}
\label{C62S1Smediumagain}
E_{2S; 1S}^{\left(\mathrm{mxd}\right)}\left(R\right) \approx & \;
- \frac{\overline M_6(2S; 1S)}{R^6} 
= - \frac{917\,504}{19\,683} E_h \left( \frac{a_0}{R} \right)^6 \,.
\end{align}
The interaction is thus still of the $R^{-6}$ form, 
as it is in the \vdw\,range, but 
the coefficient ($-46.614\,032\,414$) is different from the 
one relevant to the \vdw{} range, 
given in Eq.~\eqref{eq:WH}.

%
% $\delta$--Induced Modification of the Long--Range Interaction
%
\section{Dirac--\texorpdfstring{$\maybebm{\delta}$}{delta}
Induced Modification of the Long--Range Interaction}
\label{delta}

\subsection{Formalism and notations} \label{subsec:FormDelta}

In order to analyze the perturbation of the 
Casimir--Polder energy by an external 
potential proportional to a Dirac--$\delta$ acting 
one of the two atoms (say, atom $A$),
we now have to consider the perturbation of the 
polarizability of atom $A$ in Eq.~\eqref{eq:EntangledFinal}.
For the perturbation of the Casimir--Polder 
interaction due to a Dirac-$\delta$ potential,
we use this potential in the 
``standard normalization''~\cite{Je2003jpa},
which results in a unit prefactor in the energy shift,
\begin{align}
\label{deltaV}
& \delta V = \alpha m c^2
\left( \frac{\hbar}{m c} \right)^3 \pi\, 
\delta^{\left(3\right)}\left(\vec{r}_A\right) \,, 
\nonumber\\[0.133ex]
& \left< nS \left| \delta V \right| nS \right> =
\frac{\alpha^4 m c^2}{n^3} \,.
\end{align}
We shall consider atom $A$ (not $B$) to be perturbed.
The perturbation of the interaction energy (\ref{eq:EntangledFinal}) is
{\allowdisplaybreaks
\begin{align}
\label{deltaEreg}
& \delta_A E_{A;B}(R) = {\rm Re}\frac{\ii \hbar }%
{\pi c^4}  \int\limits_0^\infty 
{\rm d}\omega\,\frac{\ee^{2 \ii \omega R/c}}{(4 \pi \epsilon_0)^2}
\left\{\vphantom{\left[\left(\delta_A\alpha_{\underline{A}B}(\omega)\right) 
\alpha_{A\underline{B}}(\omega)+
\alpha_{\underline{A}B}(\omega) \,
\left(\delta_A\alpha_{A\underline{B}}(\omega)\right)\right]}\left( \delta_A \alpha_A(\omega) \right) 
\alpha_B(\omega) 
\right.
\nonumber\\[0.133ex]
& \left.
\pm\left[\left(\delta_A\alpha_{\underline{A}B}(\omega)\right) 
\alpha_{A\underline{B}}(\omega)+
\alpha_{\underline{A}B}(\omega) \,
\left(\delta_A\alpha_{A\underline{B}}(\omega)\right)\right]\right\}
\nonumber\\[0.133ex]
& \times 
\frac{\omega^4}{R^2} 
\left[ 1 
+ 2 \ii \frac{c}{\omega R}
- 5 \left( \frac{c}{\omega R} \right)^2 
- 6 \ii \left( \frac{c}{\omega R} \right)^3 
+ 3 \left( \frac{c}{\omega R} \right)^4 
\right].
\end{align}
Here, $\delta_A \alpha_A(\omega)$ is the Dirac-$\delta$ perturbation of the
polarizability of atom $A$ due to the potential $\delta V$, and
$\delta_A\alpha_{\underline{A}B}$ and $\delta_A\alpha_{A\underline{B}}$ are the
corrections to the mixed
polarizabilities of the type~(\ref{eq:AlphaMixedPol}). 
All of these corrections entail both an
energy as well as a wave function correction. 
We do not consider atom $B$ to be perturbed in our treatment.
We will focus in what follows on the $\delta$-corrections to the Casimir-Polder
interaction of the system $| A \rangle = | 2S\rangle$,
and $|B\rangle = | 1S \rangle$. As evident from Eq.~(\ref{deltaEreg}), and
expected from Secs.~\ref{long-range} and \ref{exchange}, we 
need to investigate the correction to the direct and exchange terms.

More concretely, in the case of the $2S$--$1S$ system, we have
\begin{align} \label{deltaEreg1S2S} 
& \delta_{2S} E_{2S;1S}(R) = {\rm Re}
\frac{\ii \hbar}{\pi \, c^4 \, (4 \pi \epsilon_0)^2} \, \int\limits_0^\infty
{\rm d}\omega \, {\rm e}^{2 \ii \omega R/c} \, \frac{\omega^4}{R^2}\,
\nonumber\\[0.133ex] 
& \; \times \left\{ \left( \delta_{2S} \alpha_{2S}(\omega) \right) \,
\alpha_{1S}(\omega) 
\pm\left[ \left(\delta_{2S}\alpha_{\underline{1S}2S}(\omega)\right) \,
\alpha_{1S\underline{2S}}(\omega) \right. \right.
\nonumber\\[0.133ex] 
& \; \left. \left. \qquad + \alpha_{\underline{1S}2S}(\omega) \,
\left(\delta_{2S}\alpha_{1S\underline{2S}}(\omega) \right)\right]\right\}
\nonumber\\[0.133ex] 
& \; \times \left[ 1 + 2 \ii \frac{c}{\omega R}
- 5 \left( \frac{c}{\omega R} \right)^2 
- 6 \ii \left( \frac{c}{\omega R} \right)^3 + 3 \left( \frac{c}{\omega R}
  \right)^4 \right]  \,, 
\end{align}
with the corrections to the various polarizabilities given by
\begin{subequations} \label{eq:Dreamland} \begin{align}
\delta_{2S}\alpha_{2S}(\omega) =& \; \delta_{2S} P_{2S}(\omega) + \delta_{2S}
P_{2S}(-\omega)\,, \nonumber\\
\delta_{2S} P_{2S}(\omega) =& \; \frac{e^2}{3}\left< 2S\right|\vec r \,
\frac{1}{\left(H-E_{2S}-\mathrm{i}\epsilon+\hbar\omega\right)^2} \, \vec r
\left|2S\right> \nonumber\\
& \times \left< 2S\right|\delta V\left|2S\right> \nonumber\\ & + \frac{2}{3}e^2 \,
\left< 2S\right|\vec r \, \frac{1}{H-E_{2S}-\mathrm{i}\epsilon+\hbar\omega} \,
\vec r \left|\delta2S\right> \,, \label{eq:PlainSortOf} \\[0.133ex]
\delta_{2S}\alpha_{\underline{1S}2S}(\omega) =& \; \delta_{2S}
P_{\underline{1S}2S}(\omega) + \delta_{2S} P_{\underline{1S}2S}(-\omega)\,,
\nonumber\\
\delta_{2S} P_{\underline{1S}2S}(\omega) =& \; \frac{e^2}{3}\left< 1S\right|\vec
r \, \frac{1}{\left(H-E_{1S}-\mathrm{i}\epsilon+\hbar\omega\right)^2} \, \vec r
\left|\delta2S\right> \,, \label{eq:OnlyWF} \\[0.133ex]
\delta_{2S}\alpha_{1S\underline{2S}}(\omega) =& \; \delta_{2S}
P_{1S\underline{2S}}(\omega) + \delta_{2S}
P_{1S\underline{2S}}(-\omega),\nonumber\\ \delta_{2S}
P_{1S\underline{2S}}(\omega) =& \; \frac{e^2}{3}\left< 1S\right|\vec r
\frac{1}{\left(H-E_{2S}-\mathrm{i}\epsilon+\hbar\omega\right)^2} \, \vec r
\left|2S\right> \nonumber\\
&\times\left< 2S\right|\delta V\left|2S\right>\nonumber\\ &+\frac{e^2}{3}\left<
1S\right|\vec r \, \frac{1}{H-E_{2S}-\mathrm{i}\epsilon+\hbar\omega} \, \vec r
\left|\delta2S\right> \label{eq:HardSortOf} \,.  \end{align} \end{subequations}
}
The first term in (\ref{eq:PlainSortOf}) and that in (\ref{eq:HardSortOf})
are identified as energy-type corrections, because they
describe
modifications to the respective polarizabilities due to the change in the $2S$
reference energy. We refer to the corresponding corrections to the respective
polarizabilities as $\delta_{2S}\alpha_{2S}^{\left(E\right)}$ and
$\delta_{2S}\alpha_{1S\underline{2S}}^{\left(E\right)}$.

Notice that
(\ref{eq:OnlyWF}) does not feature such a term, as the reference energy in the
denominator of $\alpha_{\underline{1S}2S}$ is that of the $1S$ state. The
second term in (\ref{eq:PlainSortOf}) and that in (\ref{eq:HardSortOf}), as
well as the lone term in (\ref{eq:OnlyWF}) are called wave function-type
corrections, because the corresponding terms are modifications to the
respective polarizabilities due to the change in the $2S$ state (and hence wave
function). We refer to the corresponding corrections to the respective polarizabilities as
$\delta_{2S}\alpha_{2S}^{\left(\psi\right)}$,
$\delta_{2S}\alpha_{\underline{1S}2S}^{\left(\psi\right)}$ and
$\delta_{2S}\alpha_{1S\underline{2S}}^{\left(\psi\right)}$. 
The correction $\left|\delta 2S\right>$ to the $\left|2S\right>$ 
state is given by the usual
expression
\begin{equation} 
\left|\delta2S\right> = \frac{1}{\left(E_{2S}-H\right)'} \,
\delta V\left|2S\right> \,.
\end{equation}
The corresponding wave function is
\begin{multline} 
\label{eq:DeltaPsi2S}
\delta\psi_{2S}(\vec r) =
\frac{\alpha^2}{\sqrt{2}}\frac{1}{\sqrt{4\pi}}
\left(\frac{1}{a_0}\right)^{5/2} \,
\exp\left( -\frac{r}{2 \, a_0} \right)
\\
\left[-\frac{a_0^2}{2r}-\frac{a_0\left(3-4\gamma_E-
4\ln\left(\frac{r}{a_0}\right)\right)}{4}\right.\\
\left.-\frac{r\left(-13+4\gamma_E +4 \,
\ln\left(\frac{r}{a_0}\right)\right)}{8}-\frac{r^2}{8a_0}\right] \,,
\end{multline}
where $\gamma_E \simeq 0.577\,216$ is the Euler--Mascheroni constant. Finally, the
first-order correction to the Hamiltonian due to $\delta V$ in the propagator
vanishes because the only contributing states are $P$ states whose probability
density vanishes at the origin. Namely,
\begin{equation} \label{eq:ZeroHC}
\left< nS 
\left| \vec r \, \frac{1}{(H - E + \hbar \omega)'}\delta V
\frac{1}{(H - E + \hbar \omega)'} \, \vec r \right| mS \right>=0\,
\end{equation}
regardless of the choice of $E$ and that of the principal quantum numbers $n$ and $m$.
With (\ref{deltaEreg1S2S}) and (\ref{eq:Dreamland}) we are equipped for the
investigation of the various distance regimes.

%
% Dirac-$\delta$ perturbation 
% 
\subsection{Dirac-\texorpdfstring{$\bm{\delta}$}{delta} perturbation 
in the van der Waals range \texorpdfstring{$\bm{a_0\ll R\ll a_0/\alpha}$}{}}
\label{subsec:CloseDelta}

For small separations, %REDUNDANT $a_0\ll R\ll a_0/\alpha$
the energy shift~(\ref{deltaEreg}) is approximated 
by an $R^{-6}$ interaction, as was done in
Secs.~\ref{subsec:CloseD} and~\ref{subsec:CloseM}. 
We shall use intermediate reference energies of the 
type (\ref{eq:AvgEn}) in the propagators and thus start 
from the expressions (\ref{eq:WickInt}) and (\ref{eq:WickIntM}) 
for the direct ($D_6$) and mixed ($M_6$)
coefficients, duly perturbed by the Dirac-$\delta$ potential.
This allows us to treat both nondegenerate and quasi-degenerate contributions to
these coefficients at once. 
Both energy and wave function corrections
contribute to $\delta C_6= \delta D_6\pm \delta M_6$.
This adds complexity on top of the degenerate-nondegenerate 
dichotomy, and the use of the intermediate 
reference energies in the propagator denominators 
ensures that we can avoid dealing with the 
degenerate and nondegenerate states separately.

We obtain the 
correction to the $D_6$ and $M_6$ coefficients
either by taking the short-range limit of~\eqref{deltaEreg}
and using the mean excitation energy $E_{\overline{1S 2S}}$,
or by perturbing the explicit expressions
(\ref{eq:WickInt}) and (\ref{eq:WickIntM}) by the Dirac-$\delta$.
In both approaches, the result is 
\begin{align}
\label{eq:DeltaD6Exp}
\delta  D_6(2S;1S) =& \;
\frac{3}{\pi} \frac{\hbar}{c^4 \, (4 \pi \epsilon_0)^2} \, 
\nonumber\\
& \; \times \int\limits_0^\infty {\rm d}\omega
\left[\delta_{2S}  \alpha_{1S\left(2S\right)}(\ii \omega)\, 
\alpha_{2S\left(1S\right)}(\ii \omega)\right.
\nonumber\\
& \; \left. +\alpha_{1S\left(2S\right)}(\ii \omega)\, 
\delta_{2S}\alpha_{2S\left(1S\right)}(\ii \omega)\right]
\end{align}
and
\begin{align} \label{eq:DeltaM6Exp}
\delta M_6(2S;1S) = & \;
\frac{6}{\pi} \frac{\hbar}{c^4 \, (4 \pi \epsilon_0)^2} \, 
\int\limits_0^\infty {\rm d}\omega \, 
\nonumber\\
& \; \alpha_{\overline{1S2S}}\left(\ii \omega\right) \,
\delta_{2S}\alpha_{\overline{1S2S}}\left(\ii \omega\right).
\end{align}
The Dirac-$\delta$ corrections to the polarizabilities (\ref{eq:PolGalore}) 
and (\ref{eq:PolInt}) involve the mean excitation energy
(\ref{eq:AvgEn}) in the propagator,
\begin{subequations} \label{eq:DPolGalore}
\begin{align}
\delta_{2S}\alpha_{1S\left(2S\right)}(\omega)
=& \; \delta_{2S}P_{1S\left(2S\right)}(\omega) +
\delta_{2S}P_{1S\left(2S\right)}(-\omega)\,,
\nonumber\\
\delta_{2S}P_{1S\left(2S\right)}(\omega)= & \;
\frac{e^2}{3}\left< 1S\right|\vec r \,
\frac{1}{\left(H-E_{\overline{1S2S}}+\hbar\omega\right)^2} \,
\vec r \left|1S\right>
\nonumber\\
& \; \times\frac{1}{2}\left< 2S\right|\delta V\left|2S\right>,\label{eq:OnlyEn}
\\[0.133ex]
\delta_{2S}\alpha_{2S\left(1S\right)}(\omega) = & \;
\delta_{2S}P_{2S\left(1S\right)}(\omega) +
\delta_{2S}P_{2S\left(1S\right)}(-\omega)\,,
\nonumber\\
\delta_{2S}P_{2S\left(1S\right)}(\omega) =& \;
\frac{e^2}{3}\left< 2S\right|\vec r \, 
\frac{1}{\left(H-E_{\overline{1S2S}}+\hbar\omega\right)^2} \,
\vec r \left|2S\right>
\nonumber\\
& \; \times\frac{1}{2}\left< 2S\right|\delta V\left|2S\right>
\nonumber\\
& \; + \frac{2}{3}e^2\left< 2S\right|\vec r \, 
\frac{1}{H-E_{\overline{1S2S}}+\hbar\omega} \,
\vec r\left|\delta2S\right>,
\label{eq:BothandTwice}\\[0.133ex]
\delta_{2S}\alpha_{\overline{1S2S}}(\omega)=& \;
\delta_{2S}P_{\overline{1S2S}}(\omega) +
\delta_{2S}P_{\overline{1S2S}}(-\omega)\,
\nonumber\\
\delta_{2S}P_{\overline{1S2S}}(\omega)=& \;
\frac{e^2}{3}\left< 1S\right|\vec r \, 
\frac{1}{\left(H-E_{\overline{1S2S}}+\hbar\omega\right)^2} \,
\vec r \left|2S\right>
\nonumber\\
& \; \times\frac{1}{2}
\left< 2S\right|\delta V\left|2S\right>
\nonumber\\
& \; + \frac{e^2}{3}\left< 1S\right|\vec r \, 
\frac{1}{H-E_{\overline{1S2S}}+\hbar\omega} \, 
\vec r\left|\delta2S\right>.
\label{eq:BothandOnce}
\end{align}
\end{subequations}
We recall that the use of the mean energy 
$E_{\overline{1S2S}}$ amounts to making the choice of the intermediate 
effective quantum number $n_{\mathrm{eff}}=2\sqrt{2/5}$ [see discussion below
Eq.~(\ref{eq:Q1s2s})]. Again, we distinguish the energy-type corrections, which
correspond to the first summand in (\ref{eq:BothandTwice}) and that in
(\ref{eq:BothandOnce}), as well as the lone term in (\ref{eq:OnlyEn}).
We write them as $\delta_{2S}\alpha_{2S\left(1S\right)}^{\left(E\right)}$,
$\delta_{2S}\alpha_{\overline{1S2S}}^{\left(E\right)}$, and
$\delta_{2S}\alpha_{1S\left(2S\right)}^{\left(E\right)}$ respectively.
The wave function-type corrections correspond to the second summand in
(\ref{eq:BothandTwice}) and that in (\ref{eq:BothandOnce}), and we write
them as $\delta_{2S}\alpha_{2S\left(1S\right)}^{\left(\psi\right)}$,
$\delta_{2S}\alpha_{\overline{1S2S}}^{\left(\psi\right)}$, respectively.

By a generalization of numerical techniques 
described previously~\cite{Pa1993,JePa1996},
we find that the numerical value of (\ref{eq:DeltaD6Exp}) is
\begin{align} 
\label{eq:NumDeltaC612}
\delta  D_6(2S;1S) =& \; 367.914\,605\,710\, \alpha^2\,a_0^6\,E_h\,.
\end{align}
By a similar procedure we find the numerical value of (\ref{eq:DeltaM6Exp}),
\begin{align} \label{eq:NumDeltaM612}
\delta  M_6(2S;1S) =-58.095\,351\,093\, \alpha^2\,a_0^6\,E_h\,.
\end{align}
Details on the calculation of (\ref{eq:NumDeltaC612}) and
(\ref{eq:NumDeltaM612}) are given in
Appendix~\ref{app:GiveMeMore}.

%
% Dirac-$\maybebm{\delta}$ perturbation for intermediate distances
%
\subsection{Dirac-\texorpdfstring{$\maybebm{\delta}$}{delta} 
perturbation for intermediate distance 
\texorpdfstring{$\bm{a_0/\alpha \ll R\ll \hbar c/\mathcal{L}}$}{}}
\label{subsec:MidDelta}

As was the case for the unperturbed interaction, 
it is very interesting to focus on the intermediate distance range.
Here again, as far as virtual transitions with a change in the 
principal quantum number are concerned, we are 
deeply in the Casimir--Polder regime where the result 
is given by an $R^{-7}$ interaction.
However, for virtual transitions to the quasi-degenerate states,
the frequency range is so low that the 
contribution to the Casimir-Polder integral~\eqref{deltaEreg} is not
exponentially suppressed. We therefore obtain
\begin{align}
\label{deltaEregCP}
& \delta_A E_{A;B}(R) \approx
{\rm Re} \frac{\ii \hbar}{\pi c^4 (4 \pi \epsilon_0)^2}
\int\limits_0^\infty {\rm d}\omega \, 
{\rm e}^{2 \ii \omega R/c} \, \frac{\omega^4}{R^2} \,
\nonumber\\[0.133ex]
& \; \times \bigl\{
\left( \delta_A \overline{\alpha}_A(\omega) \right) \, \alpha_B(0) 
\pm \left[\left(\delta_A\alpha_{\underline{A}B}(0)\right) \,
\overline{\alpha}_{A\underline{B}}(\omega) 
\right.
\nonumber\\[0.133ex]
& \; 
\left. 
\qquad + \alpha_{\underline{A}B}(0) \,
\left(\delta_A\overline{\alpha}_{A\underline{B}}(\omega)\right)\right]
\bigr\}
\nonumber\\[0.133ex]
& \times \left[ 1 
+ 2 \ii \frac{c}{\omega R}
- 5 \left( \frac{c}{\omega R} \right)^2 
- 6 \ii \left( \frac{c}{\omega R} \right)^3 
+ 3 \left( \frac{c}{\omega R} \right)^4 
\right] \,.
\end{align}
The rationale here is that the the quasi-resonant terms
(overlined $\alpha$'s) have to be kept in 
dynamic form (the dependence on $\omega$ is retained),
while the 
complementary terms can be taken in the static limit.

We shall treat the energy-type and wave function-type
corrections separately.
In the present Casimir-Polder range (\ref{CPrange}), 
the energy-type correction to the direct $2S$--$1S$ interaction is
{\allowdisplaybreaks
\begin{align}
\label{eq:DeltaEDirCPAlmost}
& \delta_{2S} E_{2S;1S}^{\left(\mathrm{dir}\right)\left(E\right)}(R) = 
{\rm Re} \frac{\ii \hbar \alpha_{1S}(0) }%
{\pi c^4}\int\limits_0^\infty {\rm d}\omega \,
\frac{\ee^{2 \ii \omega R/c}}{(4 \pi \epsilon_0)^2}\,
\delta_{2S} \overline{\alpha}_{2S}^{\left(E\right)}(\omega)
\nonumber\\[0.133ex]
& \times \frac{\omega^4}{R^2} \left[ 1 
+ 2 \ii \frac{c}{\omega R}
- 5 \left( \frac{c}{\omega R} \right)^2 
- 6 \ii \left( \frac{c}{\omega R} \right)^3 
+ 3 \left( \frac{c}{\omega R} \right)^4 
\right] 
\nonumber\\[0.133ex]
&=-{\rm Re} \frac{\ii \hbar \, \alpha_{1S}(0)}{\pi c^4 (4 \pi \epsilon_0)^2} \,
\left<2S\right|\delta V\left|2S\right>
\int\limits_0^\infty {\rm d}\omega \,
\ee^{2 \ii \omega R/c} \,\frac{\omega^4}{R^2}
\nonumber\\[0.133ex]
&\times\left[\frac{\partial}{\partial\mathcal{L}}
\frac{(-\mathcal{L})}{\left(-\mathcal{L}-\mathrm{i}\epsilon\right)^2 -
\left(\hbar\omega\right)^2} + 
2\frac{\partial}{\partial\mathcal{F}}
\frac{\mathcal{F}}{\left(\mathcal{F} -
\ii \epsilon\right)^2 - \left(\hbar\omega\right)^2}\right]
\nonumber\\[0.133ex]
& \times  \left[ 1 
+ 2 \ii \frac{c}{\omega R}
- 5 \left( \frac{c}{\omega R} \right)^2 
- 6 \ii \left( \frac{c}{\omega R} \right)^3 
+ 3 \left( \frac{c}{\omega R} \right)^4 
\right] \nonumber\\[0.133ex]
&\times\frac{e^2}{3} \sum_\mu
\left<2S\right| \vec r \left|2P\left(m\!=\!\mu\right)\right> \cdot 
\left<2P\left(m\!=\!\mu\right)\right| \vec r \left|2S\right> \,.
\end{align}}
Because of the pole structure of the integrand,
it is not possible to simply set the 
retardation function
\begin{multline}
R(\omega) =
\mathrm{e}^{2\mathrm{i}\omega R/c}\frac{\omega^4}{R^2} \left[ 1
+ 2 \ii \frac{c}{\omega R}
- 5 \left( \frac{c}{\omega R} \right)^2
\right.
\\
\left. - 6 \ii \left( \frac{c}{\omega R} \right)^3
+ 3 \left( \frac{c}{\omega R} \right)^4
\right]
\end{multline}
equal to unity (as was done in Secs.~\ref{subsec:MidD} and \ref{subsec:MidM});
the residue at the poles of the 
integrand in Eq.~\eqref{eq:DeltaEDirCPAlmost} 
otherwise cannot be calculated correctly.
In the $\mathcal{L}\rightarrow0$, $\mathcal{F}\rightarrow0$ limit, 
this gives
\begin{multline}
\label{eq:DeltaEDirCP}
\delta_{2S} E_{2S;1S}^{\left(\mathrm{dir}\right)\left(E\right)}(R) 
= -\frac{11}{6 \pi} \,
\frac{\alpha_{1S}(0)}{(4 \pi \epsilon_0)^2 \, \hbar c\,R^5} \,
\left<2S\right|\delta V\left|2S\right>
\\[0.133ex]
\times\frac{e^2}{3}
\sum_\mu \left<2S\right| \vec r \left|2P\left(m\!=\!\mu\right)\right>
\cdot \left<2P\left(m\!=\!\mu\right)\right| \vec r \left|2S\right>.
\end{multline}
Note that the individual terms of the retardation function 
$R(\omega)$, when used in Eq.~\eqref{eq:DeltaEDirCPAlmost},
give rise to logarithmic terms proportional to 
$\ln(2 \calF R)/R^5$ and $\ln(2 \calL R)/R^5$; these cancel in the final result.
From a similar procedure we obtain the correction to the exchange $2S$--$1S$
interaction as
\begin{align}
\label{eq:DeltaEMxdCP}
& \delta_{2S} E_{2S;1S}^{\left(\mathrm{mxd}\right)\left(E\right)}(R) =
-\frac{11}{6 \pi}\,
\frac{\alpha_{\underline{1S}2S}(0)}{(4 \pi \epsilon_0)^2 \, \hbar c\,R^5} \,
\left<2S\right|\delta V\left|2S\right>
\nonumber\\[0.133ex]
&\times \frac{e^2}{3}
\sum_\mu\left<1S\right| \vec r \left|2P\left(m\!=\!\mu\right)\right>
\cdot 
\left<2P\left(m\!=\!\mu\right)\right| \vec r \left|2S\right>.
\end{align}
The energy-type correction induces an $R^{-5}$ interaction
[see App.~\ref{ModelInt}]. The wave function-type corrections, on the other hand, are
treated in exactly the same way as the degenerate $\overline{D}_6$ and
$\overline{M}_6$ coefficients of Secs.~\ref{subsec:MidD} and \ref{subsec:MidM},
respectively. We can make use of (\ref{MASTER1}) and (\ref{MASTER2}) and obtain
\begin{multline} \label{eq:DeltaD6BarPsi}
\delta\overline{D}_6^{\left(\psi\right)}(2S;1S)= 
\frac{2}{(4 \pi \epsilon_0)^2}  \, 
\alpha_{1S}\left(0\right)\\[0.133ex]
\times\sum_\mu
\left<2S\right| \vec r \left|2P\left(m\!=\!\mu\right)\right>
\cdot 
\left<2P\left(m\!=\!\mu\right)\right| \vec r \left|\delta2S\right>
\end{multline}
and
\begin{multline} \label{eq:DeltaM6BarPsi}
\delta\overline{M}_6^{\left(\psi\right)}(2S;1S) = 
\frac{1}{(4 \pi \epsilon_0)^2}  \, 
\sum_\mu\left<1S\right| \vec r \left|2P\left(m\!=\!\mu\right)\right>
\\[0.133ex]
\cdot
\left[\delta\alpha_{\underline{1S}2S}^{\left(\psi\right)}\left(0\right) \,
\left<2P\left(m\!=\!\mu\right)\right| \vec r \left|2S\right>\right.
\\[0.133ex]
+\left.\alpha_{\underline{1S}2S}\left(0\right) \,
\left<2P\left(m\!=\!\mu\right)\right| \vec r \left|\delta2S\right>
\right] \,.
\end{multline}

From Eq.~(\ref{eq:DeltaEDirCP}), we find that for intermediate distances 
the energy-type correction to the interaction is given by
\begin{equation} \label{eq:DeltaMidDE}
\delta_{2S} E_{2S;1S}^{\left(\mathrm{dir}\right)\left(E\right)}(R) =
-\frac{891}{32} \, 
\frac{\alpha^3}{\pi}\left(\frac{a_0}{R}\right)^5 \, E_h \,.
\end{equation}
while the wave-function correction is
\begin{equation} 
\label{eq:DeltaMidDPsi}
\delta_{2S} E_{2S;1S}^{\left(\mathrm{dir}\right)\left(\psi\right)}(R)= 
-\frac{81}{4} \, \alpha^2\left(\frac{a_0}{R}\right)^6E_h\,.
\end{equation}
It is thus seen that, in the present intermediate range
(\ref{CPrange}), the dominant contribution comes from the
energy-type correction (\ref{eq:DeltaMidDE}). 

From (\ref{eq:DeltaEMxdCP}) we find that for intermediate distances the
energy-type correction to the exchange interaction is given by
\begin{equation} \label{eq:DeltaMidME}
\delta_{2S} E_{2S;1S}^{\left(\mathrm{mxd}\right)\left(E\right)}(R)
= \frac{630\,784}{59\,049} \, \frac{\alpha^3}{\pi} \,
\left(\frac{a_0}{R}\right)^5 \, E_h \,,
\end{equation}
while the wave-function correction is
\begin{multline} \label{eq:DeltaMidMPsi}
\delta_{2S} E_{2S;1S}^{\left(\mathrm{mxd}\right)\left(\psi\right)}(R)
= -\frac{8\,192}{19\,683}\\
\times\left[95+112\ln\left(\frac{3}{2}\right)\right] \,
\alpha^2\left(\frac{a_0}{R}\right)^6 \, E_h \,.
\end{multline}
As was the case for the direct interaction above, the dominant contribution
comes from the energy-type correction (\ref{eq:DeltaMidME}).

%
% Dirac-$\maybebm{\delta}$ perturbation for very large interatomic distances
%
\subsection{Very-long range Dirac-\texorpdfstring{$\maybebm{\delta}$}{delta} 
perturbation~\texorpdfstring{$\bm{R \gg \hbar c/\mathcal{L}}$}{}} 
\label{subsec:AwayDelta}

For very large interatomic separation, the 
considerations from Secs.~\ref{subsec:AwayD} and \ref{subsec:AwayM} carry over.
Perturbing the Lamb shift $\calL$ in Eq.~\eqref{estimates}
by the Dirac-$\delta$ potential~\eqref{deltaV},
one realizes that the Dirac-$\delta$ induced 
modification of the long-range interaction does not exceed 
$(\langle 2S|\delta V |2S \rangle / \calL) \times 10^{-36} \, {\rm Hz}$.
This shift is too small to be of conceivable experimental relevance and thus
not considered any further.

%
% Numerical Examples: Modification to the Hyperfine 
% Structure and Transition Frequencies
%
\section{Numerical Examples: Modification to the Hyperfine 
Structure and Transition Frequencies} 
\label{sec:Numerics}

In order to estimate the relevance of the current study,
let us recall that, e.g., the hyperfine frequency of a 
hydrogen atom in an $S$ state is determined by a Dirac-$\delta$ potential
[see Eq.~(\ref{eq:HyperFSH}) and discussion below].
Hyperfine frequencies belong to the most accurately
measured frequencies today~\cite{EsDoBaHo1971,EsDoHoBa1973,PeDeAu1980}.
Consequently, it becomes necessary to investigate
small perturbations to these frequencies caused,
e.g., by interactions with buffer gas atoms
or by interactions with other atoms in the atomic
beam (the latter would be an atom 
of the same kind as that whose hyperfine frequency is being studied).
The perturbations of hyperfine frequencies due to van der Waals interactions have
been considered in Refs.~\cite{RaLyDa1968a,RaLyDa1968b,RaDa1969,RaIkDa1970,DuDuDa1970}.
Hyperfine-perturbation coefficients in the \vdw{} range have been given in
Table~II~of Ref.~\cite{DuDuDa1970}
for H--He and H--Ne. (The hyperfine modification of the long-range 
interaction for two hydrogen atoms, however, is not 
indicated in Ref.~\cite{DuDuDa1970}.
Also, in Ref.~\cite{DuDuDa1970}, only ground-state
interactions were considered.)

In Eqs.~\eqref{eq:NumDeltaC612},~\eqref{eq:NumDeltaM612},%
%~\eqref{eq:DeltaAwayDDom},~\eqref{eq:DeltaAwayMDom},%
~\eqref{eq:DeltaMidDE},~\eqref{eq:DeltaMidME}, we had indicated results
for the Dirac-$\delta$ induced perturbation to the \vdw{} interaction, in the
close-range limit and in the intermediate range.  These
results, which are reproduced for convenience in Eqs.~\eqref{h1} and~\eqref{h2}
below, can be used directly in order to calculate the modification of the
hyperfine frequency under the influence of the long-range interaction.  As
explained in Sec.~\ref{orient}, a possible interference term due to the
non--Dirac-$\delta$ terms in the hyperfine Hamiltonian [see
Eq.~(\ref{eq:HyperFSH})], which might be assumed to influence the virtual $P$
states that are responsible for the \vdw{} interaction, vanishes after doing
the angular Racah algebra~\cite{VaMoKh1988}.  In order to interpolate between
the three asymptotic regimes, a numerical integration of Eq.~\eqref{deltaEreg}
is required.  The leading asymptotic terms indicated in
Eqs.~\eqref{h1} and~\eqref{h2} contain the essence of the changes in the
interaction in a very concise form and can be used in order to estimate the
effect of the long-range $2S$--$1S$ interaction on, e.g., the $2S$ hyperfine
frequency.

Let us now calculate the \vdw{} shift of the hyperfine frequency 
of an atom $A$ in a $1S$ or metastable $2S$ state due to 
its long-range interaction with a ground-state atom $B$.
The first summand in $H_{\mathrm{HFS}}$ in 
Eq.~(\ref{eq:HyperFSH}) is used as the perturbative
potential instead of the standard potential~(\ref{deltaV}). 
Only the term acting on atom $A$ is required. One can check that
\begin{equation} 
\label{eq:FromOneToAnother}
\delta H_{\rm HFS}=\frac{2}{3} g_s g_p \frac{m}{M}
\delta V(\vec r_A)\, \vec S_A \cdot \vec{I}_A \,.
\end{equation}
where $M$ is the proton mass. The splitting between the two hyperfine
components of the $1S_{1/2}$ level is 
given by $1\,420\,405\,751.773(1) \,
\mathrm{Hz}$~\cite{EsDoBaHo1971,EsDoHoBa1973,PeDeAu1980}.
This experimental value is very accurate (up to $10^{-3}\,\mathrm{Hz}$), which
indicates that modifications to it due to long-range interatomic interactions
could be relevant for 
future experiments and, in particular,
measurable. We can work out how this splitting is affected by the
interaction of the $1S$ atom with another $1S$ hydrogen atom. The corresponding
values are given in Table~\ref{tab:HFS1}. Note that the interaction
\emph{reduces} the energy splitting between the two hyperfine components of the
$1S_{1/2}$ level. Likewise, the splitting between the two hyperfine components
of the $2S_{1/2}$ level has been measured~\cite{KoEtAl2009} as
$177\,556\,834.3(6.7) \, \mathrm{Hz}$. 
From Eqs.~\eqref{h1} and~\eqref{h2} (as
well as numerical computations for the separations that do not clearly find
themselves either in the van der Waals or Casimir-Polder ranges)
we can work out how this splitting is affected by the interaction of the $2S$
atom with a $1S$ hydrogen atom. The corresponding values are given in
Table~\ref{tab:HFS2}. Again, the interaction \emph{reduces} the energy splitting
between the two hyperfine components of the $2S_{1/2}$ level.

\begin{center}
\begin{table} [t]
\begin{center}
\def\arraystretch{1.25}
\begin{tabular}{cl}
\hline
\hline
\multicolumn{1}{c}{distance} &
\multicolumn{1}{c}{$\delta\nu_{\mathrm{HFS}}(1S)$} \\
\hline
$20 \, \mbox{\AA}$    &$-3.387\times10^{1}\,\mathrm{Hz}$\\
$40 \, \mbox{\AA}$    &$-5.291\times10^{-1}\,\mathrm{Hz}$\\
$80 \, \mbox{\AA}$    &$-8.806\times10^{-3}\,\mathrm{Hz}$\\
$200 \, \mbox{\AA}$   &$-3.019\times10^{-5}\,\mathrm{Hz}$\\
$400 \, \mbox{\AA}$   &$-3.919\times10^{-7}\,\mathrm{Hz}$\\
$800 \, \mbox{\AA}$   &$-4.296\times10^{-9}\,\mathrm{Hz}$\\
$2\,000\, \mbox{\AA}$ &$-1.059\times10^{-12}\,\mathrm{Hz}$\\
$20\,000\, \mbox{\AA}$&$-1.059\times10^{-19}\,\mathrm{Hz}$\\
\hline
\hline
\end{tabular}
\end{center}
\caption{Numerical values of the modification
$\delta\nu_{\mathrm{HFS}\left(1S\right)}$ to the frequency splitting between
the $1S$ hyperfine components in hydrogen, due to the long-range interaction
with a $1S$ atom, as a function of the inter-atomic separation.
\label{tab:HFS1}}
\end{table}
\end{center}

\begin{table} [t]
\begin{center}
\def\arraystretch{1.25}
\begin{tabular}{cl}
\hline
\hline
\multicolumn{1}{c}{distance} &
\multicolumn{1}{c}{$\delta\nu_{\mathrm{HFS}}(2S)$} \\
\hline
$20\mbox{\AA}$&$-\left(3.592\mp 0.567\right)\times10^{2}\,\mathrm{Hz}$\\
$40\mbox{\AA}$&$-\left(5.612\mp 0.886\right)\times10^{0}\,\mathrm{Hz}$\\
$80\mbox{\AA}$&$-\left(8.769\mp 1.441\right)\times10^{-2}\,\mathrm{Hz}$\\
$200\mbox{\AA}$&$-\left(3.592\mp 0.549\right)\times10^{-4}\,\mathrm{Hz}$\\
$400\mbox{\AA}$&$-\left(5.635\mp 0.781\right)\times10^{-6}\,\mathrm{Hz}$\\
$800\mbox{\AA}$&$-\left(9.023\mp 0.861\right)\times10^{-8}\,\mathrm{Hz}$\\
$2\,000\mbox{\AA}$&$-\left(2.584\mp 1.486\right)\times10^{-10}\,\mathrm{Hz}$\\
$20\,000\mbox{\AA}$&$-\left(2.406\mp 0.973\right)\times10^{-15}\,\mathrm{Hz}$\\
\hline
\hline
\end{tabular}
\end{center}
\caption{Numerical values of the modification
$\delta\nu_{\mathrm{HFS}\left(2S\right)}$ to the frequency splitting between
the $2S$ hyperfine components in hydrogen, due to the long-range interaction
with a $1S$ atom, as a function of the inter-atomic separation. The $\mp$ sign
corresponds to the $\pm$ sign in the
$\left(\left|1S\right>\left|2S\right>\pm\left|2S\right>\left|1S\right>\right)$
superposition. \label{tab:HFS2}}
\end{table}

From the results of Secs.~\ref{long-range} and~\ref{exchange}, we can also
deduce the modifications to the $2S$--$1S$ transition frequency
due to long-range interaction with a ground state hydrogen atom. 
We indicate numerical values 
for various interatomic separations in the \vdw{} and Casimir-Polder
(intermediate) ranges in Table~\ref{tab:TF2}. The $2S$--$1S$
transition has been measured \cite{PaEtAl2011} to be 
$2\,466\,061\,413\,187\,035(10) \, \mathrm{Hz}$ (for the hyperfine centroid). 
The experimental
accuracy is thus more than sufficient for the modifications predicted here to
be relevant (see Table~\ref{tab:TF2}). For the
values $R=2\,000\mbox{\AA}$ and $R=20\,000\mbox{\AA}$ of the distance that
we choose for the Casimir-Polder (intermediate) range 
($\hbar c/\calL \gg R \gg a_0/\alpha$),
the $R^{-7}$ contribution due to the
$P$ levels that are nondegenerate with $2S$ is not quite 
negligible, in contrast to larger $R$. We therefore include
the nondegenerate contributions in the calculation of 
the numerical value of the frequency shifts.
For definiteness, the value of the mass $m$ used in the numerical 
calculations is always chosen as the electron mass,
not the reduced mass of the electron-proton system. 
If we were to choose the reduced mass instead, 
then we would have to 
differentiate in the Dirac-$\delta$ term given in 
Eq.~\eqref{deltaV} the factor $1/m^2$, 
which still goes with the electron mass, and the 
reduced mass cubed, which enters the numerator
as it is proportional to the probability density 
at the origin, $| \psi(\vec 0)|^2$.
For definiteness, and in order to 
facilitate a numerical comparison of the 
results to other (conceivably, future) investigations,
we neglect further relativistic and reduced-mass corrections,
as well as quantum electrodynamic radiative corrections.
When applied to hydrogen,
these approximations limit the accuracy of the results given in 
Tables~\ref{tab:HFS1}---\ref{tab:TF2} to a relative
accuracy of about $10^{-4} \ldots 10^{-3}$. 

\begin{table} [t]
\begin{center}
\def\arraystretch{1.25}
\begin{tabular}{cl}
\hline
\hline
\multicolumn{1}{c}{distance} &
\multicolumn{1}{c}{$\delta\nu_{1S-2S}$} \\
\hline
$20\mbox{\AA}$&$-\left(3.843\pm 0.631\right)\times10^{8}\,\mathrm{Hz}$\\
$40\mbox{\AA}$&$-\left(6.005\pm 0.987\right)\times10^{6}\,\mathrm{Hz}$\\
$80\mbox{\AA}$&$-\left(9.365\pm 1.552\right)\times10^{4}\,\mathrm{Hz}$\\
$200\mbox{\AA}$&$-\left(3.806\pm 0.651\right)\times10^{2}\,\mathrm{Hz}$\\
$400\mbox{\AA}$&$-\left(5.838\pm 1.069\right)\times10^{0}\,\mathrm{Hz}$\\
$800\mbox{\AA}$&$-\left(8.776\pm 1.821\right)\times10^{-2}\,\mathrm{Hz}$\\
$2\,000\mbox{\AA}$&$-\left(3.521\pm 1.310\right)\times10^{-4}\,\mathrm{Hz}$\\
$20\,000\mbox{\AA}$&$-\left(2.820\pm 1.078\right)\times10^{-10}\,\mathrm{Hz}$\\
\hline
\hline
\end{tabular}
\end{center}
\caption{Numerical values of the modification $\delta\nu_{1S-2S}$ to the
hydrogen $2S$--$1S$ transition frequency, due to the long-range interaction
with a $1S$ atom, as a function of the inter-atomic separation. The $\pm$ sign
corresponds to the $\pm$ sign in the
$\left(\left|1S\right>\left|2S\right>\pm\left|2S\right>\left|1S\right>\right)$
superposition. \label{tab:TF2}}
\end{table}

It is also interesting to look at how these results are modified if we consider
the positronium instead of the hydrogen atom as the system of interest. It can
be shown that the plain (unperturbed) \vdw{} interaction energies will be scaled by a
factor of approximately $2^6=64$, as a result of the
fact that the reduced mass of positronium is roughly half the 
reduced mass of hydrogen [and hence, the expectation values of 
$\vec r$ operators will scale with a factor of two, 
as will the resolvent operators $1/(H - E)$].
The latter scaling factor is due to the fact that the transition
frequencies are only half of those of the hydrogen atom
(E.g., the $2S$--$1S$ transition in positronium has been measured 
at a value of $\approx 1.233 \times10^{15}\,\mathrm{Hz}$,
see Ref.~\cite{FeEtAl1993}.)
The \vdw{} modifications to the $2S$--$1S$ transition frequency 
in positronium as compared to hydrogen will exhibit the same scaling.
The relative modification of a positronium 
transition frequency will thus be $128$ times larger 
than for hydrogen. Similar scaling arguments 
show that the modification to the hyperfine splittings will be
scaled by a factor of $2^7g_s/g_p\simeq45.845\,112$.

Likewise, the leading effective QED radiative Lamb shift Hamiltonian
for atom $A$ can be obtained as a specialization of 
the expression given in Eq.~\eqref{HLS} to atom $A$,
\begin{equation}
\label{deltaVrad}
\delta H_{\rm rad} = 
\frac{4 \alpha}{3 \pi} \ln\left(\alpha^{-2}\right) \, \delta V \,,
\end{equation}
where again we express the relevant Hamiltonian in terms of the standard
potential $\delta V$ defined in Eq.~\eqref{deltaV}. 
The ratio of the prefactors as compared to the 
hyperfine Hamiltonian~\eqref{eq:FromOneToAnother} 
is 
\begin{equation*}
\frac{2}{\pi}\,\alpha\ln\left(\alpha^{-2}\right)\frac{1}{g_s\,g_p} \,
\frac{m_p}{m}\frac{1}{\langle \vec{S}_A\cdot\vec{I}_A \rangle }\simeq
7.505\,166 \,.
\end{equation*}
The operator $\vec{S}_A\cdot\vec{I}_A$ 
assumes the numerical value $+1/4$ for an $F=1$ state,
and the numerical value $-3/4$ for an $F=0$ state.
Hence, for the hyperfine splitting, it can be replaced by unity.
We thus note that the leading logarithmic QED radiative corrections to the 
$1S$--$1S$ and $2S$--$1S$ \vdw{} interactions are larger than the 
\vdw{} modification of the hyperfine splitting, by a 
factor of roughly $7.5$. 
The results given in Tables~\ref{tab:HFS1} and~\ref{tab:HFS2}
should be multiplied by this factor to obtain the 
leading radiative term.
The QED radiative correction to the
\vdw{} interaction shifts both hyperfine components by the same frequency and in 
the same direction, and thus does not additionally modify the hyperfine 
splitting. We also note that the QED radiative correction to the 
\vdw{} interaction could be interpreted alternatively as a 
\vdw{} correction to the Lamb shift. However, 
it is not the dominant modification of atomic 
transition frequencies mediated by long-range atomic 
interactions. Namely, the main effect on an atomic 
transition frequency with a change in the principal
quantum number is caused by the direct \vdw{} effect on the 
atomic levels, which is given (for the $2S$--$1S$)
in Table~\ref{tab:TF2}.

%
% Conclusion
%
\section{Conclusion}
\label{conclu}

We have studied $2S$--$1S$ \vdw{} interactions among
hydrogen atoms in detail, and carefully differentiate 
three distance ranges given in Eqs.~\eqref{vdWrange},~\eqref{CPrange1},
and~\eqref{LSrange}. %MODIFICATION HERE
In the \vdw{} range, the interatomic interaction is 
described to good accuracy by a functional form $-C_6(A;B)/R^6$,
where $C_6(A;B)=D_6(A;B)\pm M_6(A;B)$ is the \vdw{} coefficient, 
which depends on the atomic states $\left|A\right>$ and 
$\left|B\right>$ of the two atoms.
As mentioned above, a paradigmatic example for an
interaction involving metastable atoms is the $2S$ state of 
atomic hydrogen. Indeed,
for the interaction of a $2S$ hydrogen atom with a ground-state atom,
the result of $D_6(2S;1S) = 176.752 \, E_h \, a_0^6$ has been 
obtained in Ref.~\cite{DeYo1973}. %OFF WITH THIS SENTENCE We explicitly indicate 
%here the Hartree energy $E_h = \alpha^2 m c^2$ 
%and the Bohr radius $a_0$; both of these quantities 
%are set equal to one in atomic units.

As discussed, there is an interesting discrepancy with the 
results $D_6(2S;1S) = 56.8 \, E_h \, a_0^6$ (Ref.~\cite{TaCh1986})
and $D_6(2S;1S) = (56.5 \pm 0.5) \, E_h \, a_0^6$ (Ref.~\cite{Ch1972}).
We find that the result given in Ref.~\cite{DeYo1973}
is the correct one
and trace the likely explanation for the 
discrepancy to the rather subtle treatment of the 
quasi-degenerate $2P$ levels of the excited atom
(see Sec.~\ref{subsec:CloseD}).

In an atomic beam, one typically has a few excited 
metastable $2S$ atoms interacting with a ``background''
of $1S$ atoms. The $2S$ atoms are typically of interest, and 
that is why we have chosen the sequence $2S$--$1S$ in order 
to designate their interaction in our mathematical 
formulas (the {\em first} atom mentioned 
is the one of primary spectroscopic interest).
A typical application would consist in the 
measurement of the $2S$ hyperfine interval by optical 
spectroscopy~\cite{KoFiKaHa2004,KoEtAl2009}.

For the plain interaction of a $2S$ atom with a 
ground-state hydrogen atom, we find for the 
\vdw{} regime ($a_0 \ll R \ll \tfrac{a_0}{\alpha}$) %RANGE GIVEN HERE
[see Eqs.~(\ref{C62S1Sshortnum}) and~(\ref{eq:WH})]
\begin{multline}
\label{f1}
E_{2S; 1S}\left( R \right)
\approx - 
\left(176.752\,266 \pm 27.983\,245\right)
E_h \left( \frac{a_0}{R} \right)^6\,.%END HERE
\end{multline}
The term with the $\pm$ sign depends on the 
symmetry of the wave function of the two-atom state,
as explained in Sec.~\ref{exchange}.
In Eq.~\eqref{f1}, we thus confirm the result presented in Ref.~\cite{DeYo1973} but 
add a few more significant decimal digits 
of nominal numerical accuracy [see Eq.~\eqref{C62S1Sshortnum}].
In the Casimir--Polder range 
($\tfrac{a_0}{\alpha} \ll R \ll \tfrac{\hbar c}{\cal L}$), 
we also have an interaction
of the $R^{-6}$ type, with a coefficient determined 
by the quasi-degenerate states
[see Eqs.~(\ref{C62S1Smedium}) and~(\ref{C62S1Smediumagain})]
\begin{multline}
\label{f2}
E_{2S; 1S}\left( R \right)
\approx - 
\left(\frac{243}{2}\pm \frac{917\,504}{19\,683}\right)
E_h \left( \frac{a_0}{R} \right)^6\,.%END HERE
\end{multline}
In the Lamb shift range $R\gg \hbar c/\mathcal{L}$, the plain
interaction changes to a superposition  of a Casimir-Polder term of the form
$1/R^7$, and a long-range oscillating term of the type
$\cos\left(2\mathcal{L}R/\hbar c\right)/R^2$,
while the magnitude
of the interaction is too small to be of conceivable relevance
for experiments. For details,
see Secs.~\ref{long-range} and \ref{exchange}.

For the correction $\delta E_{2S;1S}(R)$
caused by a Dirac-$\delta$ potential, due to the long-range
interaction, the evolution of the 
asymptotic behavior is interesting.
For the \vdw{} range ($a_0 \ll R \ll \tfrac{a_0}{\alpha}$), %RANGE GIVEN HERE
our leading-order result is
[see Eqs.~\eqref{eq:NumDeltaC612} and~\eqref{eq:NumDeltaM612}]
\begin{multline}
\label{h1}
\delta_{2S}E_{2S; 1S}\left( R \right)
\approx - 
\left(367.914\,605 \mp 58.095\,351 \right) \\
\times \alpha^2 E_h \left( \frac{a_0}{R} \right)^6\,.
%END HERE
\end{multline}
In the Casimir--Polder range
($\tfrac{a_0}{\alpha} \ll R \ll \tfrac{\hbar c}{\cal L}$), 
the result is [see Eqs.~\eqref{eq:DeltaMidDE} and~\eqref{eq:DeltaMidME}]
\begin{multline}
\label{h2}
\delta_{2S}E_{2S; 1S}\left( R \right)
\approx - \left(\frac{891}{32}\mp\frac{630\,784}{59\,049}\right) 
\frac{\alpha^3}{\pi}E_h \left( \frac{a_0}{R} \right)^5\,,
\end{multline}
where the coefficient is exclusively given  by the 
quasi-degenerate states.
In the Lamb shift range $R\gg \hbar c/\mathcal{L}$, the
perturbed interaction also changes to a superposition of a Casimir-Polder term
and a long-range oscillating term, while the 
overall perturbation of the interaction is negligibly small.
All these results are derived in
Secs.~\ref{long-range} and \ref{exchange}.

Recently, long-range oscillatory tails of \vdw{} interactions have
received renewed interest in the literature~\cite{SaKa2015,DoGuLa2015}.  These
oscillatory tails are caused by states with a lower energy than the
excited reference state, in an interaction of excited-state and
ground-state atoms, which can be reached from the excited state by an
allowed dipole transition. As discussed in this paper, 
for the $2S$--$1S$ interaction, the $2S\rightarrow2P_{1/2}$
Lamb shift transition provides for such a transition. However, the energy
shifts typically are proportional to the fourth power of the transition
energy (or wave number of the transition).  For the Lamb shift transition,
this transition energy is very low, and in consequence, the oscillatory
tails are suppressed in the \vdw{} interaction of the $2S$--$1S$ system
[see Eq.~\eqref{estimates}]. 

%
% Acknowledgments
%
\section*{Acknowledgments}

The high-precision experiments carried out at MPQ Garching under the guidance
of Professor T.~W.~H\"{a}nsch have been a major motivation and inspiration for
the current theoretical work.  This research was supported by the National
Science Foundation (Grant PHY--1403973).

\appendix

\begin{figure}[th]
\begin{center}
\includegraphics[width=0.81\linewidth]{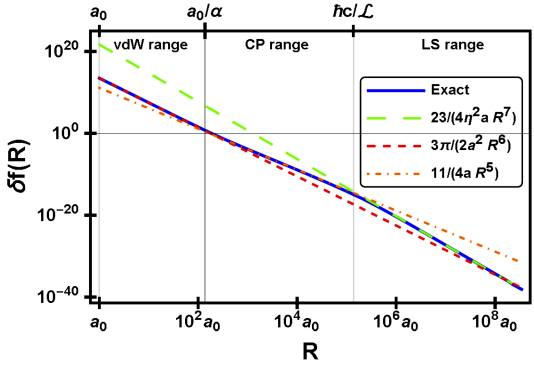}
\caption{\label{fig1} (Color online) Numerical
illustration of the model integral~\eqref{defI} in its
asymptotic regions, for the parameters given in Eq.~\eqref{param}.}
\end{center}
\end{figure}

\begin{figure}[th]
\begin{center}
\includegraphics[width=0.81\linewidth]{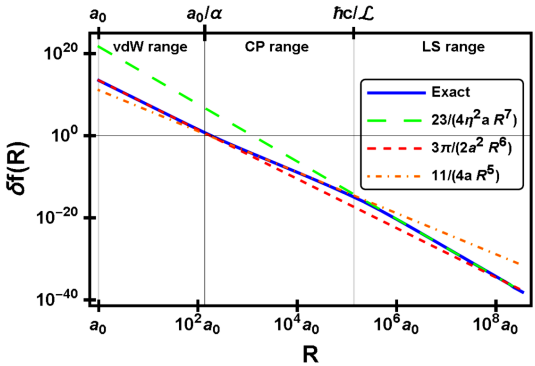}
\caption{\label{fig2} (Color online) Same as Fig.~\ref{fig1}
for  the model integral~\eqref{defJ}.}
\end{center}
\end{figure}

%
% Casimir--Polder interactions for $\bm{2S}$ states
%
\section{Model Integrals} \label{ModelInt}
\label{plots}

In order to illustrate the analytic considerations 
in Secs.~\ref{long-range},~\ref{exchange} and~\ref{delta},
we numerically study the model integrals
\begin{multline} 
\label{defI}
I\left(a,\eta,R\right)\equiv \int_0^{\infty}\mathrm{d}x\,
\frac{a}{\left(a-\mathrm{i}\epsilon\right)^2+x^2}\,
\frac{\left(-\eta\right)}{\left(-\eta-\mathrm{i}\epsilon\right)^2+x^2}\\
\times \mathrm{e}^{-2Rx}\,
\frac{x^4}{R^2}
\left[1+\frac{2}{Rx}+\frac{5}{\left(Rx\right)^2}+
\frac{6}{\left(Rx\right)^3}+\frac{3}{\left(Rx\right)^4}\right]\,,
\end{multline} 
which models the plain Casimir-Polder interaction 
as well as wave function-type corrections thereto; and
\begin{multline} 
\label{defJ}
J\left(a,\eta,R\right)\equiv\int_0^{\infty}\mathrm{d}x\,
\frac{a}{\left(a-\mathrm{i}\epsilon\right)^2+x^2}
\\
\times \frac{\partial}{\partial\eta}\frac{\left(-\eta\right)}%
{\left(-\eta-\mathrm{i}\epsilon\right)^2+x^2}
\\
\times \mathrm{e}^{-2Rx}\,
\frac{x^4}{R^2}
\left[1+\frac{2}{Rx}+\frac{5}{\left(Rx\right)^2}+
\frac{6}{\left(Rx\right)^3}+\frac{3}{\left(Rx\right)^4}\right]
\end{multline} 
which models energy-type corrections to the Casimir-Polder interaction. 
Our choices for the numerical values of the parameters are
\begin{subequations}
\label{param}
\begin{align}
\eta = & \; 10^{-3}\;,\\[0.133ex]
a = & 1\;, \\[0.133ex]
\epsilon = & 10^{-6}\; \,.
\end{align}
\end{subequations}
These values are adapted to the investigation of the quasi-degenerate
contributions to the interatomic interaction, $a$ playing the role of the
energy of a transition between quantum levels with different principal quantum
numbers, while $\eta$ corresponds to the energy of a transition between
quasi-degenerate neighbors.  These parameters and arguments are dimensionless.
The transition from the $1/R^6$ short-range asymptotics to the $1/R^7$
long-range limit is clearly displayed in Fig.~\ref{fig1}, while the
intermediate $1/R^5$ regime for $J$ is discernible in Fig.~\ref{fig2}.

%
% Correction to the $2S$--$1S$ van der Waals coefficient(s): 
% some details on the derivation
%
\section{Details on Dirac-\texorpdfstring{$\maybebm{\delta}$}{delta}
corrections to the van der Waals interaction}
\label{app:GiveMeMore}

Here we present some details on how the numerical results
(\ref{eq:NumDeltaC612}) and (\ref{eq:NumDeltaM612}) were obtained. 
We recall
that for the $2S$--$1S$ system, $\delta D_6$ is given by
\begin{multline} 
\label{eq:DeltaC612S}
\delta D_6(2S;1S) =
\frac{3}{\pi} \frac{\hbar}{(4 \pi \epsilon_0)^2} \,
\int\limits_0^\infty \dd \omega\, 
\left[\delta_{2S} \alpha_{1S\left(2S\right)}(\ii \omega) \right.
\\
\times \left.
\alpha_{2S\left(1S\right)}(\ii \omega)
+\alpha_{1S\left(2S\right)}(\ii \omega)\, 
\delta_{2S}\alpha_{2S\left(1S\right)}(\ii \omega)\right] \,,
\end{multline}
where the corrected polarizabilities read
\begin{subequations} 
\label{eq:DPolENWF}
\begin{multline}
\delta_{2S}\alpha_{1S\left(2S\right)}(\omega) =\\
\frac{e^2}{3}\sum_\pm\left< 1S\right| \vec r
\frac{1}{\left(H-E_{\overline{1S2S}} -
\mathrm{i}\epsilon\pm\hbar\omega \right)^2} 
\vec r \left|1S\right>
\\
\times\frac{1}{2}\left< 2S\right|\delta V\left|2S\right>
\equiv \delta_{2S}\alpha_{1S\left(2S\right)}^{\left(E\right)}(\omega), 
\end{multline}
because we perturb only the $2S$ energy, and
\begin{multline}
\delta_{2S} \alpha_{2S\left(1S\right)}(\omega) =\\
\frac{e^2}{3}\sum_\pm\left< 2S\right|\vec r  
\frac{1}{\left(H-E_{\overline{1S2S}}-\mathrm{i}\epsilon\pm\hbar\omega\right)^2} 
\vec r \left|2S\right> \\
\times\frac{1}{2}
\left< 2S\right|\delta V\left|2S\right>
\\
+\frac{2 e^2}{3}\sum_\pm\left< 2S\right|\vec r \,
\frac{1}{H-E_{\overline{1S2S}}-\mathrm{i}\epsilon\pm\hbar\omega} \,
\vec r \left|\delta 2S\right>
\\
=
\delta_{2S} \alpha_{2S\left(1S\right)}^{\left(E\right)}(\omega) +
\delta_{2S} \alpha_{2S\left(1S\right)}^{\left(\psi\right)}(\omega).
\end{multline}
\end{subequations} 
Here the $\left(E\right)$ superscript refers to the contribution from the energy
correction and the $\left(\psi\right)$ superscript refers to the contribution
from the wave function correction. It can be checked that the two 
summands in
\begin{multline} 
\label{eq:DeltaC612SE}
 \delta  D_6^{\left(E\right)}(2S;1S) =
\frac{3}{\pi} \frac{\hbar}{(4 \pi\epsilon_0)^2} \,
\int\limits_0^\infty {\rm d}\omega
\left[\delta_{2S} \alpha_{1S\left(2S\right)}^{\left(E\right)}(\ii \omega)
\right.
\\
\times \alpha_{2S\left(1S\right)}(\ii \omega)
\left.+\alpha_{1S\left(2S\right)}(\ii \omega)\, 
\delta_{2S}\alpha_{2S\left(1S\right)}^{\left(E\right)}(\ii \omega)\right]
\end{multline}
contribute equally. This can be traced back to the 
integral identity (\ref{eq:CUInt}). The easiest way to compute the 
energy correction to a polarizability is to notice that
\begin{align} 
\label{eq:DiffOmega}
&\left< nS\right|\vec r \, 
\frac{1}{\left(H-E-\mathrm{i}\epsilon+\hbar\omega\right)^2} \,
\vec r \left|mS\right>
\nonumber\\
& = -\frac{1}{\hbar}
\frac{\partial}{\partial\omega}\left< nS\right|\vec r \, 
\frac{1}{H-E-\mathrm{i}\epsilon+\hbar\omega} \, \vec r \left|mS\right> \,,
\end{align}
which is just the $\omega$-derivative of a typical $P$ matrix element. To see
that the summands in the integrand on the right-hand side of
(\ref{eq:DeltaC612SE}) contribute equally, however, one rather notices that
\begin{align} \label{eq:DiffE}
&\left< nS\right|\vec r \,
\frac{1}{\left(H-E-\mathrm{i}\epsilon+\hbar\omega\right)^2} \,
\vec r \left|mS\right>\nonumber\\
&=\frac{\partial}{\partial E}
\left< nS\right|\vec r \, \frac{1}{H-E-\mathrm{i}\epsilon+\hbar\omega} \,
\vec r \left|mS\right> \,,
\end{align}
and the equality follows from (\ref{eq:CUInt}). In the end, we obtain
\begin{align} 
\label{eq:NumDeltaC612E}
\delta D_6^{\left(E\right)}(2S;1S) =& \; 49.733\,193\,536\, \alpha^2\,a_0^6\,E_h\,.
\end{align}
It is considerably harder to compute the contribution to the van der Waals
coefficient from the wave function correction
{\allowdisplaybreaks
\begin{multline} \label{eq:DeltaC612SW}
\delta D_6^{\left(\psi\right)}(2S;1S) =
\frac{3 \hbar}{\pi (4 \pi\epsilon_0)^2} \\
\times\int\limits_0^\infty \dd \omega\,
\alpha_{1S\left(2S\right)}(\ii \omega)
\delta_{2S}\alpha_{2S\left(1S\right)}^{\left(\psi\right)}(\ii \omega) .
\end{multline}
The first step is to obtain the correction (\ref{eq:DeltaPsi2S}) 
to the $2S$ wave function, from which we deduce
\begin{multline} \label{eq:GrandTotal}
\delta P_{2S}^{\left(\psi\right)}\left(\omega\right)
= \frac{e^2 a_0^2}{m\,c^2}
\left( -\frac{8}{9} \, \frac{\tau^2 Q(\tau)}{\left(1-\tau\right)^7 \,
\left(1+\tau\right)^8} \right. \\
+ \frac{4\,096 \, \tau^9 \, (-2 + \tau + 7 \tau^2)}{3 \, (\tau-1)^6 \, (1+\tau)^5} 
\, \ln\left(  \frac{2 \tau}{1 + \tau} \right)
\\
+ \frac{512 \tau^7 \, (1 + \tau^2)}{(1 - \tau)^5 \, (1 + \tau)^5}
{}_2F_1\left(1, -2\tau; 1-2\tau; -\frac{1-\tau}{1+\tau}\right) \\
- \frac{512 \tau^7 \, R(\tau)}{9 \, (1 - \tau)^7 \, (1 + \tau)^7}
{}_2F_1\left(1, -2\tau; 1-2\tau; \left( \frac{1-\tau}{1+\tau}\right)^2 \right) \\
+\frac{32\,768}{3}
\frac{\tau^{10} \left(-1+4\tau^2\right)}{\left(-1+\tau\right)^2
\left(1+\tau\right)^{10}}\\
\left.  \times \sum_{k=0}^{\infty}
\left(\frac{-1+\tau}{1+\tau}\right)^k
\frac{\partial_2\,{}_2F_1\left(-k,4;4;\frac{2}{1+\tau}\right)}{\left(2+k-2\tau\right)}
\right) \,,
\end{multline}}
where
\begin{multline}
Q(\tau) = 
-123 - 123 \tau + 801 \tau^2 + 801 \tau^3 - 2\,124 \tau^4 \\
- 1\,932 \tau^5 + 4\,002 \tau^6 + 11\,234 \tau^7 + 3\,661 \tau^8 - 20\,979 \tau^9 \\
+ 2\,285 \tau^{10} + 9\,645 \tau^{11} + 26\,314 \tau^{12} + 3\,402 \tau^{13} 
\end{multline}
and 
\begin{multline}
\label{defRtau}
R(\tau) = -3 + 113 \, \tau^2 - 193 \, \tau^4 + 371 \tau^6 
\\
+ 96 \tau^2 \, (1- \tau^2) \, (1 - 4 \tau^2) \,
\ln\left( \frac{2 \tau}{1 + \tau} \right) \,.
\end{multline}
Furthermore,
\begin{equation} \label{eq:Tau}
\tau = \left( 1 + \frac{8 \hbar \omega}{\alpha^2 m c^2} \right)^{-1/2} .
\end{equation}
We can then easily deduce
$\delta_{2S}\alpha_{2S\left(1S\right)}^{\left(\psi\right)}\left(\omega\right)$ from
(\ref{eq:GrandTotal}) via
\begin{equation} 
\delta_{2S}\alpha_{2S\left(1S\right)}^{\left(\psi\right)}(\omega) =
\delta_{2S} P_{2S\left(1S\right)}^{\left(\psi\right)}(\omega) +
\delta_{2S} P_{2S\left(1S\right)}^{\left(\psi\right)}(-\omega)
\end{equation}
where $\delta_{2S} P_{2S\left(1S\right)}^{\left(\psi\right)}\left(\omega\right)$ has
the same expression as (\ref{eq:GrandTotal}), with $\tau$ replaced by
\begin{equation} \label{eq:Teff}
t_{\mathrm{eff}} = \left( 1 + \frac{16 \hbar \omega}{5\alpha^2 m c^2} \right)^{-1/2} .
\end{equation}
From all of this we obtain
\begin{align} 
\label{eq:NumDeltaC612psi}
\delta D_6^{\left(\psi\right)}(2S;1S) =& \; 318.181\,412\,174\, \alpha^2\,a_0^6\,E_h\,.
\end{align}
We now recall that for the $2S$--$1S$ system, $\delta M_6$ is given by
\begin{equation} 
\delta M_6(2S;1S) =
\frac{3}{\pi} \frac{\hbar}{(4 \pi\epsilon_0)^2} \,
\int\limits_0^\infty {\rm d}\omega\,
\delta_{2S}  \alpha_{\overline{1S2S}}(\ii \omega)\, 
\alpha_{\overline{1S2S}}(\ii \omega)
\end{equation}
where the corrected mixed polarizability reads
\label{eq:DPolENWFM}
\begin{align}
\delta_{2S} \alpha_{\overline{1S2S}}(\omega) &=
\sum_\pm\left< 1S\right| \vec r
\frac{1}{\left(H-E_{\overline{1S2S}}-\mathrm{i}\epsilon\pm\hbar\omega\right)^2}
\vec r \left|2S\right> \nonumber\\
&\times\frac{e^2}{3}\frac{1}{2}\left< 2S\right|\delta V\left|2S\right>
\nonumber\\
& + \frac{e^2}{3}
\sum_\pm\left< 1S\right| \vec r 
\frac{1}{H-E_{\overline{1S2S}}-\mathrm{i}\epsilon\pm\hbar\omega}
\vec r \left|\delta 2S\right>
\nonumber\\
&=\delta_{2S} \alpha_{\overline{1S2S}}^{\left(E\right)}(\omega) +
\delta_{2S} \alpha_{\overline{1S2S}}^{\left(\psi\right)}(\omega) \,.
\end{align}
Here again, the $\left(E\right)$ subscript refers to the contribution from the
energy correction and the $\left(\psi\right)$ subscript refers to the
contribution from the wave function correction. We again compute
\begin{multline} \label{eq:DeltaM612SE}
\delta M_6^{\left(E\right)}(2S;1S) =
\frac{6}{\pi} \frac{\hbar}{(4 \pi\epsilon_0)^2}\\
\times\int\limits_0^\infty {\rm d}\omega\, 
\delta_{2S}  \alpha_{\overline{1S2S}}^{\left(E\right)}(\ii \omega)\, 
\alpha_{\overline{1S2S}}(\ii \omega)
\end{multline}
by using (\ref{eq:DiffOmega}). For that we need 
\begin{multline} \label{eq:Q1S2S}
P_{\overline{1S2S}}\left(\omega\right)= 
\frac{e^2a_0^2}{E_h}\frac{512\sqrt{2}\,
t_{\mathrm{eff}}^2}{729\left(-1+t_{\mathrm{eff}}^2\right)^2 \, 
\left(-4+t_{\mathrm{eff}}^2\right)^3}\\
\left[
128-272t_{\mathrm{eff}}^2+120t_{\mathrm{eff}}^4+
253t_{\mathrm{eff}}^6+972t_{\mathrm{eff}}^7+419t_{\mathrm{eff}}^8\right.
\\
\left.-1944t_{\mathrm{eff}}^7\,
{}_2F_1\left(1,-t_{\mathrm{eff}};1-t_{\mathrm{eff}};
\frac{1-t_{\mathrm{eff}}}{1+t_{\mathrm{eff}}}
\frac{2-t_{\mathrm{eff}}}{2+t_{\mathrm{eff}}}\right)\right]
\end{multline}
with $t_{\mathrm{eff}}$ given by (\ref{eq:Teff}). In the end, we obtain
\begin{align} 
\label{eq:NumDeltaM612E}
\delta M_6^{\left(E\right)}(2S;1S) =& \; 12.556\,663\,547\, \alpha^2\,a_0^6\,E_h\,.
\end{align}
Finally, we calculate the wave-function contribution to the 
mixing coefficient for the Dirac-$\delta$ correction,
\begin{multline} 
\label{eq:DeltaM612SW}
\delta M_6^{\left(\psi\right)}(2S;1S) =
\frac{6}{\pi} \frac{\hbar}{(4 \pi\epsilon_0)^2}\\
\times\int\limits_0^\infty {\rm d}\omega\,\delta_{2S} \,
\alpha_{\overline{1S2S}}^{\left(\psi\right)}(\ii \omega)\, 
\alpha_{\overline{1S2S}}(\ii \omega) \,.
\end{multline} 
From (\ref{eq:DeltaPsi2S}), we deduce
{\allowdisplaybreaks
\begin{multline} \label{eq:DeltaPsi1S2S}
\delta P_{1S\underline{2S}}^{\left(\psi\right)}\left(\omega\right)
= \frac{e^2a_0^2}{m\,c^2} \, \left( -
\frac{128 \sqrt{2} \tau^2 \, S(\tau)}{2187 \,
\left(1-\tau^2\right)^4 \, \left(1-4\tau^2\right)^3} 
\right.
\\
- \frac{2\,048 \, \sqrt{2} \, \tau^2 \, T(\tau)}
{729 \, (1 - \tau)^3 \, (1 + \tau)^2 \, (1 - 4 \tau^2)^2} \,
\ln\left( \frac{2 \tau}{1 + \tau} \right)
\\
- \frac{2\,048 \, \sqrt{2} \, \tau^2 \, (1 + 2 \tau^2) \, U(\tau)}
{729 \, (1 - \tau^2)^2 \, (1 - 4 \tau^2)^3} \,
\ln\left( \frac{3 \tau}{1 + \tau} \right)
\\
+ \frac{1\,024\, \sqrt{2} \, \tau^7 \, (1 + \tau^2)}
{(1 - \tau^2)^2 \, (1 - 4 \tau^2)^3} \,
{}_2 F_1\left(1,-2\tau;1-2\tau;- \frac{1 - 2 \tau}{1 + 2 \tau} \right)
\\
+ \frac{1\,024\, \sqrt{2} \, \tau^7 \, R(\tau)}%
{9 (1 - \tau^2)^4 \, (1 - 4 \tau^2)^3} \\
\times{}_2 F_1\left(1,-2\tau;1-2\tau;- \frac{1 - \tau}{1 + \tau} \,
\frac{1 - 2 \tau}{1 + 2 \tau} \right) \\
+ \frac{65\,536\sqrt{2}}{3} \,
\frac{\tau^{10}}{\left(-1+\tau\right)\left(1+\tau\right)^5\left(1+2\tau\right)^4} \\
\left. \times\sum_{k=0}^{\infty}\left(-\frac{1-2\tau}{1+2\tau}\right)^k
\frac{\partial_2\,_2F_1\left(-k,4;4;\frac{2}{1+\tau}\right)}{\left(2+k-2\tau\right)} 
\right) \,,
\end{multline}
}
with $\tau$ given by (\ref{eq:Tau}),
and $R(\tau)$ is defined in Eq.~\eqref{defRtau}. 
The function $S(\tau)$ is given as follows,
\begin{multline}
S(\tau) = -157 + 2\,436 \tau^2 - 13\,326 \tau^4 + 5\,832 \tau^5 + 58\,868 \tau^6 \\
+ 225\,504 \tau^7 - 283\,245 \tau^8 + 99\,144 \tau^9 \\
- 431\,184 \tau^{10} + 695\,952 \tau^{11} + 200\,048 \tau^{12} \,,
\end{multline}
while $T(\tau)$ reads as 
\begin{multline}
T(\tau) = 2 - 2 \tau - 15 \tau^2 + 15 \tau^3 + 15 \tau^4 \\
- 15 \tau^5 + 268 \tau^6 + 1\,676 \tau^7 \,,
\end{multline}
and $U(\tau)$ is
\begin{equation}
U(\tau) = -2 + 27 \tau^2 - 129 \tau^4 + 50 \tau^6 \,.
\end{equation}
We can then easily deduce
$\delta_{2S}\alpha_{\overline{1S2S}}^{\left(\psi\right)}\left(\omega\right)$ from
(\ref{eq:DeltaPsi1S2S}) via
\begin{equation} 
\label{eq:Deduce}
\delta_{2S}\alpha_{\overline{1S2S}}^{\left(\psi\right)}(\omega) =
\delta_{2S} P_{\overline{1S2S}}^{\left(\psi\right)}(\omega) +
\delta_{2S} P_{\overline{1S2S}}^{\left(\psi\right)}(-\omega)
\end{equation}
where again $\delta_{2S} P_{\overline{1S2S}}^{\left(\psi\right)}\left(\omega\right)$ has
the same expression as (\ref{eq:DeltaPsi1S2S}), with $\tau$ replaced by
$t_{\mathrm{eff}}$ [see (\ref{eq:Teff})]. From all of this we obtain
\begin{align} 
\delta M_6^{\left(\psi\right)}(2S;1S) =& \; 
-70.652\,014\,640\, \alpha^2\,a_0^6\,E_h\,.
\end{align}


\begin{thebibliography}{49}%
\makeatletter
\providecommand \@ifxundefined [1]{%
 \@ifx{#1\undefined}
}%
\providecommand \@ifnum [1]{%
 \ifnum #1\expandafter \@firstoftwo
 \else \expandafter \@secondoftwo
 \fi
}%
\providecommand \@ifx [1]{%
 \ifx #1\expandafter \@firstoftwo
 \else \expandafter \@secondoftwo
 \fi
}%
\providecommand \natexlab [1]{#1}%
\providecommand \enquote  [1]{``#1''}%
\providecommand \bibnamefont  [1]{#1}%
\providecommand \bibfnamefont [1]{#1}%
\providecommand \citenamefont [1]{#1}%
\providecommand \href@noop [0]{\@secondoftwo}%
\providecommand \href [0]{\begingroup \@sanitize@url \@href}%
\providecommand \@href[1]{\@@startlink{#1}\@@href}%
\providecommand \@@href[1]{\endgroup#1\@@endlink}%
\providecommand \@sanitize@url [0]{\catcode `\\12\catcode `\$12\catcode
  `\&12\catcode `\#12\catcode `\^12\catcode `\_12\catcode `\%12\relax}%
\providecommand \@@startlink[1]{}%
\providecommand \@@endlink[0]{}%
\providecommand \url  [0]{\begingroup\@sanitize@url \@url }%
\providecommand \@url [1]{\endgroup\@href {#1}{\urlprefix }}%
\providecommand \urlprefix  [0]{URL }%
\providecommand \Eprint [0]{\href }%
\providecommand \doibase [0]{http://dx.doi.org/}%
\providecommand \selectlanguage [0]{\@gobble}%
\providecommand \bibinfo  [0]{\@secondoftwo}%
\providecommand \bibfield  [0]{\@secondoftwo}%
\providecommand \translation [1]{[#1]}%
\providecommand \BibitemOpen [0]{}%
\providecommand \bibitemStop [0]{}%
\providecommand \bibitemNoStop [0]{.\EOS\space}%
\providecommand \EOS [0]{\spacefactor3000\relax}%
\providecommand \BibitemShut  [1]{\csname bibitem#1\endcsname}%
\let\auto@bib@innerbib\@empty
%</preamble>
\bibitem [{\citenamefont {Jentschura}\ and\ \citenamefont
  {Yerokhin}(2006)}]{JeYe2006}%
  \BibitemOpen
  \bibfield  {author} {\bibinfo {author} {\bibfnamefont {U.~D.}\ \bibnamefont
  {Jentschura}}\ and\ \bibinfo {author} {\bibfnamefont {V.~A.}\ \bibnamefont
  {Yerokhin}},\ }\bibfield  {title} {\enquote {\bibinfo {title} {\relax{Quantum
  electrodynamic corrections to the hyperfine structure of excited $S$
  states}},}\ }\href@noop {} {\bibfield  {journal} {\bibinfo  {journal} {Phys.
  Rev. A}\ }\textbf {\bibinfo {volume} {73}},\ \bibinfo {pages} {062503}
  (\bibinfo {year} {2006})}\BibitemShut {NoStop}%
\bibitem [{\citenamefont {Bethe}(1947)}]{Be1947}%
  \BibitemOpen
  \bibfield  {author} {\bibinfo {author} {\bibfnamefont {H.~A.}\ \bibnamefont
  {Bethe}},\ }\bibfield  {title} {\enquote {\bibinfo {title} {\relax{The
  Electromagnetic Shift of Energy Levels}},}\ }\href@noop {} {\bibfield
  {journal} {\bibinfo  {journal} {Phys. Rev.}\ }\textbf {\bibinfo {volume}
  {72}},\ \bibinfo {pages} {339--341} (\bibinfo {year} {1947})}\BibitemShut
  {NoStop}%
\bibitem [{\citenamefont {Chibisov}(1972)}]{Ch1972}%
  \BibitemOpen
  \bibfield  {author} {\bibinfo {author} {\bibfnamefont {M.~I.}\ \bibnamefont
  {Chibisov}},\ }\bibfield  {title} {\enquote {\bibinfo {title}
  {\relax{Dispersion Interaction of Neutral Atoms}},}\ }\href@noop {}
  {\bibfield  {journal} {\bibinfo  {journal} {Opt. Spectrosc.}\ }\textbf
  {\bibinfo {volume} {32}},\ \bibinfo {pages} {1--3} (\bibinfo {year}
  {1972})}\BibitemShut {NoStop}%
\bibitem [{\citenamefont {Deal}\ and\ \citenamefont {Young}(1973)}]{DeYo1973}%
  \BibitemOpen
  \bibfield  {author} {\bibinfo {author} {\bibfnamefont {W.~J.}\ \bibnamefont
  {Deal}}\ and\ \bibinfo {author} {\bibfnamefont {R.~H.}\ \bibnamefont
  {Young}},\ }\bibfield  {title} {\enquote {\bibinfo {title}
  {\relax{Long--Range Dispersion Interactions Involving Excited Atoms; the
  H(1s)---H(2s) Interaction}},}\ }\href@noop {} {\bibfield  {journal} {\bibinfo
   {journal} {Int. J. Quantum Chem.}\ }\textbf {\bibinfo {volume} {7}},\
  \bibinfo {pages} {877--892} (\bibinfo {year} {1973})}\BibitemShut {NoStop}%
\bibitem [{\citenamefont {Tang}\ and\ \citenamefont {Chan}(1986)}]{TaCh1986}%
  \BibitemOpen
  \bibfield  {author} {\bibinfo {author} {\bibfnamefont {A.~Z.}\ \bibnamefont
  {Tang}}\ and\ \bibinfo {author} {\bibfnamefont {F.~T.}\ \bibnamefont
  {Chan}},\ }\bibfield  {title} {\enquote {\bibinfo {title} {\relax{Dynamic
  Multipole polarizability of atomic hydrogen}},}\ }\href@noop {} {\bibfield
  {journal} {\bibinfo  {journal} {Phys. Rev. A}\ }\textbf {\bibinfo {volume}
  {33}},\ \bibinfo {pages} {3671--3678} (\bibinfo {year} {1986})}\BibitemShut
  {NoStop}%
\bibitem [{\citenamefont {Kolachevsky}\ \emph {et~al.}(2004)\citenamefont
  {Kolachevsky}, \citenamefont {Fischer}, \citenamefont {Karshenboim},\ and\
  \citenamefont {H\"{a}nsch}}]{KoFiKaHa2004}%
  \BibitemOpen
  \bibfield  {author} {\bibinfo {author} {\bibfnamefont {N.}~\bibnamefont
  {Kolachevsky}}, \bibinfo {author} {\bibfnamefont {M.}~\bibnamefont
  {Fischer}}, \bibinfo {author} {\bibfnamefont {S.~G.}\ \bibnamefont
  {Karshenboim}}, \ and\ \bibinfo {author} {\bibfnamefont {T.~W.}\ \bibnamefont
  {H\"{a}nsch}},\ }\bibfield  {title} {\enquote {\bibinfo {title}
  {\relax{High-Precision Optical Measurement of the $2S$ Hyperfine Interval in
  Atomic Hydrogen}},}\ }\href@noop {} {\bibfield  {journal} {\bibinfo
  {journal} {Phys. Rev. Lett.}\ }\textbf {\bibinfo {volume} {92}},\ \bibinfo
  {pages} {033003} (\bibinfo {year} {2004})}\BibitemShut {NoStop}%
\bibitem [{\citenamefont {Kolachevsky}\ \emph {et~al.}(2009)\citenamefont
  {Kolachevsky}, \citenamefont {Matveev}, \citenamefont {Alnis}, \citenamefont
  {Parthey}, \citenamefont {Karshenboim},\ and\ \citenamefont
  {H\"{a}nsch}}]{KoEtAl2009}%
  \BibitemOpen
  \bibfield  {author} {\bibinfo {author} {\bibfnamefont {N.}~\bibnamefont
  {Kolachevsky}}, \bibinfo {author} {\bibfnamefont {A.}~\bibnamefont
  {Matveev}}, \bibinfo {author} {\bibfnamefont {J.}~\bibnamefont {Alnis}},
  \bibinfo {author} {\bibfnamefont {C.~G.}\ \bibnamefont {Parthey}}, \bibinfo
  {author} {\bibfnamefont {S.~G.}\ \bibnamefont {Karshenboim}}, \ and\ \bibinfo
  {author} {\bibfnamefont {T.~W.}\ \bibnamefont {H\"{a}nsch}},\ }\bibfield
  {title} {\enquote {\bibinfo {title} {\relax{Measurement of the $2S$ Hyperfine
  Interval in Atomic Hydrogen}},}\ }\href@noop {} {\bibfield  {journal}
  {\bibinfo  {journal} {Phys. Rev. Lett.}\ }\textbf {\bibinfo {volume} {102}},\
  \bibinfo {pages} {213002} (\bibinfo {year} {2009})}\BibitemShut {NoStop}%
\bibitem [{\citenamefont {Dalgarno}\ and\ \citenamefont
  {Davidson}(1966)}]{DaDa1966}%
  \BibitemOpen
  \bibfield  {author} {\bibinfo {author} {\bibfnamefont {A.}~\bibnamefont
  {Dalgarno}}\ and\ \bibinfo {author} {\bibfnamefont {W.~D.}\ \bibnamefont
  {Davidson}},\ }\bibfield  {title} {\enquote {\bibinfo {title} {\relax{The
  Calculation of Van Der Waals Interactions}},}\ }\href@noop {} {\bibfield
  {journal} {\bibinfo  {journal} {Adv. At. Mol. Opt. Phys.}\ }\textbf {\bibinfo
  {volume} {2}},\ \bibinfo {pages} {1--32} (\bibinfo {year}
  {1966})}\BibitemShut {NoStop}%
\bibitem [{\citenamefont {Dalgarno}(1967)}]{Da1967}%
  \BibitemOpen
  \bibfield  {author} {\bibinfo {author} {\bibfnamefont {A.}~\bibnamefont
  {Dalgarno}},\ }\bibfield  {title} {\enquote {\bibinfo {title} {\relax{New
  Methods for Calculating Long--Range Intermolecular Forces}},}\ }\href@noop {}
  {\bibfield  {journal} {\bibinfo  {journal} {Adv. Chem. Phys.}\ }\textbf
  {\bibinfo {volume} {12}},\ \bibinfo {pages} {143--166} (\bibinfo {year}
  {1967})}\BibitemShut {NoStop}%
\bibitem [{\citenamefont {Ray}\ \emph {et~al.}(1968{\natexlab{a}})\citenamefont
  {Ray}, \citenamefont {Lyons},\ and\ \citenamefont {Das}}]{RaLyDa1968a}%
  \BibitemOpen
  \bibfield  {author} {\bibinfo {author} {\bibfnamefont {S.}~\bibnamefont
  {Ray}}, \bibinfo {author} {\bibfnamefont {J.~D.}\ \bibnamefont {Lyons}}, \
  and\ \bibinfo {author} {\bibfnamefont {T.~P.}\ \bibnamefont {Das}},\
  }\bibfield  {title} {\enquote {\bibinfo {title} {\relax{Hyperfine Pressure
  Shift and van der Waals Interactions. I. Hydrogen--Helium System}},}\
  }\href@noop {} {\bibfield  {journal} {\bibinfo  {journal} {Phys. Rev.}\
  }\textbf {\bibinfo {volume} {174}},\ \bibinfo {pages} {104--112} (\bibinfo
  {year} {1968}{\natexlab{a}})},\ \bibinfo {note} {erratum Phys. Rev.~{\bf
  181}, 465 (1969)]}\BibitemShut {NoStop}%
\bibitem [{\citenamefont {Ray}\ \emph {et~al.}(1968{\natexlab{b}})\citenamefont
  {Ray}, \citenamefont {Lyons},\ and\ \citenamefont {Das}}]{RaLyDa1968b}%
  \BibitemOpen
  \bibfield  {author} {\bibinfo {author} {\bibfnamefont {S.}~\bibnamefont
  {Ray}}, \bibinfo {author} {\bibfnamefont {J.~D.}\ \bibnamefont {Lyons}}, \
  and\ \bibinfo {author} {\bibfnamefont {T.~P.}\ \bibnamefont {Das}},\
  }\bibfield  {title} {\enquote {\bibinfo {title} {\relax{Hyperfine Pressure
  Shift and van der Waals Interactions. II. Nitrogen--Helium System}},}\
  }\href@noop {} {\bibfield  {journal} {\bibinfo  {journal} {Phys. Rev.}\
  }\textbf {\bibinfo {volume} {174}},\ \bibinfo {pages} {112--118} (\bibinfo
  {year} {1968}{\natexlab{b}})},\ \bibinfo {note} {erratum Phys. Rev.~{\bf
  181}, 465 (1969)]}\BibitemShut {NoStop}%
\bibitem [{\citenamefont {Rao}\ and\ \citenamefont {Das}(1969)}]{RaDa1969}%
  \BibitemOpen
  \bibfield  {author} {\bibinfo {author} {\bibfnamefont {B.~K.}\ \bibnamefont
  {Rao}}\ and\ \bibinfo {author} {\bibfnamefont {T.~P.}\ \bibnamefont {Das}},\
  }\bibfield  {title} {\enquote {\bibinfo {title} {\relax{Hyperfine Pressure
  Shift and van der Waals Interactions. III. Temperature Dependence}},}\
  }\href@noop {} {\bibfield  {journal} {\bibinfo  {journal} {Phys. Rev.}\
  }\textbf {\bibinfo {volume} {185}},\ \bibinfo {pages} {95--97} (\bibinfo
  {year} {1969})}\BibitemShut {NoStop}%
\bibitem [{\citenamefont {Rao}\ and\ \citenamefont {Das}(1970)}]{RaIkDa1970}%
  \BibitemOpen
  \bibfield  {author} {\bibinfo {author} {\bibfnamefont {B.~K.}\ \bibnamefont
  {Rao}}\ and\ \bibinfo {author} {\bibfnamefont {T.~P.}\ \bibnamefont {Das}},\
  }\bibfield  {title} {\enquote {\bibinfo {title} {\relax{Hyperfine Pressure
  Shift and van der Waals Interactions. IV. Hydrogen--Rare--Gas Systems}},}\
  }\href@noop {} {\bibfield  {journal} {\bibinfo  {journal} {Phys. Rev. A}\
  }\textbf {\bibinfo {volume} {2}},\ \bibinfo {pages} {1411--1421} (\bibinfo
  {year} {1970})}\BibitemShut {NoStop}%
\bibitem [{\citenamefont {Dalgarno}\ \emph {et~al.}(1968)\citenamefont
  {Dalgarno}, \citenamefont {Drake},\ and\ \citenamefont
  {Victor}}]{DaDrVi1968}%
  \BibitemOpen
  \bibfield  {author} {\bibinfo {author} {\bibfnamefont {A.}~\bibnamefont
  {Dalgarno}}, \bibinfo {author} {\bibfnamefont {G.~W.~F.}\ \bibnamefont
  {Drake}}, \ and\ \bibinfo {author} {\bibfnamefont {G.~A.}\ \bibnamefont
  {Victor}},\ }\bibfield  {title} {\enquote {\bibinfo {title}
  {\relax{Nonadiabatic Long--Range Forces}},}\ }\href@noop {} {\bibfield
  {journal} {\bibinfo  {journal} {Phys. Rev.}\ }\textbf {\bibinfo {volume}
  {176}},\ \bibinfo {pages} {194--197} (\bibinfo {year} {1968})}\BibitemShut
  {NoStop}%
\bibitem [{\citenamefont {Dutta}\ \emph {et~al.}(1970)\citenamefont {Dutta},
  \citenamefont {Dutta},\ and\ \citenamefont {Das}}]{DuDuDa1970}%
  \BibitemOpen
  \bibfield  {author} {\bibinfo {author} {\bibfnamefont {C.~M.}\ \bibnamefont
  {Dutta}}, \bibinfo {author} {\bibfnamefont {N.~C.}\ \bibnamefont {Dutta}}, \
  and\ \bibinfo {author} {\bibfnamefont {T.~P.}\ \bibnamefont {Das}},\
  }\bibfield  {title} {\enquote {\bibinfo {title} {\relax{Many--Body Approach
  to the Hyperfine Pressure Shift in Optical--Pumping Experiments}},}\
  }\href@noop {} {\bibfield  {journal} {\bibinfo  {journal} {Phys. Rev. A}\
  }\textbf {\bibinfo {volume} {2}},\ \bibinfo {pages} {30--37} (\bibinfo {year}
  {1970})}\BibitemShut {NoStop}%
\bibitem [{\citenamefont {Yan}\ \emph {et~al.}(1996)\citenamefont {Yan},
  \citenamefont {Babb}, \citenamefont {Dalgarno},\ and\ \citenamefont
  {Drake}}]{YaBaDaDr1996}%
  \BibitemOpen
  \bibfield  {author} {\bibinfo {author} {\bibfnamefont {Z.~C.}\ \bibnamefont
  {Yan}}, \bibinfo {author} {\bibfnamefont {J.~F.}\ \bibnamefont {Babb}},
  \bibinfo {author} {\bibfnamefont {A.}~\bibnamefont {Dalgarno}}, \ and\
  \bibinfo {author} {\bibfnamefont {G.~W.~F.}\ \bibnamefont {Drake}},\
  }\bibfield  {title} {\enquote {\bibinfo {title} {Variational calculations of
  dispersion coefficients for interactions among h, he, and li atoms},}\
  }\href@noop {} {\bibfield  {journal} {\bibinfo  {journal} {Phys. Rev. A}\
  }\textbf {\bibinfo {volume} {54}},\ \bibinfo {pages} {2824--2833} (\bibinfo
  {year} {1996})}\BibitemShut {NoStop}%
\bibitem [{\citenamefont {Yan}\ \emph {et~al.}(1997)\citenamefont {Yan},
  \citenamefont {Dalgarno},\ and\ \citenamefont {Babb}}]{YaDaBa1997}%
  \BibitemOpen
  \bibfield  {author} {\bibinfo {author} {\bibfnamefont {Z.~C.}\ \bibnamefont
  {Yan}}, \bibinfo {author} {\bibfnamefont {A.}~\bibnamefont {Dalgarno}}, \
  and\ \bibinfo {author} {\bibfnamefont {J.~F.}\ \bibnamefont {Babb}},\
  }\bibfield  {title} {\enquote {\bibinfo {title} {\relax{Long-range
  interactions of lithium atoms}},}\ }\href@noop {} {\bibfield  {journal}
  {\bibinfo  {journal} {Phys. Rev. A}\ }\textbf {\bibinfo {volume} {55}},\
  \bibinfo {pages} {2882--2887} (\bibinfo {year} {1997})}\BibitemShut {NoStop}%
\bibitem [{\citenamefont {Jamieson}\ \emph {et~al.}(1995)\citenamefont
  {Jamieson}, \citenamefont {Drake},\ and\ \citenamefont
  {Dalgarno}}]{JaDrDa1995}%
  \BibitemOpen
  \bibfield  {author} {\bibinfo {author} {\bibfnamefont {M.~J.}\ \bibnamefont
  {Jamieson}}, \bibinfo {author} {\bibfnamefont {G.~W.~F.}\ \bibnamefont
  {Drake}}, \ and\ \bibinfo {author} {\bibfnamefont {A.}~\bibnamefont
  {Dalgarno}},\ }\bibfield  {title} {\enquote {\bibinfo {title} {Retarded
  dipole-dipole dispersion interaction potential for helium},}\ }\href@noop {}
  {\bibfield  {journal} {\bibinfo  {journal} {Phys. Rev. A}\ }\textbf {\bibinfo
  {volume} {51}},\ \bibinfo {pages} {3358--3361} (\bibinfo {year}
  {1995})}\BibitemShut {NoStop}%
\bibitem [{\citenamefont {Chen}\ and\ \citenamefont {Chung}(1996)}]{ChCh1996}%
  \BibitemOpen
  \bibfield  {author} {\bibinfo {author} {\bibfnamefont {M.~K.}\ \bibnamefont
  {Chen}}\ and\ \bibinfo {author} {\bibfnamefont {K.~T.}\ \bibnamefont
  {Chung}},\ }\bibfield  {title} {\enquote {\bibinfo {title}
  {\relax{Retardation long-range potentials between two helium atoms}},}\
  }\href@noop {} {\bibfield  {journal} {\bibinfo  {journal} {Phys. Rev. A}\
  }\textbf {\bibinfo {volume} {53}},\ \bibinfo {pages} {1439--1446} (\bibinfo
  {year} {1996})}\BibitemShut {NoStop}%
\bibitem [{\citenamefont {Marinescu}\ \emph {et~al.}(1994)\citenamefont
  {Marinescu}, \citenamefont {Sadeghpour},\ and\ \citenamefont
  {Dalgarno}}]{MaSaDa1994}%
  \BibitemOpen
  \bibfield  {author} {\bibinfo {author} {\bibfnamefont {M.}~\bibnamefont
  {Marinescu}}, \bibinfo {author} {\bibfnamefont {H.~R.}\ \bibnamefont
  {Sadeghpour}}, \ and\ \bibinfo {author} {\bibfnamefont {A.}~\bibnamefont
  {Dalgarno}},\ }\bibfield  {title} {\enquote {\bibinfo {title}
  {\relax{Dispersion coefficients for alkali-metal dimers}},}\ }\href@noop {}
  {\bibfield  {journal} {\bibinfo  {journal} {Phys. Rev. A}\ }\textbf {\bibinfo
  {volume} {49}},\ \bibinfo {pages} {982--988} (\bibinfo {year}
  {1994})}\BibitemShut {NoStop}%
\bibitem [{\citenamefont {Marinescu}\ and\ \citenamefont
  {Dalgarno}(1995)}]{MaDa1995}%
  \BibitemOpen
  \bibfield  {author} {\bibinfo {author} {\bibfnamefont {M.}~\bibnamefont
  {Marinescu}}\ and\ \bibinfo {author} {\bibfnamefont {A.}~\bibnamefont
  {Dalgarno}},\ }\bibfield  {title} {\enquote {\bibinfo {title} {Dispersion
  forces and long-range electronic transition dipole moments of alkali-metal
  dimer excited states},}\ }\href@noop {} {\bibfield  {journal} {\bibinfo
  {journal} {Phys. Rev. A}\ }\textbf {\bibinfo {volume} {52}},\ \bibinfo
  {pages} {311--328} (\bibinfo {year} {1995})}\BibitemShut {NoStop}%
\bibitem [{\citenamefont {Pachucki}(2005)}]{Pa2005longrange}%
  \BibitemOpen
  \bibfield  {author} {\bibinfo {author} {\bibfnamefont {K.}~\bibnamefont
  {Pachucki}},\ }\bibfield  {title} {\enquote {\bibinfo {title} {Relativistic
  corrections to the long-range interaction between closed-shell atoms},}\
  }\href@noop {} {\bibfield  {journal} {\bibinfo  {journal} {Phys. Rev. A}\
  }\textbf {\bibinfo {volume} {72}},\ \bibinfo {pages} {062706} (\bibinfo
  {year} {2005})}\BibitemShut {NoStop}%
\bibitem [{\citenamefont {Berestetskii}\ \emph {et~al.}(1982)\citenamefont
  {Berestetskii}, \citenamefont {Lifshitz},\ and\ \citenamefont
  {Pitaevskii}}]{BeLiPi1982vol4}%
  \BibitemOpen
  \bibfield  {author} {\bibinfo {author} {\bibfnamefont {V.~B.}\ \bibnamefont
  {Berestetskii}}, \bibinfo {author} {\bibfnamefont {E.~M.}\ \bibnamefont
  {Lifshitz}}, \ and\ \bibinfo {author} {\bibfnamefont {L.~P.}\ \bibnamefont
  {Pitaevskii}},\ }\href@noop {} {\emph {\bibinfo {title} {\relax{Quantum
  Electrodynamics, Volume 4 of the Course on Theoretical Physics}}}},\ \bibinfo
  {edition} {2nd}\ ed.\ (\bibinfo  {publisher} {Pergamon Press},\ \bibinfo
  {address} {Oxford, UK},\ \bibinfo {year} {1982})\BibitemShut {NoStop}%
\bibitem [{\citenamefont {Karshenboim}\ and\ \citenamefont
  {Ivanov}(1997)}]{KaIv1997optspec}%
  \BibitemOpen
  \bibfield  {author} {\bibinfo {author} {\bibfnamefont {S.~G.}\ \bibnamefont
  {Karshenboim}}\ and\ \bibinfo {author} {\bibfnamefont {V.~G.}\ \bibnamefont
  {Ivanov}},\ }\bibfield  {title} {\enquote {\bibinfo {title} {\relax{Radiative
  Corrections to the $2E1$ Decay Rate of the $2s$-State in Hydrogen-Like
  Atoms}},}\ }\href@noop {} {\bibfield  {journal} {\bibinfo  {journal} {Opt.
  Spectrosc.}\ }\textbf {\bibinfo {volume} {83}},\ \bibinfo {pages} {1--5}
  (\bibinfo {year} {1997})}\BibitemShut {NoStop}%
\bibitem [{\citenamefont {Jentschura}(2004)}]{Je2004rad}%
  \BibitemOpen
  \bibfield  {author} {\bibinfo {author} {\bibfnamefont {U.~D.}\ \bibnamefont
  {Jentschura}},\ }\bibfield  {title} {\enquote {\bibinfo {title}
  {\relax{Self--Energy Correction to the Two--Photon Decay Width in
  Hydrogenlike Atoms}},}\ }\href@noop {} {\bibfield  {journal} {\bibinfo
  {journal} {Phys. Rev. A}\ }\textbf {\bibinfo {volume} {69}},\ \bibinfo
  {pages} {052118} (\bibinfo {year} {2004})}\BibitemShut {NoStop}%
\bibitem [{\citenamefont {Itzykson}\ and\ \citenamefont
  {Zuber}(1980)}]{ItZu1980}%
  \BibitemOpen
  \bibfield  {author} {\bibinfo {author} {\bibfnamefont {C.}~\bibnamefont
  {Itzykson}}\ and\ \bibinfo {author} {\bibfnamefont {J.~B.}\ \bibnamefont
  {Zuber}},\ }\href@noop {} {\emph {\bibinfo {title} {\relax{Quantum Field
  Theory}}}}\ (\bibinfo  {publisher} {McGraw-Hill},\ \bibinfo {address} {New
  York},\ \bibinfo {year} {1980})\BibitemShut {NoStop}%
\bibitem [{\citenamefont {Mohr}\ \emph {et~al.}(2012)\citenamefont {Mohr},
  \citenamefont {Taylor},\ and\ \citenamefont {Newell}}]{MoTaNe2012}%
  \BibitemOpen
  \bibfield  {author} {\bibinfo {author} {\bibfnamefont {P.~J.}\ \bibnamefont
  {Mohr}}, \bibinfo {author} {\bibfnamefont {B.~N.}\ \bibnamefont {Taylor}}, \
  and\ \bibinfo {author} {\bibfnamefont {D.~B.}\ \bibnamefont {Newell}},\
  }\bibfield  {title} {\enquote {\bibinfo {title} {\relax{CODATA Recommended
  Values of the Fundamental Physical Constants: 2010}},}\ }\href@noop {}
  {\bibfield  {journal} {\bibinfo  {journal} {Rev. Mod. Phys.}\ }\textbf
  {\bibinfo {volume} {84}},\ \bibinfo {pages} {1527--1605} (\bibinfo {year}
  {2012})}\BibitemShut {NoStop}%
\bibitem [{\citenamefont {Power}\ and\ \citenamefont
  {Thirunamachandran}(1995)}]{PoTh1995}%
  \BibitemOpen
  \bibfield  {author} {\bibinfo {author} {\bibfnamefont {E.~A.}\ \bibnamefont
  {Power}}\ and\ \bibinfo {author} {\bibfnamefont {T.}~\bibnamefont
  {Thirunamachandran}},\ }\bibfield  {title} {\enquote {\bibinfo {title}
  {\relax{Dispersion forces between molecules with one or both molecules
  excited}},}\ }\href@noop {} {\bibfield  {journal} {\bibinfo  {journal} {Phys.
  Rev. A}\ }\textbf {\bibinfo {volume} {51}},\ \bibinfo {pages} {3660--3666}
  (\bibinfo {year} {1995})}\BibitemShut {NoStop}%
\bibitem [{\citenamefont {Safari}\ and\ \citenamefont
  {Karimpour}(2015)}]{SaKa2015}%
  \BibitemOpen
  \bibfield  {author} {\bibinfo {author} {\bibfnamefont {H.}~\bibnamefont
  {Safari}}\ and\ \bibinfo {author} {\bibfnamefont {M.~R.}\ \bibnamefont
  {Karimpour}},\ }\bibfield  {title} {\enquote {\bibinfo {title}
  {\relax{Body-Assisted van der Waals Interaction between Excited Atoms}},}\
  }\href@noop {} {\bibfield  {journal} {\bibinfo  {journal} {Phys. Rev. Lett.}\
  }\textbf {\bibinfo {volume} {114}},\ \bibinfo {pages} {013201} (\bibinfo
  {year} {2015})}\BibitemShut {NoStop}%
\bibitem [{\citenamefont {Donaire}\ \emph {et~al.}(2015)\citenamefont
  {Donaire}, \citenamefont {Gu\'{e}rout},\ and\ \citenamefont
  {Lambrecht}}]{DoGuLa2015}%
  \BibitemOpen
  \bibfield  {author} {\bibinfo {author} {\bibfnamefont {M.}~\bibnamefont
  {Donaire}}, \bibinfo {author} {\bibfnamefont {R.}~\bibnamefont
  {Gu\'{e}rout}}, \ and\ \bibinfo {author} {\bibfnamefont {A.}~\bibnamefont
  {Lambrecht}},\ }\bibfield  {title} {\enquote {\bibinfo {title}
  {\relax{Quasiresonant van der Waals Interaction between Nonidentical
  Atoms}},}\ }\href@noop {} {\bibfield  {journal} {\bibinfo  {journal} {Phys.
  Rev. Lett.}\ }\textbf {\bibinfo {volume} {115}},\ \bibinfo {pages} {033201}
  (\bibinfo {year} {2015})}\BibitemShut {NoStop}%
\bibitem [{\citenamefont {Jentschura}\ and\ \citenamefont
  {Debierre}(2016)}]{JeDe2016vdW}%
  \BibitemOpen
  \bibfield  {author} {\bibinfo {author} {\bibfnamefont {U.~D.}\ \bibnamefont
  {Jentschura}}\ and\ \bibinfo {author} {\bibfnamefont {V.}~\bibnamefont
  {Debierre}},\ }\href@noop {} {}\bibinfo {howpublished} {{\em Long--Range
  Tails in \vdw{} Interactions of Excited--State and Ground--State Atoms}, in
  preparation} (\bibinfo {year} {2016})\BibitemShut {NoStop}%
\bibitem [{\citenamefont {Bethe}\ and\ \citenamefont
  {Salpeter}(1957)}]{BeSa1957}%
  \BibitemOpen
  \bibfield  {author} {\bibinfo {author} {\bibfnamefont {H.~A.}\ \bibnamefont
  {Bethe}}\ and\ \bibinfo {author} {\bibfnamefont {E.~E.}\ \bibnamefont
  {Salpeter}},\ }\href@noop {} {\emph {\bibinfo {title} {\relax{Quantum
  Mechanics of One- and Two-Electron Atoms}}}}\ (\bibinfo  {publisher}
  {Springer},\ \bibinfo {address} {Berlin},\ \bibinfo {year}
  {1957})\BibitemShut {NoStop}%
\bibitem [{\citenamefont {Gavrila}\ and\ \citenamefont
  {Costescu}(1970)}]{GaCo1970}%
  \BibitemOpen
  \bibfield  {author} {\bibinfo {author} {\bibfnamefont {M.}~\bibnamefont
  {Gavrila}}\ and\ \bibinfo {author} {\bibfnamefont {A.}~\bibnamefont
  {Costescu}},\ }\bibfield  {title} {\enquote {\bibinfo {title}
  {\relax{Retardation in the Elastic Scattering of Photons by Atomic
  Hydrogen}},}\ }\href@noop {} {\bibfield  {journal} {\bibinfo  {journal}
  {Phys. Rev. A}\ }\textbf {\bibinfo {volume} {2}},\ \bibinfo {pages}
  {1752--1758} (\bibinfo {year} {1970})}\BibitemShut {NoStop}%
\bibitem [{\citenamefont {Pachucki}(1993)}]{Pa1993}%
  \BibitemOpen
  \bibfield  {author} {\bibinfo {author} {\bibfnamefont {K.}~\bibnamefont
  {Pachucki}},\ }\bibfield  {title} {\enquote {\bibinfo {title}
  {\relax{Higher-\relax{O}rder \relax{B}inding \relax{C}}orrections to the
  \relax{L}amb \relax{S}hift},}\ }\href@noop {} {\bibfield  {journal} {\bibinfo
   {journal} {Ann. Phys. (N.Y.)}\ }\textbf {\bibinfo {volume} {226}},\ \bibinfo
  {pages} {1--87} (\bibinfo {year} {1993})}\BibitemShut {NoStop}%
\bibitem [{\citenamefont {Jentschura}\ and\ \citenamefont
  {Pachucki}(1996)}]{JePa1996}%
  \BibitemOpen
  \bibfield  {author} {\bibinfo {author} {\bibfnamefont {U.}~\bibnamefont
  {Jentschura}}\ and\ \bibinfo {author} {\bibfnamefont {K.}~\bibnamefont
  {Pachucki}},\ }\bibfield  {title} {\enquote {\bibinfo {title}
  {\relax{Higher-order binding corrections to the Lamb shift of $2\relax{P}$
  states}},}\ }\href@noop {} {\bibfield  {journal} {\bibinfo  {journal} {Phys.
  Rev. A}\ }\textbf {\bibinfo {volume} {54}},\ \bibinfo {pages} {1853--1861}
  (\bibinfo {year} {1996})}\BibitemShut {NoStop}%
\bibitem [{\citenamefont {Swainson}\ and\ \citenamefont
  {Drake}(1991)}]{SwDr1991b}%
  \BibitemOpen
  \bibfield  {author} {\bibinfo {author} {\bibfnamefont {R.~A.}\ \bibnamefont
  {Swainson}}\ and\ \bibinfo {author} {\bibfnamefont {G.~W.~F.}\ \bibnamefont
  {Drake}},\ }\bibfield  {title} {\enquote {\bibinfo {title} {\relax{A unified
  treatment of the non-relativistic and relativistic hydrogen atom II: the
  Green functions}},}\ }\href@noop {} {\bibfield  {journal} {\bibinfo
  {journal} {J. Phys. A}\ }\textbf {\bibinfo {volume} {24}},\ \bibinfo {pages}
  {95--120} (\bibinfo {year} {1991})}\BibitemShut {NoStop}%
\bibitem [{\citenamefont {Bateman}(1953)}]{Ba1953vol1}%
  \BibitemOpen
  \bibfield  {author} {\bibinfo {author} {\bibfnamefont {H.}~\bibnamefont
  {Bateman}},\ }\href@noop {} {\emph {\bibinfo {title} {\relax{Higher
  Transcendental Functions}}}},\ Vol.~\bibinfo {volume} {1}\ (\bibinfo
  {publisher} {McGraw-Hill},\ \bibinfo {address} {New York},\ \bibinfo {year}
  {1953})\BibitemShut {NoStop}%
\bibitem [{\citenamefont {Adhikari}\ \emph {et~al.}(2016)\citenamefont
  {Adhikari}, \citenamefont {Kawasaki},\ and\ \citenamefont
  {Jentschura}}]{AdKaJe2016}%
  \BibitemOpen
  \bibfield  {author} {\bibinfo {author} {\bibfnamefont {C.~M.}\ \bibnamefont
  {Adhikari}}, \bibinfo {author} {\bibfnamefont {A.}~\bibnamefont {Kawasaki}},
  \ and\ \bibinfo {author} {\bibfnamefont {U.~D.}\ \bibnamefont {Jentschura}},\
  }\bibfield  {title} {\enquote {\bibinfo {title} {\relax{Magic Wavelength for
  the hydrogen $1S$--$2S$ transition: Contribution of the continuum and the
  reduced-mass correction}},}\ }\href@noop {} {\bibfield  {journal} {\bibinfo
  {journal} {Phys. Rev. A}\ }\textbf {\bibinfo {volume} {94}},\ \bibinfo
  {pages} {032510} (\bibinfo {year} {2016})}\BibitemShut {NoStop}%
\bibitem [{\citenamefont {Salomonson}\ and\ \citenamefont
  {\"{O}ster}(1989)}]{SaOe1989}%
  \BibitemOpen
  \bibfield  {author} {\bibinfo {author} {\bibfnamefont {S.}~\bibnamefont
  {Salomonson}}\ and\ \bibinfo {author} {\bibfnamefont {P.}~\bibnamefont
  {\"{O}ster}},\ }\bibfield  {title} {\enquote {\bibinfo {title} {Solution of
  the pair equation using a finite discrete spectrum},}\ }\href@noop {}
  {\bibfield  {journal} {\bibinfo  {journal} {Phys. Rev. A}\ }\textbf {\bibinfo
  {volume} {40}},\ \bibinfo {pages} {5559--5567} (\bibinfo {year}
  {1989})}\BibitemShut {NoStop}%
\bibitem [{\citenamefont {Ko\l{}os}(1967)}]{Ko1967}%
  \BibitemOpen
  \bibfield  {author} {\bibinfo {author} {\bibfnamefont {W.}~\bibnamefont
  {Ko\l{}os}},\ }\bibfield  {title} {\enquote {\bibinfo {title}
  {\relax{Long-range interaction between $1S$ and $2S$ or $2p$ hydrogen
  atoms}},}\ }\href@noop {} {\bibfield  {journal} {\bibinfo  {journal} {Int. J.
  Quantum Chem.}\ }\textbf {\bibinfo {volume} {1}},\ \bibinfo {pages}
  {169--186} (\bibinfo {year} {1967})}\BibitemShut {NoStop}%
\bibitem [{\citenamefont {Casimir}\ and\ \citenamefont
  {Polder}(1948)}]{CaPo1948}%
  \BibitemOpen
  \bibfield  {author} {\bibinfo {author} {\bibfnamefont {H.~B.~G.}\
  \bibnamefont {Casimir}}\ and\ \bibinfo {author} {\bibfnamefont
  {D.}~\bibnamefont {Polder}},\ }\bibfield  {title} {\enquote {\bibinfo {title}
  {\relax{The Influence of Radiation on the London-van-der-Waals Forces}},}\
  }\href@noop {} {\bibfield  {journal} {\bibinfo  {journal} {Phys. Rev.}\
  }\textbf {\bibinfo {volume} {73}},\ \bibinfo {pages} {360--372} (\bibinfo
  {year} {1948})}\BibitemShut {NoStop}%
\bibitem [{\citenamefont {Jentschura}(2015)}]{Je2015rapid}%
  \BibitemOpen
  \bibfield  {author} {\bibinfo {author} {\bibfnamefont {U.~D.}\ \bibnamefont
  {Jentschura}},\ }\bibfield  {title} {\enquote {\bibinfo {title}
  {\relax{Long-range atom-wall interactions and mixing terms: Metastable
  hydrogen}},}\ }\href@noop {} {\bibfield  {journal} {\bibinfo  {journal}
  {Phys. Rev. A}\ }\textbf {\bibinfo {volume} {91}},\ \bibinfo {pages}
  {010502(R)} (\bibinfo {year} {2015})}\BibitemShut {NoStop}%
\bibitem [{\citenamefont {Jentschura}(2003)}]{Je2003jpa}%
  \BibitemOpen
  \bibfield  {author} {\bibinfo {author} {\bibfnamefont {U.~D.}\ \bibnamefont
  {Jentschura}},\ }\bibfield  {title} {\enquote {\bibinfo {title}
  {\relax{Corrections to Bethe Logarithms induced by Local Potentials}},}\
  }\href@noop {} {\bibfield  {journal} {\bibinfo  {journal} {J. Phys. A}\
  }\textbf {\bibinfo {volume} {36}},\ \bibinfo {pages} {L229} (\bibinfo {year}
  {2003})}\BibitemShut {NoStop}%
\bibitem [{\citenamefont {Essen}\ \emph {et~al.}(1971)\citenamefont {Essen},
  \citenamefont {Donaldson}, \citenamefont {Bangham},\ and\ \citenamefont
  {Hope}}]{EsDoBaHo1971}%
  \BibitemOpen
  \bibfield  {author} {\bibinfo {author} {\bibfnamefont {L.}~\bibnamefont
  {Essen}}, \bibinfo {author} {\bibfnamefont {R.~W.}\ \bibnamefont
  {Donaldson}}, \bibinfo {author} {\bibfnamefont {M.~J.}\ \bibnamefont
  {Bangham}}, \ and\ \bibinfo {author} {\bibfnamefont {E.~G.}\ \bibnamefont
  {Hope}},\ }\bibfield  {title} {\enquote {\bibinfo {title} {\relax{Frequency
  of the Hydrogen Maser}},}\ }\href@noop {} {\bibfield  {journal} {\bibinfo
  {journal} {Nature (London)}\ }\textbf {\bibinfo {volume} {229}},\ \bibinfo
  {pages} {110} (\bibinfo {year} {1971})}\BibitemShut {NoStop}%
\bibitem [{\citenamefont {Essen}\ \emph {et~al.}(1973)\citenamefont {Essen},
  \citenamefont {Donaldson}, \citenamefont {Hope},\ and\ \citenamefont
  {Bangham}}]{EsDoHoBa1973}%
  \BibitemOpen
  \bibfield  {author} {\bibinfo {author} {\bibfnamefont {L.}~\bibnamefont
  {Essen}}, \bibinfo {author} {\bibfnamefont {R.~W.}\ \bibnamefont
  {Donaldson}}, \bibinfo {author} {\bibfnamefont {E.~G.}\ \bibnamefont {Hope}},
  \ and\ \bibinfo {author} {\bibfnamefont {M.~J.}\ \bibnamefont {Bangham}},\
  }\bibfield  {title} {\enquote {\bibinfo {title} {\relax{Hydrogen Maser Work
  at the National Physical Laboratory}},}\ }\href@noop {} {\bibfield  {journal}
  {\bibinfo  {journal} {Metrologia}\ }\textbf {\bibinfo {volume} {9}},\
  \bibinfo {pages} {128} (\bibinfo {year} {1973})}\BibitemShut {NoStop}%
\bibitem [{\citenamefont {Petit}\ \emph {et~al.}(1980)\citenamefont {Petit},
  \citenamefont {Descaintfuscien},\ and\ \citenamefont {Audoin}}]{PeDeAu1980}%
  \BibitemOpen
  \bibfield  {author} {\bibinfo {author} {\bibfnamefont {P.}~\bibnamefont
  {Petit}}, \bibinfo {author} {\bibfnamefont {M.}~\bibnamefont
  {Descaintfuscien}}, \ and\ \bibinfo {author} {\bibfnamefont {C.}~\bibnamefont
  {Audoin}},\ }\bibfield  {title} {\enquote {\bibinfo {title}
  {\relax{Temperature Dependence of the Hydrogen Maser Wall Shift in the
  Temperature Range 295--395K}},}\ }\href@noop {} {\bibfield  {journal}
  {\bibinfo  {journal} {Metrologia}\ }\textbf {\bibinfo {volume} {16}},\
  \bibinfo {pages} {7--14} (\bibinfo {year} {1980})}\BibitemShut {NoStop}%
\bibitem [{\citenamefont {Varshalovich}\ \emph {et~al.}(1988)\citenamefont
  {Varshalovich}, \citenamefont {Moskalev},\ and\ \citenamefont
  {Khersonskii}}]{VaMoKh1988}%
  \BibitemOpen
  \bibfield  {author} {\bibinfo {author} {\bibfnamefont {D.~A.}\ \bibnamefont
  {Varshalovich}}, \bibinfo {author} {\bibfnamefont {A.~N.}\ \bibnamefont
  {Moskalev}}, \ and\ \bibinfo {author} {\bibfnamefont {V.~K.}\ \bibnamefont
  {Khersonskii}},\ }\href@noop {} {\emph {\bibinfo {title} {\relax{Quantum
  Theory of Angular Momentum}}}}\ (\bibinfo  {publisher} {World Scientific},\
  \bibinfo {address} {Singapore},\ \bibinfo {year} {1988})\BibitemShut
  {NoStop}%
\bibitem [{\citenamefont {Parthey}\ \emph {et~al.}(2011)\citenamefont
  {Parthey}, \citenamefont {Matveev}, \citenamefont {Alnis}, \citenamefont
  {Bernhardt}, \citenamefont {Beyer}, \citenamefont {Holzwarth}, \citenamefont
  {Maistrou}, \citenamefont {Pohl}, \citenamefont {Predehl}, \citenamefont
  {Udem}, \citenamefont {Wilken}, \citenamefont {Kolachevsky}, \citenamefont
  {Abgrall}, \citenamefont {Rovera}, \citenamefont {Salomon}, \citenamefont
  {Laurent},\ and\ \citenamefont {H\"{a}nsch}}]{PaEtAl2011}%
  \BibitemOpen
  \bibfield  {author} {\bibinfo {author} {\bibfnamefont {C.~G.}\ \bibnamefont
  {Parthey}}, \bibinfo {author} {\bibfnamefont {A.}~\bibnamefont {Matveev}},
  \bibinfo {author} {\bibfnamefont {J.}~\bibnamefont {Alnis}}, \bibinfo
  {author} {\bibfnamefont {B.}~\bibnamefont {Bernhardt}}, \bibinfo {author}
  {\bibfnamefont {A.}~\bibnamefont {Beyer}}, \bibinfo {author} {\bibfnamefont
  {R.}~\bibnamefont {Holzwarth}}, \bibinfo {author} {\bibfnamefont
  {A.}~\bibnamefont {Maistrou}}, \bibinfo {author} {\bibfnamefont
  {R.}~\bibnamefont {Pohl}}, \bibinfo {author} {\bibfnamefont {K.}~\bibnamefont
  {Predehl}}, \bibinfo {author} {\bibfnamefont {T.}~\bibnamefont {Udem}},
  \bibinfo {author} {\bibfnamefont {T.}~\bibnamefont {Wilken}}, \bibinfo
  {author} {\bibfnamefont {N.}~\bibnamefont {Kolachevsky}}, \bibinfo {author}
  {\bibfnamefont {M.}~\bibnamefont {Abgrall}}, \bibinfo {author} {\bibfnamefont
  {D.}~\bibnamefont {Rovera}}, \bibinfo {author} {\bibfnamefont
  {C.}~\bibnamefont {Salomon}}, \bibinfo {author} {\bibfnamefont
  {P.}~\bibnamefont {Laurent}}, \ and\ \bibinfo {author} {\bibfnamefont
  {T.~W.}\ \bibnamefont {H\"{a}nsch}},\ }\bibfield  {title} {\enquote {\bibinfo
  {title} {\relax{Improved Measurement of the Hydrogen 1S--2S Transition
  Frequency}},}\ }\href@noop {} {\bibfield  {journal} {\bibinfo  {journal}
  {Phys. Rev. Lett.}\ }\textbf {\bibinfo {volume} {107}},\ \bibinfo {pages}
  {203001} (\bibinfo {year} {2011})}\BibitemShut {NoStop}%
\bibitem [{\citenamefont {Fee}\ \emph {et~al.}(1993)\citenamefont {Fee},
  \citenamefont {Mills}, \citenamefont {Chu}, \citenamefont {Shaw},
  \citenamefont {Danzmann}, \citenamefont {Chichester},\ and\ \citenamefont
  {Zuckerman}}]{FeEtAl1993}%
  \BibitemOpen
  \bibfield  {author} {\bibinfo {author} {\bibfnamefont {M.~S.}\ \bibnamefont
  {Fee}}, \bibinfo {author} {\bibfnamefont {A.~P.}\ \bibnamefont {Mills}},
  \bibinfo {author} {\bibfnamefont {S.}~\bibnamefont {Chu}}, \bibinfo {author}
  {\bibfnamefont {E.~D.}\ \bibnamefont {Shaw}}, \bibinfo {author}
  {\bibfnamefont {K.}~\bibnamefont {Danzmann}}, \bibinfo {author}
  {\bibfnamefont {R.~J.}\ \bibnamefont {Chichester}}, \ and\ \bibinfo {author}
  {\bibfnamefont {D.~M.}\ \bibnamefont {Zuckerman}},\ }\bibfield  {title}
  {\enquote {\bibinfo {title} {\relax{Measurement of the positronium $1{}^3
  S_1$--$2{}^3S_1$ interval by continuous-wave two-photon excitation}},}\
  }\href@noop {} {\bibfield  {journal} {\bibinfo  {journal} {Phys. Rev. Lett.}\
  }\textbf {\bibinfo {volume} {70}},\ \bibinfo {pages} {1397--1400} (\bibinfo
  {year} {1993})}\BibitemShut {NoStop}%
\end{thebibliography}
\end{document}